\documentclass[twoside]{article}

\usepackage{color}
\usepackage{amssymb}
\usepackage{dsfont}
\usepackage{amssymb,amsmath,ulem,cancel}

\numberwithin{equation}{section}

\newtheorem{theorem}{Theorem}[section]
\newtheorem{lem}{Lemma}[section]
\newtheorem{pro}{Proposition}[section]
\newtheorem{cor}{Corollary}[section]
\newtheorem{rem}{Remark}[section]
\newtheorem{rems}{Remarks}[section]
\newtheorem{ex}{Example}[section]
\newtheorem{defi}{Definition}[section]
\newtheorem{hyp}{Assumption}[section]
\newtheorem{con}{Conjecture}[section]

\newcommand{\ssc}{\subsection}
\newcommand{\sssc}{\subsubsection}

\newcommand{\bt}{\begin{theorem}}
\newcommand{\et}{\end{theorem}}
\newcommand{\bl}{\begin{lem}}
\newcommand{\el}{\end{lem}}
\newcommand{\bp}{\begin{pro}}
\newcommand{\ep}{\end{pro}}
\newcommand{\bcor}{\begin{cor}}
\newcommand{\ecor}{\end{cor}}
\newcommand{\bcon }{\begin{con} \rm }
\newcommand{\econ }{\end{con}}
\newcommand{\lab }{\label }
\newcommand{\bd}{\begin{defi} \rm }
\newcommand{\ed}{\end{defi}}
\newcommand{\brem }{\begin{rem} \rm }
\newcommand{\erem }{\end{rem}}
\newcommand{\brems }{\begin{rems} \rm }
\newcommand{\erems }{\end{rems}}
\newcommand{\bhyp }{\begin{hyp} \rm }
\newcommand{\ehyp }{\end{hyp}}
\newcommand{\bex}{\begin{ex} \rm }
\newcommand{\eex}{\end{ex}}
\newcommand{\be}{\begin{equation}}
\newcommand{\ee}{\end{equation}}
\newcommand{\bde}{\begin{displaymath}}
\newcommand{\ede}{\end{displaymath}}
\newcommand{\beq}{\begin{eqnarray*}}
\newcommand{\eeq}{\end{eqnarray*}}
\newcommand{\beqa}{\begin{eqnarray}}
\newcommand{\eeqa}{\end{eqnarray}}
\newcommand{\bea}{\begin{align*}}
\newcommand{\eea}{\end{align*}}

\def\proof{\noindent {\it Proof. $\, $}}
\def\endproof{\hfill $\Box$ \vskip 5 pt}
\def\xxx{x}
\def\I{\mathds{1}}
\def\wh{\widehat}
\def\wt{\widetilde}
\def\phi{\varphi }

\newcommand{\cac}{\alpha }
\newcommand{\ppff}{\bar{p}_0}

\newcommand{\dPFA}{\mbox{\sf PFA}}
\newcommand{\dCFA}{\mbox{\sf CFA}}
\newcommand{\dTFA}{\mbox{\sf TFA}}

\newcommand{\pA}{A}
\newcommand{\pC}{C}

\newcommand{\pCc}{c}

\newcommand{\etab}{\eta^b}
\newcommand{\etal}{\eta^l}

\newcommand{\VCc}{V^c}
\newcommand{\pD}{D}
\newcommand{\pDTbh}{\bar A^{c,h}}
\newcommand{\pDThh}{\widehat A^{c,h}}
\newcommand{\pDTsh}{\widehat A^{c,s}}

\newcommand{\pACh}{A^{h}}
\newcommand{\pACs}{A^{s}}
\newcommand{\pFCs}{F^{s}}

\newcommand{\pFChb}{\bar{F}^{h}}
\newcommand{\pFCh}{F^{h}}
\newcommand{\whpFCs}{\wh{F}^{s}}
\newcommand{\whpFCh}{\wh{F}^{h}}

\newcommand{\Blr}{B^{l}}
\newcommand{\Bbr}{B^{b}}

\newcommand{\Bilr}{B^{i,l}}
\newcommand{\Bibr}{B^{i,b}}

\newcommand{\rll}{r^{l}}
\newcommand{\rbb}{r^{b}}

\newcommand{\rill}{r^{i,l}}
\newcommand{\ribb}{r^{i,b}}

\newcommand{\Vnet}{{V}^{\textrm{net}}}

\newcommand{\Vnett}{{\widetilde{V}}^{\textrm{net}}}
\newcommand{\Vnettl}{{\widetilde{V}}^{l,\textrm{net}}}
\newcommand{\Vnettb}{{\widetilde{V}}^{b,\textrm{net}}}

\def\t1{\tau_{(1)}}

\def\rr{\mathbb R}
\def\ff{{\mathbb F}}

\def\gg{{\mathbb G}}

\def\F{{\cal F}}

\def\G{{\cal G}}

\def\VLL{V^0}
\def\P{\mathbb P}

\def\PH{\wh {\mathbb P}}
\def\PT{\wt {\mathbb P}}

\def\Q{\mathbb Q}

\def\E{{\mathbb E}}

\def\EPT{{\mathbb E}_{\widetilde{\mathbb  P}}}
\def\EPG{{\mathbb E}_{\mathbb  P^{\gamma }}}

\def\FCVA{{\rm FCVA}}

\def\C-FVA{{\rm C-FVA}}

\newcommand{\Keywords}[1]{\par\noindent{\small{\bf Keywords\/}: #1}}
\newcommand{\Class}[1]{\par\noindent{\small{\bf Mathematics Subjects Classification (2010)\/}: #1}}

\title{{\Large \bf VALUATION AND HEDGING OF CONTRACTS WITH
\\ \vskip 1 pt FUNDING COSTS AND COLLATERALIZATION}\vskip 50 pt}

\author{Tomasz R. Bielecki\footnote{The research of Tomasz R. Bielecki was supported under NSF grant DMS-1211256.}
\\ Department of Applied Mathematics \\
 Illinois Institute of Technology \\
 Chicago, IL 60616, USA \\ \\
Marek Rutkowski\footnote{The research of Marek Rutkowski was supported under Australian Research Council's Discovery Projects funding scheme (DP120100895).}
\\ School of Mathematics and Statistics \\ University of Sydney
\\ Sydney, NSW 2006, Australia}

\date{\vskip 25 pt Version of 1 December 2014}

\begin{document}
\maketitle
\vskip 20 pt
\begin{abstract}
The research presented in this work is motivated by recent papers by Brigo et al. \cite{BCP12,BCPP11},
 Burgard and Kjaer \cite{BK,BK11a,BK13}, Cr\'epey \cite{SC1,SC2}, Fujii and Takahashi \cite{FT}, Piterbarg \cite{PV10}
and Pallavicini et al. \cite{PPB12}. Our goal is to provide a sound theoretical underpinning for some results presented in these papers by developing a unified framework for the non-linear approach to hedging and pricing of OTC financial contracts.
The impact that various funding bases and margin covenants exert on the values and hedging strategies for OTC contracts is examined. The relationships between our research and papers by other authors, with an exception of
Pallavicini et al. \cite{PPB12} and Piterbarg \cite{PV10}, are not discussed in this part of our research. More detailed studies of these relationships, as well as the issue of the counterparty credit risk, are examined in the follow-up paper.
\vskip 20 pt
\Keywords{hedging, funding costs, counterparty risk, margin agreement}
\vskip 20 pt
\Class{91G40,$\,$60J28}
\end{abstract}

\newpage
$\,$ \thispagestyle{empty}
\newpage

 \tableofcontents

\newpage

%%%%%%%%%%%%%%%%%%%%%%%%%%%%%%%%     SECTION 1     %%%%%%%%%%%%%%%%%%%%%%%%%%%%%%%%%%%%%%
%%%%%%%%%%%%%%%%%%%%%%%%%%%%%%%%%%%%%%%%%%%%%%%%%%%%%%%%%%%%%%%%%%%%%%%%%%%%%%%%%%%%%%%%%
%%%%%%%%%%%%%%%%%%%%%%%%%%%%%%%%%%%%%%%%%%%%%%%%%%%%%%%%%%%%%%%%%%%%%%%%%%%%%%%%%%%%%%%%%
\section{Introduction}
%%%%%%%%%%%%%%%%%%%%%%%%%%%%%%%%%%%%%%%%%%%%%%%%%%%%%%%%%%%%%%%%%%%%%%%%%%%%%%%%%%%%%%%%%
%%%%%%%%%%%%%%%%%%%%%%%%%%%%%%%%%%%%%%%%%%%%%%%%%%%%%%%%%%%%%%%%%%%%%%%%%%%%%%%%%%%%%%%%%
%%%%%%%%%%%%%%%%%%%%%%%%%%%%%%%%%%%%%%%%%%%%%%%%%%%%%%%%%%%%%%%%%%%%%%%%%%%%%%%%%%%%%%%%%

Let us consider a vanilla CDS contract with cumulative cash flow process, seen from the perspective of a protection buyer -- the hedger, given as $\pD =\pD^1- \pD^2$, where we set
\bde
\pD^1_t=(1-R)\I_{\{t\geq \tau\}}, \quad \pD^2_t= \kappa (t\wedge \tau), \quad t\in[0,T],
\ede
and where we use the pre Big-Bang convention of zero strike (see \cite{Markit}).
This is thus a payer CDS maturing at time $T$, with contractual spread $\kappa$, and referencing a given obligor, say $O$, whose default time is denoted by $\tau.$ By a \textit{vanilla} CDS contract we mean, in particular, that the counterparty risk and collateral are ignored in the cumulative cash flow process $D$ given above. The classical way to discount the above cash flows was to apply the same discount factor, say $\beta^0=(B^0)^{-1}$, where $B^0$ was the unique (and thus the same for all market participants), locally risk-free, money market (cash) account. Thus, the discounted cumulative cash flows would take the form
\bde
\widetilde \pD_t=\int _{(0,t]}\beta^0_u\, d\pD_u, \quad t\in [0,T].
\ede
This choice of discounting was consistent with the classical approach to hedging the position in this CDS contract, which hinged on creation of a self-financing {\it trading strategy}, say $\phi  = (\xi^1,\dots ,\xi^d, \psi^0)$, with the corresponding wealth process
\bde  % \lab{rtf01}
V_t (\phi) =  \sum_{i=1}^d \xi^i_tS^i_t + \psi^0_t B^0_t
\ede
where $S^1, \dots , S^d$ are some relevant traded assets, such as relevant CDS options or relevant equity and/or equity options. In particular, this meant that all trades of the hedger were fully funded by the same money market account, denoted here as $B^0$.
Furthermore, it was assumed that both parties had access to the same traded risky assets, money market account, and market information. Since, in such a classical setup, the discounted cash flows were symmetric from the perspective of two parties in the contract, that is, the discounted cash flows, as seen from the perspective of one party, were the negative of the discounted cash flows as seen from the perspective of the other party, the hedging and pricing exercise was symmetric in an analogous way.
Alas, things are not vanilla anymore. In particular: \hfill \break
$\bullet \, $ contracts now tend to be collateralized,\hfill \break
$\bullet \, $ parties may need to account for different funding rates, \hfill \break
$\bullet \, $ counterparty and systemic risks need to be accounted for, \hfill \break
$\bullet \, $ netting of portfolio positions becomes an important issue.

Consider, for example, the unilateral counterparty risk of the protection seller defaulting. The cumulative cash flows of a CDS contract would now become (see, for instance, Bielecki et al.  \cite{BCJZ10})
\be \label{flowki}
{\cal D}_t = \I_{\{t<\tau^{c} \}} \pD_t +\I_{\{t\geq \tau^{c} \}}\pD_{\tau^c-}
+  \I_{\{\tau^c\leq T\} }( \pC_{\tau^c}+R^c\chi^+-\chi^-), \quad t\in [0,T],
\ee
where:\hfill \break
$\bullet \, $ $\tau^c$ is the default time of the protection seller (the counterparty to the hedger in this contract),
\hfill \break $\bullet \, $ $R^c$ is the recovery rate upon default of the protection seller,
\hfill \break $\bullet \, $ ${\cal P}_{\tau^c}$ is the replacement value of the CDS contract upon default of the protection seller,
\hfill \break $\bullet \, $ $C$ is the collateral process, which may depend on the strategy that the hedger uses to dynamically hedge his position in the CDS contract,
\hfill \break $\bullet \, $ $\chi^+=\max(0,\chi)$ and $\chi^-=\max(0,-\chi)$ where
$\chi={\cal P}_{\tau^c}+\I_{\{\tau=\tau^c\}}(1-R)- \pC_{\tau^c}$.

In practice, separate parts of these cash flows would now typically be discounted by different rates that, in particular, may depend on the currency in which funding is being executed. Formally, a hedging portfolio would now refer to multiple funding accounts, denoted hereafter as $B^1,\dots, B^d$. Furthermore, the discounted cash flows (and thus also prices) will typically be asymmetric relative to the parties in the contract since their funding sources are no longer assumed to be identical.

 Accordingly, the current way to hedge the position in a CDS contract would be to create a  {\it trading strategy}, say $\phi = (\xi,\psi) = (\xi^1,\dots ,\xi^d, \psi^0,\dots ,\psi^d)$  composed of risky securities $S^i,\, i=1,2,\ldots,d$, the cash account $B^0$ used for unsecured lending/borrowing,  and funding accounts $B^i,\, i=1,2,\ldots,d,$ used for (unsecured or secured) funding of the $i$th asset, with the corresponding wealth process
\be  \lab{rtf02}
V_t (\phi) :=  \sum_{i=1}^d \xi^i_tS^i_t +  \sum_{j=0}^d \psi^j_t B^j_t.
\ee
In fact, a trading strategy represented by  \eqref{rtf02} is merely a special case of more general portfolios examined in this paper. In particular, the hedger needs also to account for various netting possibilities of short/long positions in the assets comprising the hedging portfolio. Hence the classical form of the {\it self-financing} condition no longer holds and thus one needs to analyze a suitably modified version of this condition. Moreover, the collateral posted by a party as part of the cash flows of the contract, may depend on the hedging strategy employed by this party directly, or indirectly -- through the wealth process of the hedging portfolio. This feature, in particular, makes the contract cash flows asymmetric relative to the two parties in the contract; it also makes valuation and hedging problem implicit and non-linear. For an extensive discussion and study of this aspect of valuation and hedging in the context of valuation and hedging of counterparty risk, the reader may consult Bielecki et al. \cite{BCJZ10} and the monograph by Cr\'epey et al. \cite{CBB}.

In view of the above mentioned complexities, the problem of marking to market and hedging the CDS contract in the current market environment is not longer as straightforward as it was the case in the past. Yet, as we shall argue, both the classic and novel approaches are rooted in the same principles of self-financing trading and no arbitrage, appropriately adapted to ways in which cash flows are now modified and the ways in which hedging portfolios are now formed. The aim of this work is thus to provide a framework for systematic analysis of the presence/absence of arbitrage, as well as a systematic analysis of hedging and valuation with regard to OTC contracts, whose cash flows account for additional features analogous to the additional features that modify the cash flows of the CDS contract as in \eqref{flowki} above, and the hedging portfolios $\varphi$ with wealth processes given by either \eqref{rtf02} or its suitable extension. Note that that collateral process may also be present in the security cash flows even if counterparty risk is not explicitly accounted for. Accordingly, the goals of this paper are:
\hfill \break $\bullet \, $ to provide a blueprint for derivation of dynamics of the wealth process corresponding to self-financing trading strategy and to examine such dynamics under various trading covenants,
 % in particular, we identify the funding and liquidity adjustments as components of these dynamics,
\hfill \break $\bullet \, $  to introduce and discuss the relevant concepts of arbitrage and no-arbitrage valuation,
\hfill \break $\bullet \, $  to highlight the so-called additive martingale property (see Remark \ref{amp}) and its role in non-linear and implicit pricing via BSDEs,
\hfill \break $\bullet \, $  to examine how our abstract model-free framework relates to some previous works, specifically, the papers by Pallavicini et al. \cite{PPB12} and Piterbarg \cite{PV10}.

 For other related work, the interested reader may also consult Bianchetti \cite{Bia10}, Brigo et al. \cite{BBM12,BCP12,BCPP11}, Burgard and Kjaer \cite{BK,BK11a,BK11b,BK13}, Castagna \cite{Cast1,Cast2}, Cr\'epey \cite{SC1,SC2}, Fujii and Takahashi \cite{FT}, Fujii et al.  \cite{FST10}, Henrard \cite{Hen13}, Hull and White \cite{HW12a,HW12b}, Kenyon \cite{KC10a,KC10b}, Kijima et al. \cite{KTW08} and Mercurio \cite{Mer10,Mer12,Mer13}. Since we are not specifically concerned here with the analysis of counterparty risky cash flows, we do not study in detail the related adjustments, such as the counterparty risk adjustment. We argue that this adjustment should also naturally emerge as a part of the dynamics of the relevant wealth process, in analogy to the funding adjustment and the liquidity adjustment. Therefore, we make an attempt to identify funding, liquidity, and  counterparty risk adjustments with relevant parts of the wealth dynamics of the hedging portfolio. By contrast, we are not interested in valuation and hedging of these adjustments in separation from valuation and hedging of the entire contract, although, once portions of the dynamics of the wealth process  corresponding to the adjustments are identified, then the adjustments can, in principle, be valued, leading to respective (deal) valuation adjustments: funding valuation adjustment (FVA), liquidity valuation adjustment (LVA), credit valuation adjustment (CVA), etc..  In practice, such valuation adjustments are typically done by successive adjustments to tentative  ``risk-neutral prices'' of the cash flows $\pD$, but this is not what we have set forth in the present paper.

\newpage

In this regard, let us mention that in the banks' practice the counterparty risky OTC contract is valued and hedged in at least two pieces. Typically, the counterparty risk clean and uncollateralized part is valued and hedged by the trading desk, whereas the funding and credit valuation adjustment (FCVA) is computed and hedged by the CVA desk. Accordingly, the price of the contract is then taken as the sum of the FCVA and the price of the counterparty risk clean and uncollateralized part of the deal. This
leads to the following price decomposition
\be\label{cacy0}
 \wh \pi    = \pi  + \FCVA + \textrm{additional adjustments (if needed)}
\ee
where $\pi $ is the fair value of the uncollateralized contract between non-defaultable counterparties and $\wh \pi $ is the `value' for the investor of the contract between two defaultable parties with idiosyncratic funding costs, collateral, and other relevant costs and/or risks. The abstract methodology presented will hopefully shed light on the arbitrage aspects of valuation and hedging of the entire OTC contract versus separate valuation and hedging of such two its components.

In conclusion, this paper contributes to the existing literature in at least the following ways:
\hfill \break $\bullet \, $ We introduce a systematic approach to valuation and hedging in ``nonlinear markets,'' that is, in markets where cash flows of the financial contracts (may) depend on the hedging strategies.
\hfill \break $\bullet \, $  Our systematic approach allows to identify primary sources of and quantify various adjustment to valuation and hedging, primarily the funding and liquidity adjustment and credit risk adjustment.
\hfill \break $\bullet \, $ We propose a way to define no-arbitrage in such ``nonlinear markets,'' and we provide conditions that imply absence of arbitrage in some specific market trading models.
 \hfill \break $\bullet \, $ Accordingly, we formulate a concept of no-arbitrage price, and we provide relevant (non-linear) BSDE that produces the no-arbitrage price in case when the cash flows can be replicated.

The paper is organized as follows. In Section \ref{mm}, we start by introducing a generic market model with several risky assets and multiple funding accounts. We then derive alternative representations for the dynamics of the wealth process of a self-financing trading strategy for a given process representing all cash flows of a contract. We first solve this problem in the
basic model and subsequently extend to more advanced models with various forms of netting.

In Section \ref{Sec2}, we introduce the concept of an arbitrage-free model, by proposing an essential extension
of the classic definition, and we provide sufficient conditions for the no-arbitrage property of a market model under alternative assumptions about trading and netting. Surprisingly, this crucial issue was completely neglected in most existing papers that dealt with funding costs and collateralization. The authors focused instead on the `risk-neutral valuation' under
a vaguely specified martingale measure, which was assumed a priori to exist.
By contrast, we propose a precise definition of the hedger's price via either replication or a suitable form of super-hedging.
Moreover, we show that the problem of arbitrage under funding costs is non-trivial, but it can indeed be dealt with using a judicious definition of arbitrage opportunities and a specific form of a martingale measure.

As was already mentioned, collateralization of contracts became a widespread market practice. For this reason, we
examine in Section \ref{seccoll} various conventions regarding margin account and we study the impact of collateralization
on the dynamics of the hedger's portfolio. In our stylized approach to costs of margining, we consider both
the case of segregated margin accounts and the case of rehypothecation. Moreover, we acknowledge that collateral posted or received is either in the form of cash or shares of a risky asset.

In Section \ref{Sec21}, we deal with the fair pricing under funding costs and collateralization first in an abstract
setup and then for a generic diffusion-type model.
Let us stress that the pricing functional for the hedger will be typically non-linear,
since hedging strategies are typically non-additive when a collection of contracts, rather than single deal, is studied.
For instance, in the case of a market model with partial netting, the pricing problem can be represented
in terms of a non-linear BSDE, which is shown to admit a unique solution under mild assumptions on the underlying model.
For further results in this vein, the interested reader is referred to Nie and Rutkowski \cite{NR2,NR4,NR5,NR3}.
To put our framework into perspective, we also analyze valuation methods proposed by
Piterbarg \cite{PV10} and Pallavicini et al. \cite{PPB12}. It appears that our approach covers
as a special case the pricing results established in \cite{PV10}. By contrast, we argue that the method developed in \cite{PPB12} is somewhat opaque and it is not fully consistent with our approach.

\newpage

%%%%%%%%%%%%%%%%%%%%%%%%%%%%%%%%     SECTION 2     %%%%%%%%%%%%%%%%%%%%%%%%%%%%%%%%%%%%%%
%%%%%%%%%%%%%%%%%%%%%%%%%%%%%%%%%%%%%%%%%%%%%%%%%%%%%%%%%%%%%%%%%%%%%%%%%%%%%%%%%%%%%%%%%
%%%%%%%%%%%%%%%%%%%%%%%%%%%%%%%%%%%%%%%%%%%%%%%%%%%%%%%%%%%%%%%%%%%%%%%%%%%%%%%%%%%%%%%%%
\section{Trading under Funding Costs}\label{mm}
%%%%%%%%%%%%%%%%%%%%%%%%%%%%%%%%%%%%%%%%%%%%%%%%%%%%%%%%%%%%%%%%%%%%%%%%%%%%%%%%%%%%%%%%%
%%%%%%%%%%%%%%%%%%%%%%%%%%%%%%%%%%%%%%%%%%%%%%%%%%%%%%%%%%%%%%%%%%%%%%%%%%%%%%%%%%%%%%%%%
%%%%%%%%%%%%%%%%%%%%%%%%%%%%%%%%%%%%%%%%%%%%%%%%%%%%%%%%%%%%%%%%%%%%%%%%%%%%%%%%%%%%%%%%%

Let us first introduce the notation for market models considered in this work.

\noindent {\bf Probability space.} Throughout the paper, we fix a finite trading horizon $T$ for our model of the financial market. All processes introduced in what follows are implicitly assumed to be $\gg$-adapted and defined on the underlying probability space $(\Omega, \G, \gg , \P)$ where the filtration $\gg = (\G_t)_{t \in [0,T]}$ models the flow of information available to all traders (in particular, any semimartingale is assumed to be c\`adl\`ag). For convenience, we assume that the initial $\sigma$-field
${\cal G}_0$ is trivial, although this assumption can be easily relaxed.

\noindent {\bf Risky assets.} We denote by $S^i$ the {\it ex-dividend price} (or simply the {\it price}) of the $i$th risky asset with the {\it cumulative dividend stream} after time 0 represented by the process $\pA^i$. Note that we do not postulate that
processes $S^i,\, i=1,2, \dots, d$ are positive. Hence by the {\it long cash position} (resp. {\it short cash position}), we mean
the situation when $\xi^i_t S^i_t \leq 0$ (resp.  $\xi^i_t S^i_t \geq 0$) where $\xi^i_t$ is the number of hedger's positions
in asset $S^i$ at time $t$.

\noindent {\bf Funding accounts.}  The {\it cash account} $B^0 = B$ is used for unsecured lending or borrowing of cash. In the case when the borrowing and lending cash rates are different, we use symbols $\Blr$ (resp. $\Bbr$)
to denote the process modeling the unsecured {\it lending} (resp. {\it borrowing}) cash account. Notation $B^i$ stands for the  {\it funding account}, which may represent either unsecured or secured funding for the $i$th risky asset.
A similar convention applies to this account: in case when borrowing/lending rates differ, we use symbols $\Bilr$ and $\Bibr$ to denote lending/borrowing accounts associated with the $i$th risky asset.
As a general rule, we will assume the position of the hedger. Hence the superscripts $l$ (resp. $b$) will refer to rates
applied to deposits (resp. loans) from the viewpoint of the hedger. Observe that funding accounts are sometimes
referred to as {\it non-risky assets.}  Unless explicitly stated otherwise, we assume that
$\Blr=\Bbr =B$ and $\Bilr=\Bibr= B^i$ for all $i$.

A more detailed mathematical and financial interpretation of funding accounts will be presented in what follows. Let us only mention
here that $S^i$ is aimed to represent the price of any traded security, such as, stock, stock option, interest rates swap, currency option, cross-currency swap, CDS, CDO, etc.  In essence, the rate $r^{i,l}$ (resp. $r^{i,b}$) corresponding to the {\it lending} (resp. {\it borrowing}) account $\Bilr$ (resp. $\Bibr$) represents the incremental cost of maintaining the long cash position (resp. short cash position) in asset $S^i$  (for a more precise interpretation of this statement, see Remark \ref{amp}). Hence the actual interpretation of `borrowing' and `lending' accounts $\Bilr$ and $\Bibr$ will depend on a contract at hand and the relevant features of financial environment. In particular, the rates denoted here as $r^{i,l}$ and $r^{i,b}$ may in turn depend on multiple yield curves in several economies and/or other funding arrangements of a particular party (for instance, the
hedger's internal funding costs).

\bhyp \lab{asss2.1}
It is assumed throughout  that the price processes of {\it primary assets} satisfy: \hfill \break
(i) ex-dividend prices $S^i$ for $i=1,2,\dots , d$ are semimartingales, \hfill \break
(ii) cumulative dividend streams $\pA^i$ for $i=1,2,\dots , d$ are processes of finite variation with $\pA^i_{0}=0$, \hfill \break
(iii) funding accounts $B^j$ for $j=0,1,\dots , d$ are strictly positive and continuous processes of finite variation with $B^j_0=1$.
\ehyp

\bd
The {\it cumulative dividend price} $S^{i,\textrm{cld}}$ is given as
\be \lab{pri1}
S^{i,\textrm{cld}}_t := S^i_t+ B^{i}_t\int_{(0,t]} (B^i_u)^{-1} \, d\pA^i_u ,\quad t\in [0,T],
\ee
and thus the {\it discounted cumulative dividend price}  $\wh S^{i,\textrm{cld}}:=  (B^i)^{-1} S^{i,\textrm{cld}}$ satisfies
\be \lab{pri2}
\wh S^{i,\textrm{cld}}_t = \wh S^i_t+\int_{(0,t]} (B^i_u)^{-1} \, d\pA^i_u ,\quad t\in [0,T] ,
\ee
where we denote $\wh S^i :=  (B^i)^{-1} S^i$.
\ed

If the $i$th traded asset does not pay any dividend up to time $T$, then the equality $ S^{i,\textrm{cld}}_t = S^{i}_t$ holds for every $t \in [0,T]$. Note that the process $S^{i,\textrm{cld}}$ (hence also the process $\wh S^{i,\textrm{cld}}$) is c\`adl\`ag.

%\brem
%Under the convention that $S^i$ is the cum-dividend price the interval $(0,t]$ should be replaced by $[0,t)$
%in equalities (\ref{pri1}) and (\ref{pri2}).
%\erem

\brem
Note that formula \eqref{pri1} hinges on an implicit assumption that positive (resp. negative) dividends from the $i$th asset are invested in (resp. funded from) the $i$th funding account $B^i$. Since the main valuation and hedging results for derivative securities obtained in this section are represented in terms of primitive processes $S^i,B^i$ and $\pA^i$, rather than $S^{i,\textrm{cld}}$, the choice of a particular convention regarding reinvestment of dividends associated with the
$i$th risky asset is in fact immaterial. The implicit choice made in equation \eqref{pri1} was motivated by the mathematical convenience only.
\erem

\brem
We adopt the following notational conventions: \\
(i) for any random variable $\chi $, the equality $ \chi = \chi^+ - \chi^-$
is the unique decomposition of $\chi$ into its positive and negative parts, \\
(ii) for any stochastic process $A$ of finite variation, the equality $A = A^+ - A^-$ represents the
unique decomposition of $A$ where $A^+$ and $A^-$ are increasing processes with $A_0 = A^+_0 - A^-_0$.
\erem

%%%%%%%%%%%%%%%%%%%%%%%%%%%%%%%%%%%%%%%%%%%%%%%%%%%%%%%%%%%%%%%%%%%%%%%%
\ssc{Contracts and Trading Strategies} \lab{secfun}
%%%%%%%%%%%%%%%%%%%%%%%%%%%%%%%%%%%%%%%%%%%%%%%%%%%%%%%%%%%%%%%%%%%%%%%%

We are in a position to introduce trading strategies based on a finite family of primary assets satisfying Assumption \ref{asss2.1}. In Sections \ref{secfun} and \ref{secelem}, we consider a dynamic portfolio denoted as $\phi = (\xi,\psi) = (\xi^1,\dots ,\xi^d, \psi^0,\dots ,\psi^d)$, which is composed of risky securities $S^i,\, i=1,2,\ldots,d$, the cash account $B$ used for unsecured lending/borrowing, and funding accounts $B^i,\, i=1,2,\ldots,d,$ used for (either unsecured or secured) funding of the $i$th asset.
Let us first formally define the class of contracts under our consideration.

\bd \lab{defcocc}
By a {\it bilateral financial contract}, or simply a {\it contract}, we mean an arbitrary c\`adl\`ag process $\pA$ of finite variation. The process $A$ is aimed to represent the {\it cumulative cash flows} of a given contract from time 0 till its maturity date $T$. By convention, we set $\pA_{0-}=0$.
\ed

The process $\pA$ is assumed to model all cash flows of a given contract, which are either paid out from the wealth or added to the wealth, as seen from the perspective of the {\it hedger} (recall that the other party is referred to as the {\it counterparty}).
Note that the process $A$ includes the initial cash flow $A_0$ of a contract at its inception date $t_0=0$.
For instance, if a contract has the initial {\it price} $p$ and stipulates that the hedger will receive cash flows  $\bar{\pA}_1,
\bar{\pA}_2, \dots , \bar{\pA}_k$ at future dates $t_1, t_2, \dots , t_k \in (0,T]$, then we set $A_0=p$ so that
\bde
\pA_t = p + \sum_{l=1}^k \I_{[t_l,T]}(t) \bar{\pA}_l .
\ede
If a unique future cash flow associated with a contract is the terminal payment at time $T$, which is
denoted as $X$, then the process $A$ for this security takes form $A_t = p \, \I_{[0,T]}(t) + X \I_{[T]}(t)$.
For instance, if the hedger sells at time 0 a European call option on the risky asset $S^i$, then
the terminal payoff equals $X = - (S^i_T-K)^+$ and thus $A_t = p \, \I_{[0,T]}(t) - (S^i_T-K)^+  \I_{[T]}(t)$.
The symbol $p$ is frequently used to emphasize that all future cash flows $\bar{\pA}_l$ for $l=1,2, \dots, k$ are explicitly specified by the contract's covenants, but the initial cash flow $A_0$ is yet to be formally defined and evaluated.
%Obviously, if the assumption that  $\sigma$-field ${\cal G}_0$ is trivial is relaxed, then $A_0$ becomes a ${\cal G}_0$-measurable %random variable, rather than a real number, as will be the case hereafter.
Valuation of a contract $A$ means, in particular, searching for the range of {\it fair values} $p$ at time $0$ from the viewpoint of either the hedger or the counterparty. Although the valuation paradigm will be the same for the two parties, due either to the asymmetry in their trading costs and opportunities, or the non-linearity of the wealth dynamics, they will typically obtain different sets of fair prices for~$A$.

By a {\it trading strategy} associated with a contract $\pA$, we mean the triplet $(x, \phi , \pA )$. The wealth process
$V(x, \phi , \pA )$ of a trading strategy depends on the {\it initial endowment} $x$ of the hedger,  represented by an arbitrary real number $x$,  his {\it hedging portfolio} $\phi $ and {\it contractual cash flows} $\pA$. Note that by the hedger's initial endowment, we mean his exogenously given wealth before the initial price $p$ was received or paid by him at time 0. This means that $V_0(x,\phi ,0)=x$, whereas for a given contract $A$, the initial wealth of the hedger's strategy at time 0 equals $V_0(x,\phi ,A)=x+A_0 = x+p$. Before stating the definition of a self-financing trading strategy, we formulate the standing assumption regarding the integrability of stochastic processes.

\bhyp
We assume that $\xi^i$ for $i=1,2,\ldots,d$ (resp. $\psi^j$ for $j=0,1,\dots ,d$) are arbitrary $\gg$-predictable
(resp. $\gg$-adapted) processes such that the stochastic integrals used in what follows are well defined.
\ehyp

\bd  \lab{ts1}
For the hedger's initial endowment $x$, we say that a trading strategy $(x, \phi , \pA )$, associated with a contract $\pA$,
is {\it self-financing} whenever the wealth process $V(x, \phi , \pA )$, which is given by the formula
\be  \lab{rtf1}
V_t(x, \phi , \pA ) =  \sum_{i=1}^d \xi^i_tS^i_t + \sum_{j=0}^d \psi^j_t B^j_t ,
\ee
satisfies, for every $t \in [0,T]$,
\be \lab{portf2}
V_t(x , \phi , \pA ) = x + \sum_{i=1}^d \int_{(0,t]} \xi^i_u \, d(S^i_u + \pA^i_u )
+  \sum_{j=0}^d \int_0^t \psi^j_u \, dB^j_u + \pA_t .
\ee
\ed

It is important to stress that a hedging portfolio $\phi $ and contractual cash flows $\pA$ cannot be dealt with separately since,
in general, the wealth dynamics is obtained by a non-linear superposition of $\phi $ and $\pA$.
A possibility of separation of intertemporal cash flows of $A$ and a hedging portfolio $\phi $ for $A$ is, of course, a well known (and very handy) feature of the classic (that is, linear) arbitrage pricing theory, which in turn results in price additivity in a frictionless market model.

\brem
Obviously, the wealth process always depends on the initial endowment $x$, a portfolio $\phi $ and contractual cash flows $A$,
so that the notation $V(x, \phi , A)$ is adequate. However, for the sake of brevity, the shorthand
notation $V(\phi ,A)$ (or even $V(\phi )$) will sometimes be used in the remaining part of Section \ref{mm}
if there is no danger of confusion.
\erem

\brem
Formula \eqref{portf2} yields the following wealth decomposition
\be \lab{c2a}
V_t(x, \phi,\pA ) = x + G_t(x,\phi,\pA ) + F_t(x,\phi,\pA ) + \pA_t
\ee
where
\be \lab{gg66}
G_t(x,\phi,\pA ):=\sum_{i=1}^d \int _{(0,t]}\xi^i_u \,(dS^i_u + d\pA^i_u)
\ee
represents the gains/losses associated with holding long/short positions in risky assets $S^1, S^2, \dots , S^d$ and
\be  \lab{gg666}
F_t (x, \phi , \pA ) := \sum_{j=0}^d \int_0^t \psi^j_u \, dB^j_u
\ee
represents the portfolio's {\it funding costs}. This additive decomposition of the wealth process
will no longer hold when more constraints will be imposed on trading.
\erem

\brem \label{rem:cash}
In some related papers (see, for instance, \cite{PV10}), the process $\gamma$, which is given by, for all $t\in [0,T]$,
\bde
\gamma_t = x + F_t(x,\phi,\pA )+\sum_{i=1}^d  \int _{(0,t]} \xi^i_u \,d\pA^i_u+A_t ,
\ede
is referred to as the \textsl{cash} process financing the portfolio $\phi$.
In this context, it is important to stress that the equality
\bde
V_t(x,\phi,\pA) = \sum_{i=1}^d  \int _{(0,t]} \xi^i_u \, dS^i_u+\gamma_t,
\ede
holds but, in general, we have that $V_t(x,\phi,\pA) \ne \sum_{i=1}^d  \xi^i_t S^i_t + \gamma_t$.
%\bde
%V_t(x,\phi,\pA) \ne \sum_{i=1}^d  \xi^i_t S^i_t + \gamma_t.
%\ede
\erem

%%%%%%%%%%%%%%%%%%%%%%%%%%%%%%%%%%%%%%%%%%%%%%%%%%%%%%%%%%%%%%%%%%%
\ssc{Basic Model with Funding Costs} \lab{secelem}
%%%%%%%%%%%%%%%%%%%%%%%%%%%%%%%%%%%%%%%%%%%%%%%%%%%%%%%%%%%%%%%%%%%

Let us first describe a preliminary setting, which henceforth will  be referred to as the basic model with funding costs or,
simply, the {\it basic model.}

\bd
By the {\it basic model with funding costs,} we mean a market model in which the lending and borrowing accounts coincide,
so that $B = \Blr = \Bbr $, the equalities $B^i = \Bilr = \Bibr $ hold for all $i=1,2, \dots ,d$, and trading in funding accounts $B^i$ and risky assets $S^i$ is a priori unconstrained.
\ed

A thorough analysis of the basic model is merely a first step towards more realistic models with various trading and/or funding constraints. We will show that explicit formulae for the wealth dynamics under various constraints can be derived from results for the basic model by progressively refining the computations involving the wealth process and funding costs.
For reasons that will be explained later, we are interested not only in dynamics of the wealth process, but also in dynamics of the {\it netted wealth}, as given by Definition \ref{nhnhnh}. Let us only mention here that
the concept of the netted wealth will be a convenient tool to examine the no-arbitrage features of a market model under funding costs and collateralization. To be a bit more specific, the concept of a martingale measure should now be applied to the discounted netted wealth, rather than to the discounted wealth of a trading strategy since the latter process includes the cash flows of $A$, whereas in the former case they are in some sense counterbalanced by the cash flows of $-A$.

\bd  \lab{nhnhnh}
The {\it netted wealth} $\Vnet(x, \phi , \pA )$ of a trading strategy $(x, \phi, \pA)$ is given by the equality
$\Vnet(x, \phi , \pA ) = V(x, \phi , \pA ) + V(0, \wt \phi ,-\pA )$ where $(0, \wt \phi ,-A)$ is the unique self-financing strategy such that $\xi^i_t = \psi^i_t = 0$ for every $i=1,2,\dots ,d$ and all $t \in [0,T]$.
\ed

We have the following lemma (for its extension to the case of different lending and borrowing accounts, see Lemma~\ref{nhnhnx1}).

\bl \lab{nhnhn}
If $B = \Blr = \Bbr $ then the following equality holds, for all $t \in [0,T]$, {\rm
\be \lab{portf2c}
\Vnet_t (x, \phi , \pA ) =  V_t (x, \phi , \pA ) - B_t \int_{[0,t]} B_u^{-1} \, d\pA_u .
\ee }
\el

\proof
By setting  $\xi^i_t = \psi^i_t = 0$ in  \eqref{rtf1} and \eqref{portf2} , we obtain $V_t(0, \wt \phi ,-A) = \wt \psi^0_t B_t$ and
\bde
V_t(0,\wt \phi , -A ) = \int_0^t \wt \psi^0_u \, dB_u - \pA_t .
\ede
Since $V_0(0, \wt \phi ,-A)= - A_0 $, we obtain \eqref{portf2c}.
\endproof

Note that
\bde
\Vnet_0 (x, \phi , \pA ) = V_0(x, \phi , \pA ) + V_0(0,\wt \phi ,-\pA )= x+ A_0 - A_0 =x,
\ede
so that the initial netted wealth is independent of $A_0$. Nevertheless, the process $\Vnet (x, \phi , \pA )$ may depend on $A_0$, in general, if the dynamics of the wealth processes $V(x, \phi , \pA )$ and $V(0, \wt \phi ,-\pA )$ are non-linear. Intuitively, the netted wealth $\Vnet(x, \phi , \pA )$ represents the wealth of the hedger, who takes the back-to-back long and short positions in $A$, uses a dynamic portfolio $\phi $ with the initial endowment $x$ to hedge the long position, and leaves the short position {\it unhedged}, meaning that no investments in risky assets is undertaken to hedge the short position. In particular, the initial cash flows $A_0$ and $-A_0$ obviously cancel out, meaning that the initial price received from (or paid to) a counterparty in contract $A$ is immediately passed on to a counterparty in contract $-A$. Therefore, in the context of the computation of the netted wealth process, the value of $A_0$ should be immaterial for the hedger. This observation motivates us to make the following natural assumption, which ensures that the netted wealth is independent of $A_0$.

\bhyp \lab{hyp23}
When computing the netted wealth process $\Vnet(x, \phi , \pA )$, we set $A_0 =0$.
\ehyp

With Assumption \ref{hyp23} in force, the representation \eqref{portf2c} takes the form
\be \lab{portf2c_T}
\Vnet_t (x, \phi , \pA ) =  V_t (x, \phi , \pA ) - B_t \int_{(0,t]} B_u^{-1} \, d\pA_u.
\ee
In practice, the offset of cash flows at time 0 is only possible when the {\it market prices} of $A$ and $-A$ at time 0 satisfy $p^m_0(-A)= -p^m_0(A)$. Obviously, by the market price of $A$ (resp. $-A$), we mean here the initial price at time 0 of future
cash flows of $A$ (resp. $-A$) on the time interval $(0,T]$. Then the financial interpretation of the netted wealth at time $T$ can be restated as follows: in order to assess a potential profitability of a given contract $A$ with respect to his market model, the hedger, who has the initial endowment $x$, enters at time 0 into a contract $A$ at its market price $p^m_0(A)$, and simultaneously takes a short position in the same contract at its market price $- p^m_0(A)$, so that the net cost of his two positions in the contract at time 0 is null. Subsequently, starting from his initial endowment $x$, he implements a dynamic hedging portfolio $\phi $ for the long position and, concurrently, uses only the cash account (in general, the borrowing and lending accounts) to reinvest the incoming and outgoing cash flows associated with the short position. At terminal date $T$, the hedger aggregates the terminal wealth of a dynamically hedged long position in some contract with the outcome of the unhedged short position in the same contract. Independently of the level of the initial price of the contract, this gives him an indication whether entering into this contract could lead to an arbitrage opportunity for him. For the precise statement of this property and a detailed discussion, we refer to Section \ref{tyht} (see, in particular, Definition \ref{abop0}).

%%%%%%%%%%%%%%%%%%%%%%%%%%%%%%%%%%%%%%%%%%%%%%%%%%%%%%%%%
\sssc{A Preliminary Result}
%%%%%%%%%%%%%%%%%%%%%%%%%%%%%%%%%%%%%%%%%%%%%%%%%%%%%%%%%

Let introduce the following notation, for $i=1,2,\dots ,d$,
\be \lab{portf2a}
 K^i_t :=  \int_{(0,t]} B^i_u \, d\wh S^{i}_u + \pA^i_{t} =  \int_{(0,t]} B^i_u \, d\wh S^{i,\textrm{cld}}_u
\ee
where the second equality is an immediate consequence of \eqref{pri2}, and
\be \lab{portf2b}
K^{\phi }_t :=  \int_{(0,t]} B_u \, d\wt V_u (x,\phi ,A) - (\pA_{t} - \pA_0) = \int_{(0,t]} B_u \, d\Vnett_u (x,\phi ,A)
\ee
where we set $\Vnett (x,\phi ,A ) :=  B^{-1}\Vnet(x,\phi ,A)$ and
$\wt  V(x, \phi , A) :=B^{-1}  V (x, \phi ,A)$, so that the second equality follows from \eqref{portf2c}.
Obviously,
\be \lab{hyhy}
\Vnett_t(x,\phi ,A ) = x + \int_{(0,t]} B_u^{-1} \, dK^{\phi }_u .
\ee

The process $K^i$ is equal to the wealth, discounted by the funding account $B^i$, of a self-financing strategy
that uses the risky security $S^i$ and the associated funding account $B^i$, where $B^i_t$ units of
the cumulative dividend price of the $i$th asset are held at time $t$.

The following preliminary result is primarily tailored to cover the valuation and hedging of an {\it unsecured} contract. We thus mainly focus here on funding costs associated with trading in risky assets.
We will argue later on that Proposition \ref{prop1.1} is a convenient starting point to analyze a wide spectrum of practically appealing situations. To achieve our goals, it will be enough to impose later specific constraints on trading strategies, which will reflect particular market conditions faced by the hedger (such as: different lending, borrowing and funding rates) and/or additional covenants of an OTC contract under study (such as: a margin account, closeout  payoffs, or benefits stemming from defaults).
For a detailed study of trading strategies involving a {\it secured} (that is, {\it collateralized}$\, $) contract, we refer to Section \ref{seccoll}.

\bp \lab{prop1.1}
(i) For any  self-financing strategy $\phi$ we have that, for every $t \in [0,T]$,
\be  \lab{portf3b}
K^{\phi }_t= \sum_{i=1}^d \int_{(0,t]} \xi^i_u \, dK^i_u + \sum_{i=1}^d \int_0^t ( \psi^i_u B^i_u + \xi^i_u S^{i}_u ) (\wt B^i_u)^{-1}  \, d\wt B^i_u
\ee
where we set $\wt B^i := B^{-1} B^i$. \hfill \break
(ii) The equality
\be  \lab{portf3bc}
K^{\phi }_t= \sum_{i=1}^d \int_{(0,t]} \xi^i_u \, dK^i_u ,  \quad t\in [0,T] ,
\ee
holds if and only if
\be \lab{conee}
\sum_{i=1}^d \int_0^t ( \psi^i_u B^i_u + \xi^i_u S^{i}_u ) (\wt B^i_u)^{-1}  \, d\wt B^i_u  = 0 , \quad t\in [0,T].
\ee
(iii) In particular, if for each $i=1,2,\dots ,d$ we have that: either $B^i_t=B_t$ for all $t \in [0,T]$ or
\be \lab{portf3a}
\zeta^i_t := \psi^i_t B^i_t + \xi^i_t  S^{i}_t = 0, \quad t\in [0,T],
\ee
then \eqref{conee} is valid and thus \eqref{portf3bc} holds. \hfill \break
(iv) Assume that $B^i=B$ for every $i=1,2,\dots ,d$ and denote $\wt S^{i,\textrm{cld}} = B^{-1}S^{i,\textrm{cld}} $. Then
\be \lab{class2}
d\Vnett_t (x,\phi ,A)= \sum_{i=1}^d\xi^i_t \, d\wt S^{i,\textrm{cld}}_t .
\ee
\ep

\proof
Recall that  (see \eqref{portf2})
\bde
dV_t (x, \phi ,A) =  \sum_{i=1}^d  \xi^i_t \, d(S^i_t + \pA^i_t) + \sum_{j=0}^d \psi^j_t \, dB^j_t  + d\pA_t .
\ede
Using \eqref{rtf1},  for the discounted wealth $\wt V (\phi ,A) = B^{-1} V (\phi ,A)$ we obtain
\begin{align*}
d\wt V_t (x, \phi ,A) &=  \sum_{i=1}^d  \xi^i_t \, d((B_t)^{-1}S^{i}_t )
+ \sum_{i=1}^d  \xi^i_t  (B_t)^{-1} \, d\pA^i_t + \sum_{i=1}^d \psi^i_t \, d( (B_t)^{-1}B^i_t)
+ (B_t)^{-1} \, d\pA_t \\  &=  \sum_{i=1}^d  \xi^i_t \, d\wt S^{i,\textrm{cld}}_t +
\sum_{i=1}^d \psi^i_t \, d\wt B^i_t + (B_t)^{-1} \, d\pA_t
\end{align*}
where $\wt B^i =B^{-1} B^i$ and
\bde
\wt S^{i,\textrm{cld}}_t = S^{i}_t B_t^{-1} + \int_{(0,t]} B_u^{-1} \, d\pA^i_u
= \wt S^{i}_t + \int_{(0,t]} B_u^{-1} \, d\pA^i_u.
\ede
Consequently,
\begin{align*}
dK^{\phi}_t &= B_t \, d\wt V_t (x, \phi ,A) - d\pA_t  =  \sum_{i=1}^d  B_t \xi^i_t \, d\wt S^{i,\textrm{cld}}_t +\sum_{i=1}^d B_t \psi^i_t \, d\wt B^i_t
\\ &=  \sum_{i=1}^d  B_t \xi^i_t \, d( \wh S^{i}_t \wt B^i_t ) + \sum_{i=1}^d  B_t \xi^i_t \, B_t^{-1} d\pA^i_t +\sum_{i=1}^d B_t \psi^i_t \, d\wt B^i_t  \\
&=  \sum_{i=1}^d  B_t \xi^i_t \wh S^{i}_t \, d\wt B^i_t +
\sum_{i=1}^d  B_t \wt B^i_t \xi^i_t \, d\wh S^{i}_t  + \sum_{i=1}^d \xi^i_t \, d\pA^i_t +\sum_{i=1}^d B_t \psi^i_t \, d\wt B^i_t  \\
&=\sum_{i=1}^d  B_t \xi^i_t \wh S^{i}_t \, d\wt B^i_t +
\sum_{i=1}^d \xi^i_t \, (B^i_t \, d\wh S^{i}_t + d\pA^i_t ) +  \sum_{i=1}^d B_t \psi^i_t \, d\wt B^i_t \\
&= \sum_{i=1}^d \xi^i_t \, dK^i_t + \sum_{i=1}^d B_t ( \psi^i_t + \xi^i_t \wh S^{i}_t ) \, d\wt B^i_t .
\end{align*}
This completes the proof of part (i). Parts (ii) and (iii) now follow easily.
By combining formulae \eqref{portf2a} and \eqref{portf3b}, we obtain part (iv). Note that \eqref{class2} is the classic
condition for a market with a single cash account $B$.
\endproof

\brem \lab{rem2.6}
Note that equality $B^i=B$ (resp. equality \eqref{portf3a}) may correspond to unsecured (resp. secured) funding of the $i$th stock, where unsecured funding means that a risky security is not posted as collateral.
In this financial interpretation, condition \eqref{portf3a} would mean that at any date $t$ the value of the long or short position in the $i$th stock should be exactly offset by the value of the $i$th secured funding account.
Although this condition is aimed to cover the case of the fully secured funding of the $i$th risky asset using the corresponding repo rate, it is fair to acknowledge that it is rather restrictive and thus not always practical.
It would be suitable for repo contracts with the daily resettlement, but it would not cover the case of long term
repo contracts.

Note also that if condition \eqref{portf3a} holds for all $i=1,2,\dots, d$, then the wealth process satisfies $V_t (x,\phi ,A) = \psi^0_t B_t$ for every $t \in [0,T]$. This is consistent with the interpretation that
all gains/losses are immediately reinvested in the cash account $B$. To make this setup more realistic, we
need, in particular, to introduce different borrowing and lending rates and add more constraints on trading.
More generally, the $i$th risky security can be funded in part
using $B^i$ and using $B$ for another part, so that condition \eqref{portf3a} may fail to hold.
However, this case can also be covered by  the model in which condition \eqref{portf3a} is met
by artificially splitting the $i$th asset into two `sub-assets' that are subject to different funding rules.
Needless to say that the valuation and hedging results for a derivative security
will depend on the way in which risky assets used for hedging are funded.
\erem

%%%%%%%%%%%%%%%%%%%%%%%%%%%%%%%%%%%%%%%%%%%%%%%%%%%%%%%%%%%%%%%%%%%%%%%%%%%%%%%
\sssc{Wealth Dynamics in the Basic Model}
%%%%%%%%%%%%%%%%%%%%%%%%%%%%%%%%%%%%%%%%%%%%%%%%%%%%%%%%%%%%%%%%%%%%%%%%%%%%%%%

To obtain some useful representations for the wealth dynamics in the basic model, we first prove an auxiliary lemma.
From equality \eqref{ortf2a}, one can  deduce that the increment $dK^i_t$ represents the change
in the price of the $i$th asset net of funding cost. For the lack of the better terminology, we propose to
call $K^i$ the \textit{netted realized cash flow} of the $i$th asset.

\bl \lab{lmm1}
The following equalities hold, for all $t \in [0,T]$,
\be \lab{ortf2a}
K^i_t = S^i_t -S^i_0 +\pA^i_t - \int_0^t \wh S^i_u \, dB^i_u
\ee
and
\begin{align} \lab{ortf2b}
K^{\phi}_t &= V_t (x,\phi ,A ) - V_0 (x,\phi ,A) - (\pA_t - A_0) - \int_0^t \wt  V_u (x,\phi ,A) \, dB_u \nonumber \\
&= F_t (x,\phi ,A ) +G_t (x,\phi ,A )- \int_0^t \wt  V_u (x,\phi ,A) \, dB_u.
\end{align}
\el

\proof
The It\^o formula, \eqref{pri2} and \eqref{portf2a} yield
\begin{align} \lab{44rr}
 \int_{(0,t]} B^i_u \, d\wh S^{i,\textrm{cld}}_u
&= \int_{(0,t]} B^i_u \, d\wh S^{i}_u + \pA^i_t
=  B^i_t \wh S^{i}_t - B^i_0 \wh S^{i}_0-
\int_0^t \wh S^{i}_u \, dB^i_u + \pA^i_t \\
&= S^i_t -S^i_0 + \pA^i_t - \int_0^t \wh S^i_u \, dB^i_u.  \nonumber
\end{align}
The proof of the second equality is analogous.
\endproof

\brem  \lab{amp}
For each $i$, the differential $K^i_t$ admits both ``multiplicative'' decomposition
\be \lab{mult}
K^i_t= \int_{(0,t]} B^i_u \, d\wh S^{i,\textrm{cld}}_u,
\ee
and ``additive'' decomposition
\be \lab{add}
K^i_t= S^i_t -S^i_0 +\pA^i_t - \int_0^t \wh S^i_u \, dB^i_u.
\ee
If under some probability measure, say $\PH$,  the process $\wh S^{i,\textrm{cld}}$ is a (local) martingale, then, in view of \eqref{mult}, process $K^i$ is also a (local) martingale under the same measure. We call this the multiplicative martingale property of $K^i$ (under $\PH$). Because of \eqref{add}, we also say that $K^i$ enjoys the additive martingale property (under $\PH$). The financial meaning of this property is rather intuitive considering that that $dK^{i }_t= d(S^i_t + \pA^i_{t})-  \wh S^{i}_t\, dB^i_t$ represents change in capital gains/losses of the $i$th asset, net of local holding income/holding cost of the asset. Likewise, note that if equality \eqref{portf3bc} is valid, then the process $K^{\phi }$ is a (local) martingale. In view of  \eqref{ortf2b}, we call this additive martingale property of $K^{\phi }$ (under $\PH$). Again, the financial meaning of this property is fairly clear,
since  $dK^{\phi }_t =  dG_t (x,\phi ,A) + dF_t (x,\phi ,A) - \wt V_t (x,\phi ,A) \, dB_t$ represents total local change in gains/losses and funding costs, net of local wealth reinvestment income/wealth service charge.
\erem

In view of Lemma \ref{lmm1}, the following corollary to Proposition \ref{prop1.1} is immediate.
Since the funding costs in the basic model may depend on funding accounts $B^0, B^1, \dots , B^d$, we
emphasize this dependence by writing $F(\phi ) = F(\phi; B^0, B^1, \dots , B^d)$.
Recall that the processes $\zeta^i,\, i=1,2, \dots , d$ are given by \eqref{portf3a}.

\bcor \lab{correx}
 Formula \eqref{portf3b} is equivalent to the following expressions
\begin{align} \lab{class1}
&d\Vnett_t (x,\phi ,A)=  \sum_{i=1}^d\xi^i_t \wt B^i_t \, d\wh S^{i,\textrm{cld}}_t
+ \sum_{i=1}^d \zeta^i_t (B^i_t)^{-1}  \, d\wt B^i_t,
\\ \lab{clacss1} &d\wt{V}_t (x,\phi ,A )=  \sum_{i=1}^d\xi^i_t \wt{B}^i_t \,
d\wh S^{i,\textrm{cld}}_t + \sum_{i=1}^d  \zeta^i_t (B^i_t)^{-1} \, d\wt B^i_t
 + (B_t)^{-1}\, d\pA_t ,\\
 \lab{portf3c1} &dV_t (x,\phi ,A ) =  \wt V_t (x,\phi ,A ) \, dB_t + \sum_{i=1}^d \xi^i_t \, dK^i_t
 +\sum_{i=1}^d \zeta^i_t (\wt B^i_t)^{-1}  \, d\wt B^i_t + d\pA_t .
\end{align}
Hence the funding costs of $\phi $ satisfy
\be \lab{ff66}
F_t (\phi; B^0, B^1, \dots , B^d) = \int_0^t \wt V_u(x,\phi , A )\, dB_u + \sum_{i=1}^d \int_0^t  \zeta^i_u
 (\wt B^i_u)^{-1}  \, d\wt B^i_u - \sum_{i=1}^d  \int_0^t \xi^i_u \wh S^i_u \, dB^i_u .
\ee
\ecor

\brem \lab{remind}
Formula \eqref{class1} may suggest that in the basic model with funding costs the dynamics of the process $\Vnet(x, \phi , A)$ do not depend on $A$. To this end, one could argue as follows:
suppose that we take any processes $\xi = (\xi^1, \dots , \xi^d)$ and $\psi = (\psi^1, \dots , \psi^d)$. Then, under the present assumptions, for any two external cash flows, say $A$ and $\wh{A}$, we may compute
the unique wealth processes $V(x,\phi , A)$ and $V(x, \wh{\phi}, \wh{A})$ from \eqref{clacss1} and, consequently,
using \eqref{rtf1}, also the unique processes $\psi^0$ and $\wh{\psi}^0$ such that the full strategies $\phi$ and $\wh{\phi}$
are self-financing. Then the wealth processes $V(x,\phi , A)$ and $V(x, \wh{\phi}, \wh{A})$ will be manifestly different,
but from \eqref{class1} we see that the netted wealth processes $\Vnet (x,\phi , A)$ and $\Vnet (x,\wh{\phi }, \wh{A})$
coincide and thus they do not depend on $A$. This argument is in fact flawed since, typically,
the processes $\xi = (\xi^1, \dots , \xi^d)$ and $\psi = (\psi^1, \dots , \psi^d)$ may also depend on future cash flows of $A$.
This feature is rather obvious when one addresses the issue of replication of a contract formally represented by the process $A$.

To illustrate this remark, let us consider a toy model with the cash account $B$ and one risky asset, namely, the
unit discount bond maturing at $T$ with the price process $S^1_t = B(t,T)$. Let $0<t_0<T$ and let $\eta$ be a positive ${\cal G}_{t_0}$-measurable random variable. Recall that when dealing with the netted wealth, we may and do assume that $A_0=0$.
We set $A_t = \eta B(t_0,T) \I_{[t_0,T]} - \eta \I_{[T]}$ and we consider the portfolio $\phi = (\xi^1 , \psi^0)$ where $\xi^1_t = \eta \I_{[t_0,T]}(t)$ and $\psi^0_t = 0$ for all $t$, meaning that at time $t_0$ the incoming
cash flow $\eta$ is invested in the discount bond. If we assume that $x=0$, then the wealth process $V(0,\phi ,A)$ satisfies $V_t (0, \phi ,A) =  \eta B(t,T) \I_{[t_0,T[}(t) $ so that, in particular, $V_T(0, \phi ,A)=0$. By contrast, equation \eqref{portf2c} yields
\be \lab{uubb}
\Vnet_T (0, \phi , A) =  V_T (0, \phi , \pA ) - B_T \int_{(0,T]} B_t^{-1} \, d\pA_t =
\eta \bigg( 1 -  B(t_0,T)\, \frac{B_T}{B_{t_0}} \bigg)
\ee
and thus the netted wealth manifestly depends on $\eta $, that is, on the contract $A$. Note that by modifying $\xi^1_{t_0}$ so that $\psi^0_{t_0} \ne 0$, one can easily produce an example in which $V_T(0, \phi ,A)$ is non-zero and it also depends on $A$. The advantage of the netted wealth lies in the fact that it is suitable when one wishes to identify arbitrage opportunities.
For instance, from \eqref{uubb}, we may deduce that an arbitrage opportunity arises for the hedger if the inequality
$B(t_0,T) < \frac{B_{t_0}}{B_T}$ holds $\P$-a.s., that is, when the bond price is too low with respect to the cash account.
Needless to say that this conclusion is trivial and it can be obtained without difficulty through other means.
Our goal in this example was simply to illustrate the potential of the netted wealth as a handy tool, which can be also applied when non-linear constraints on trading are imposed, in particular, under netting or collateralization.
\erem

\bex \lab{ex1}
Suppose that the processes $B^j,\, j=0,1,\dots ,d$ are absolutely continuous, so that they can be represented as
$dB^j_t = r^j_t B^j_t \, dt $ for some $\gg$-adapted processes $r^j,\, j=0,1,\dots ,d$. Then \eqref{clacss1} yields
\be  \lab{pof3d1}
dV_t(\phi)= r_t V_t (\phi) \, dt + \sum_{i=1}^d \zeta^i_t (r^i_t-r_t)\, dt
+ \sum_{i=1}^d \xi^i_t \big(dS^i_t- r^i_t S^i_t \, dt + d\pA^i_t  \big) + d\pA_t
\ee
where, for brevity, we write $V_t(\phi)=V_t(x, \phi , A)$.  Equation \eqref{pof3d1} yields
\be  \lab{pof6d1}
dV_t(\phi)=  \sum_{j=0}^d r^j_t \psi^j_t B^j_t \, dt + \sum_{i=1}^d \xi^i_t \big(dS^i_t + d\pA^i_t  \big) + d\pA_t,
\ee
which can also be seen as an immediate consequence of \eqref{portf2}.
We note that the dynamics of funding costs of $\phi $ are given by
\be \lab{pof8d1}
dF_t (\phi; B^0, B^1, \dots , B^d) =\sum_{j=0}^d r^j_t \psi^j_t B^j_t \, dt .
\ee
\eex

%%%%%%%%%%%%%%%%%%%%%%%%%%%%%%%%%%%%%%%%%%%%%%%%%%%%%%%%%%%%%%%%%%%%%%%%%%%%%%%
\sssc{A Common Unsecured Account for Risky Assets} \lab{sec213}
%%%%%%%%%%%%%%%%%%%%%%%%%%%%%%%%%%%%%%%%%%%%%%%%%%%%%%%%%%%%%%%%%%%%%%%%%%%%%%%

Let us analyze a special case of the basic model with a common unsecured account for risky assets.
To this end, we assume that $B^i= B$ for $i=1,2,\dots ,k$ for some $k \leq d$.
This means that all unsecured accounts $B^1, B^2, \dots ,B^k$ collapse in a single cash account,
denoted as $B$, but the secured accounts $B^{k+1}, B^{k+2}, \dots ,B^d$ driven by the repo rates may vary from one asset to another. Formally, it is now convenient to postulate that $\psi^i=0$ for $i=1,2,\dots ,k$, so that a portfolio $\phi $ may be represented as $\phi = (\xi^1,\dots ,\xi^d, \psi^0, \psi^{k+1},\dots ,\psi^d)$. Hence formula \eqref{rtf1} reduces to
\bde
V_t (\phi ) = \sum_{i=1}^d \xi^i_t S^i_t + \psi^0_t B_t + \sum_{i=k+1}^d \psi^i_t B^i_t
\ede
where we write $V_t(\phi)=V_t(x, \phi , A)$ and the self-financing condition  \eqref{portf2} becomes
\bde
V_t (\phi ) = V_0(\phi )+ \sum_{i=1}^d \int_{(0,t]} \xi^i_u \, d(S^i_u + \pA^i_u )
+ \int_0^t \psi^0_u \, dB_u + \sum_{i=k+1}^d \int_0^t \psi^i_u \, dB^i_u
 + \pA_t .
\ede
Consequently, equality \eqref{clacss1} takes the following form
\be \lab{oo99}
dV_t (\phi )= \wt V_t (\phi )\, dB_t+
\sum_{i=1}^k \xi^i_t B_t \, d\wt S^{i,\textrm{cld}}_t
+ \sum_{i=k+1}^d\xi^i_t B^i_t \, d\wh S^{i,\textrm{cld}}_t
+ \sum_{i=k+1}^d  \zeta^i_t (\wt B^i_t)^{-1}  \, d\wt B^i_t
 + d\pA_t
\ee
where we denote
\bde % \lab{pbri2}
\wt S^{i,\textrm{cld}}_t := \wt S^i_t+\int_{(0,t]} B_u^{-1} \, d\pA^i_u ,\quad t\in [0,T] ,
\ede
where in turn $\wt S^i := B^{-1} S^i$.

\bex \lab{ex2} If all accounts  $B^j,\, j=0,1,\dots ,d$ are absolutely continuous so that,
in particular, $r^i = r$ for $i=1,2,\dots ,k$, then
\be \lab{t6u}
dV_t (\phi)= \Big( r_t \psi^0_t B_t+ \sum_{i=k+1}^d r^i_t \psi^i_t B^i_t \Big)\, dt + \sum_{i=1}^d \xi^i_t \big(dS^i_t + d\pA^i_t  \big) + d\pA_t .
\ee
If, in addition, $\zeta^i_t=0$ for $i=k+1, k+2, \dots ,d$, then $V_t (\phi ) = \sum_{i=1}^k\xi^i_t S^i_t + \psi^0_t B_t $
and \eqref{t6u} yields
\bde
dF_t(\phi ) = r_t \Big( V_t (\phi) - \sum_{i=1}^k \xi^i_t S^i_t \Big) \, dt - \sum_{i=k+1}^d \xi^i_t r^i_t S^i_t \, dt.
\ede
\eex

%%%%%%%%%%%%%%%%%%%%%%%%%%%%%%%%%%%%%%%%%%%%%%%%%%%%%%%%%%%%%%%%%%%%%%%%%%%%%%%
\ssc{Different Lending and Borrowing Cash Rates}  \lab{sec2.2.1}
%%%%%%%%%%%%%%%%%%%%%%%%%%%%%%%%%%%%%%%%%%%%%%%%%%%%%%%%%%%%%%%%%%%%%%%%%%%%%%%

In the first extension of the basic model, we assume that the unsecured borrowing and lending cash rates are different. Recall that $\Blr$ and $\Bbr$ stand for the account processes corresponding to the lending and borrowing rates, respectively.
This can be seen as a first example of a {\it generic market model}, in the sense explained in Section \ref{GMM}; further
examples are given in the foregoing subsections.

It is now natural to represent a portfolio $\phi $ as $\phi = (\xi^1,\dots ,\xi^d,\psi^{l}, \psi^{b}, \psi^{1},\dots ,\psi^d )$ where, by assumption, $\psi^{l}_t \geq 0$ and $\psi^{b}_t \leq 0$ for all $t \in [0,T]$. Since simultaneous lending and borrowing of cash is either precluded or not efficient (if $\rbb \geq \rll$), we also postulate that $\psi^{l}_t \psi^{b}_t =0$ for all $t \in [0,T]$. The wealth process of a trading strategy $( \phi , A)$ now equals (recall that we denote $V_t(\phi)=V_t(x,\phi , A)$)
\be\label{vifi}
V_t (\phi ) =  \sum_{i=1}^d \xi^i_t S^i_t + \sum_{i=1}^d \psi^i_t B^i_t + \psi^{l}_t \Blr_t + \psi^{b}_t \Bbr_t ,
\ee
and the self-financing condition reads
\begin{align}   \lab{sfvifi}
V_t (\phi ) =\, \, & V_0(\phi )+ \sum_{i=1}^d \int_{(0,t]} \xi^i_u \, d(S^i_u + \pA^i_u ) + \sum_{i=1}^d \int_0^t \psi^i_u \, dB^i_u \\
&+ \int_0^t \psi^{l}_u \, d\Blr_u + \int_0^t \psi^{b}_u \, d\Bbr_u + \pA_t . \nonumber
\end{align}
It is worth noting that $\psi^{l}_t$ and  $\psi^{b}_t$ satisfy
\bde
\psi^{l}_t = (\Blr_t)^{-1} \Big( V_t (\phi ) -  \sum_{i=1}^d \xi^i_t S^i_t -\sum_{i=1}^d \psi^i_t B^i_t\Big)^+
\ede
and
\bde
\psi^{b}_t = - (\Bbr_t)^{-1} \Big( V_t (\phi ) -  \sum_{i=1}^d \xi^i_t S^i_t- \sum_{i=1}^d \psi^i_t B^i_t \Big)^-.
\ede
The following corollary furnishes the wealth dynamics under the present assumptions.

\bcor \lab{corcc}
(i) Assume that $\Blr$ and $\Bbr$ are account processes corresponding to the lending and borrowing rates.
Let $\phi $ be any self-financing strategy such that $\psi^{l}_t \geq 0,\, \psi^{b}_t \leq 0$ and $\psi^{l}_t \psi^{b}_t =0$ for all $t \in [0,T]$. Then the wealth process $V(\phi )$, which is given by \eqref{vifi}, has the following dynamics
\begin{align} \lab{oo44}
dV_t (\phi ) = \, \, & \sum_{i=1}^d\xi^i_t B^i_t \, d\wh S^{i,\textrm{cld}}_t
+ \sum_{i=1}^d  \zeta^i_t (B^i_t)^{-1}  \, dB^i_t
 + d\pA_t \nonumber \\
&+ \Big( V_t (\phi ) -  \sum_{i=1}^d \xi^i_t S^i_t -\sum_{i=1}^d \psi^i_t B^i_t\Big)^+  (\Blr_t)^{-1} \, d\Blr_t
\\&- \Big( V_t (\phi ) -  \sum_{i=1}^d \xi^i_t S^i_t -\sum_{i=1}^d \psi^i_t B^i_t\Big)^- (\Bbr_t)^{-1}\, d\Bbr_t . \nonumber
\end{align}
(ii) If, in addition, $\psi^i_t=0$ for $i=1,2, \dots ,k$ and $\zeta^i_t=0$ for $i=k+1,k+2, \dots , d$
for all $t \in [0,T]$, then
\begin{align} \lab{oo4n4}
dV_t (\phi ) =\, \, & \sum_{i=1}^k \xi^i_t \, d(S^i_t + \pA^i_t )
 + \sum_{i=k+1}^d \xi^i_t B^i_t \, d\wh S^{i,\textrm{cld}}_t + d\pA_t
\\ &+ \Big( V_t (\phi )  -  \sum_{i=1}^k \xi^i_t S^i_t \Big)^+  (\Blr_t)^{-1} \, d\Blr_t
- \Big( V_t (\phi )  -  \sum_{i=1}^k \xi^i_t S^i_t \Big)^- (\Bbr_t)^{-1}\, d\Bbr_t . \nonumber
\end{align}
\ecor

\proof
Formula \eqref{oo4n4} can be derived from \eqref{oo44}, using also the following equality (see \eqref{44rr})
\bde
B^i_t \, d\wh S^{i,\textrm{cld}}_t =dS^i_t   - \wh S^i_t \, dB^i_t+ d\pA^i_t .
\ede
The details are left to the reader.
\endproof

\bex \lab{ex3}
Under the assumptions of part (ii) in Corollary \ref{corcc} if, in addition,
the accounts $B^i$ for $i=k+1,k+2, \dots , d$ as well as $B^l$ and $B^b$ are absolutely continuous, then \eqref{oo4n4} becomes
\begin{align}  \lab{oo33}
dV_t(\phi ) =\, \, &\sum_{i=1}^k \xi^i_t \big(dS^i_t + d\pA^i_t  \big) +
\sum_{i=k+1}^d \xi^i_t \big(dS^i_t- r^i_t S^i_t \, dt + d\pA^i_t  \big) + d\pA_t
\\&+ \rll_t  \Big( V_t (\phi) - \sum_{i=1}^k \xi^i_t S^i_t \Big)^+ \, dt
- \rbb_t  \Big( V_t (\phi) - \sum_{i=1}^k \xi^i_t S^i_t \Big)^- \, dt \nonumber
\end{align}
and thus the funding costs satisfy
\bde
dF_t(\phi ) =\rll_t  \Big( V_t (\phi) - \sum_{i=1}^k \xi^i_t S^i_t \Big)^+ \, dt
- \rbb_t  \Big( V_t (\phi) - \sum_{i=1}^k \xi^i_t S^i_t \Big)^- \, dt
- \sum_{i=k+1}^d  r^i_t \xi^i_t S^i_t \, dt .
\ede
In particular, by setting $k=0$, we obtain
\be \lab{oov33}
dV_t(\phi ) =\sum_{i=1}^d \xi^i_t \big(dS^i_t- r^i_t S^i_t \, dt + d\pA^i_t  \big) + d\pA_t
+ \rll_t  \big( V_t (\phi)  \big)^+ \, dt
- \rbb_t  \big( V_t (\phi) \big)^- \, dt .
\ee
\eex

%%%%%%%%%%%%%%%%%%%%%%%%%%%%%%%%%%%%%%%%%%%%%%%%%%%%%%%%%%%%%%%%%%%%%%%%%%%%%%%
\ssc{Trading Strategies with Funding Costs and Netting}  \lab{sec2.2}
%%%%%%%%%%%%%%%%%%%%%%%%%%%%%%%%%%%%%%%%%%%%%%%%%%%%%%%%%%%%%%%%%%%%%%%%%%%%%%%

So far, long and short positions in funding accounts $B^j,\, j=0,1, \dots , d$ were assumed to bear the same interest. This assumption will be now relaxed, so that in this section, besides postulating that $\Blr\ne \Bbr$ we also postulate that $B^{i,l}\ne B^{i,b}\, j=0,1, \dots , d.$ Accordingly, we consider trading portfolio $\phi = (\xi^1,\dots ,\xi^d, \psi^{l}, \psi^{b},\psi^{1,l},\psi^{1,b},\dots ,\psi^{d,l},\psi^{d,b})$
whenever this is needed, and we define the corresponding wealth process as
\be  \lab{rtf1-lb}
V_t(x, \phi , \pA ) = \psi^{l}_t \Blr_t+\psi^{b}_t\Bbr_t+  \sum_{i=1}^d ( \xi^i_tS^i_t + \psi^{i,l}_t B^{i,l}_t+\psi^{i,b}_t B^{i,b}_t).
\ee
Consequently, we will now deal with the extended framework in which the issue of aggregating long and short positions in risky assets becomes crucial. The concept of aggregation of long and short positions can be introduced at various levels of inclusiveness, from the
 total absence of offsetting and netting to the most encompassing case of netting of all positions, whenever this is possible.

Let us explain the offsetting/netting terminology adopted in this work.
By {\it offsetting}, we mean the compensation of long and short positions either for a given risky asset or for the non-risky asset.
This concept is irrelevant unless the borrowing and lending rates are different for at least one risky asset or for the cash account.
By {\it netting}, we mean the aggregation of long or short cash positions across various risky assets, which share
some funding accounts. Therefore, the possibility of netting becomes relevant when they exist
some risky assets, say $S^i$ and $S^j$, for which at least one of the following equalities holds: $B^{i,b} = B^{j,b},\, B^{i,b} = B^{j,l},\, B^{i,l} = B^{j,b}$ or $B^{i,l} = B^{j,l}$. Needless to say that several variants of models with netting can be introduced and
examined. To illustrate this concept, we will study here only one particular instance of a market model with netting (see Section \ref{sec2.2.3}).

For our further purposes, it will be enough to distinguish between the following cases: \hfill \break
(a) the complete absence of offsetting and netting of long/short positions, \hfill \break
(b) the offsetting of long/short positions for every risky asset, but no netting, \hfill \break
(c) the offsetting of long/short positions for every risky asset combined with some form of netting of long/short cash positions
    for all risky assets that are funded from common funding accounts.

%%%%%%%%%%%%%%%%%%%%%%%%%%%%%%%%%%%%%%%%%%%%%%%%%%%%%%%%%%%%%%%%%%%%%%%%%%%%%%%
\sssc{Absence of Offsetting}  \lab{sec22.2}
%%%%%%%%%%%%%%%%%%%%%%%%%%%%%%%%%%%%%%%%%%%%%%%%%%%%%%%%%%%%%%%%%%%%%%%%%%%%%%%

To describe the case (a) of the total absence of offsetting and netting of long and short
positions in all risky assets, one can postulate that for all $i=1,2, \dots ,d$ and $t \in [0,T]$,
\be\lab{costuniegra}
\xi^{i,b}_t S^i_t+ \psi^{i,l}_t \Bilr_t = 0 , \quad
\xi^{i,l}_t S^i_t + \psi^{i,b}_t \Bibr_t = 0
\ee
where $\xi^{i,b}_t S^i_t  \leq 0, \,  \xi^{i,l}_t S^i_t \geq 0$,
so that $\psi^{i,l}_t \geq 0$ and $\psi^{i,b}_t \leq 0$ for all $t \in [0,T]$. In particular, even when the
equality $\xi^{i,l}_t + \xi^{i,b}_t = 0$ holds for all $t$, meaning that the net
position in the $i$th asset is null at any time,  an incremental cost
of holding open both positions may still arise, due to the spread between the rates implicit in accounts $\Bilr$ and $\Bibr$.
It is clear that this case is very restrictive and not practically appealing and thus it will not be analyzed in what follows.

%%%%%%%%%%%%%%%%%%%%%%%%%%%%%%%%%%%%%%%%%%%%%%%%%%%%%%%%%%%%%%%%%%%%%%%%%%%%%%%
\sssc{Offsetting of Positions in Risky Assets} \lab{sec2.2.2}
%%%%%%%%%%%%%%%%%%%%%%%%%%%%%%%%%%%%%%%%%%%%%%%%%%%%%%%%%%%%%%%%%%%%%%%%%%%%%%%

Let us now examine the netting convention (b). For this purpose, we postulate that $V(\phi)=V(x, \phi , A)$ satisfies
\bde
V_t(\phi ) = \psi^{l}_t \Blr_t + \psi^{b}_t \Bbr_t
+ \sum_{i=1}^d ( \xi^i_t S^i_t  + \psi^{i,l}_t \Bilr_t + \psi^{i,b}_t \Bibr_t )
= \psi^{l}_t \Blr_t + \psi^{b}_t \Bbr_t
\ede
where $\psi^{i,l}_t \geq 0$ and $\psi^{i,b}_t \leq 0$ for $t \in [0,T]$ and, for $i=1,2, \dots ,d$ and $t \in [0,T]$,
\be \lab{xntt6}
\xi^i_t S^i_t + \psi^{i,l}_t \Bilr_t +  \psi^{i,b}_t \Bibr_t = 0 .
\ee
The present netting mechanism can be interpreted as follows: for the purpose of hedging, it would be pointless to hold
simultaneously long and short positions in any asset $i$; it is enough to look at the net position in the $i$th asset. For example, if the hedger already holds the short position in some asset and the need to take the long position
of the same size arises, it is natural to postulate that the short position is first closed.

Note also that condition  \eqref{xntt6} is fairly restrictive, since it prevents netting of short
and long cash positions across all risky assets which share the same
long and short funding accounts. By definition, the {\it long} (resp. {\it short}) {\it cash position} in the $i$th asset corresponds to the positive (resp. negative) sign of $\xi^i_tS^i_t$. Recall that we did not postulate that the prices processes $S^i$ of risky assets are non-negative. See also Remark \ref{rem2.6} for general comments regarding condition \eqref{portf3a}, which also apply to condition \eqref{xntt6}.

Since a simultaneous lending and borrowing of cash from the funding account $i$
is precluded (or not efficient, if $\ribb \geq \rill$),
we also postulate that $\psi^{i,l}_t \psi^{i,b}_t =0$ for all $t \in [0,T]$, and for $i=0,1,\ldots,d.$
This implies that
\be \lab{viigy1}
\psi^{l}_t = (\Blr_t)^{-1} ( V_t (\phi ))^+, \quad
\psi^{b}_t = - (\Bbr_t)^{-1} ( V_t (\phi ))^-
\ee
and, for every $i=1,2,\dots , d$,
\be  \lab{viigy2}
\psi^{i,l}_t = (\Bilr_t)^{-1} (\xi^i_t S^i_t)^- , \quad \psi^{i,b}_t = - (\Bibr_t)^{-1} (\xi^i_t S^i_t)^+.
\ee
Note the essential difference between the present setup and the situation outlined in Section \ref{sec22.2}
where it was not postulated that the offsetting equality $\psi^{i,l}_t \psi^{i,b}_t =0$ holds for all $t \in [0,T]$.
The self-financing condition now reads
\begin{align}\lab{portvf2}
V_t (\phi ) = &V_0(\phi )+ \sum_{i=1}^d \int_{(0,t]} \xi^i_u \, d(S^i_u + \pA^i_u ) + \int_0^t \psi^{l}_u \, d\Blr_u + \int_0^t \psi^{b}_u \, d\Bbr_u \\ & + \sum_{i=0}^d \int_0^t \psi^{i,l}_u \, d\Bilr_u + \sum_{i=0}^d \int_0^t \psi^{i,b}_u \, d\Bibr_u
+ \pA_t  \nonumber
\end{align}
and thus the following result is straightforward.

\bcor \lab{cortcc}
Assume that $\Bilr$ and $\Bibr$ are account processes corresponding to the lending and borrowing rates.
We postulate that $\psi^{i,l}_t \geq 0,\, \psi^{i,b}_t \leq 0$ and $\psi^{i,l}_t \psi^{i,b}_t =0$ for all
$i=0,1, \dots ,d$ and $t \in [0,T]$, and equality \eqref{xntt6} holds for all $i=1,2,\dots ,d$.
Then the wealth process $V(\phi)=V(\phi , A)$ equals, for all $t \in [0,T]$,
\bde
V_t(\phi ) =  \psi^{l}_t \Blr_t + \psi^{b}_t \Bbr_t
\ede
and the wealth dynamics are
\begin{align}  \lab{oo4v4}
dV_t(\phi ) = \, \, &\sum_{i=1}^d \xi^i_t \, (dS^i_t + d\pA^i_t)
+ \sum_{i=1}^d ( \xi^i_t S^i_t )^-  (\Bilr_t)^{-1} \, d\Bilr_t
- \sum_{i=1}^d (\xi^i_t S^i_t )^+ (\Bibr_t)^{-1}\, d\Bibr_t \\
&+( V_t (\phi ) )^+  (\Blr_t)^{-1} \, d\Blr_t
- ( V_t (\phi ))^- (\Bbr_t)^{-1}\, d\Bbr_t + d\pA_t . \nonumber
\end{align}
\ecor

\brem
When the equality $\Bilr=\Bibr= B^i$ holds for all $i=1,2, \dots ,d$, then formula  \eqref{oo4v4} can be seen as
a special case of formula  \eqref{oo44} with $\zeta^i_t=0$ for all $i$ and $t \in [0,T]$ (see also dynamics \eqref{oov33}).
\erem

\bex \lab{ex3a} Under the assumptions of Corollary \ref{cortcc} if, in addition, the processes $\Bilr$
and $\Bibr$ for $i=0,1, \dots ,d$ are absolutely continuous, then \eqref{oo4v4} becomes
 (note that \eqref{oox33} extends \eqref{oov33})
 \begin{align}  \lab{oox33}
dV_t(\phi ) =\, \, &\sum_{i=1}^k \xi^i_t \big(dS^i_t + d\pA^i_t  \big)
+ \sum_{i=1}^d \rill_t( \xi^i_t S^i_t )^-  \, dt -  \sum_{i=1}^d \ribb_t (\xi^i_t S^i_t )^+ dt
\\ &+   \rll_t ( V_t (\phi) )^+ \, dt - \rbb_t ( V_t (\phi))^- \, dt + d\pA_t \nonumber
\end{align}
and thus the funding costs satisfy
\bde
dF_t(\phi ) = \rll_t ( V_t (\phi) )^+ \, dt - \rbb_t ( V_t (\phi))^- \, dt
+ \sum_{i=1}^d \rill_t( \xi^i_t S^i_t )^-  \, dt
- \sum_{i=1}^d \ribb_t (\xi^i_t S^i_t )^+ dt .
\ede
\eex

%%%%%%%%%%%%%%%%%%%%%%%%%%%%%%%%%%%%%%%%%%%%%%%%%%%%%%%%%%%%%%%%%%%%%%%%%%%%%%%
\sssc{Model with Partial Netting}  \lab{sec2.2.3}
%%%%%%%%%%%%%%%%%%%%%%%%%%%%%%%%%%%%%%%%%%%%%%%%%%%%%%%%%%%%%%%%%%%%%%%%%%%%%%%

We will examine here a special case of netting convention (c), which seems to be of some interest in practice.
We now assume that $\Bilr = \Blr$ for all $i =1,2, \dots ,d$ and we postulate
that all short cash positions in risky assets $S^1, S^2, \dots , S^d$ are aggregated. Intuitively, this means
that all positive cash flows, inclusive of proceeds from short-selling of risky assets,
are included in the wealth and transferred to the cash account $\Blr$ or $\Bbr$.
By contrast, long cash positions in risky assets $S^i$ are assumed to be funded
from respective funding accounts $\Bibr$. We thus deal here with the case of the partial netting of positions
across risky assets. The trading framework introduced in this subsection will be henceforth referred to as the {\it market model
with partial netting.}

The present setup is formalized by postulating that the wealth process $V(\phi ) = V(x,\phi ,A)$ equals
\be \lab{biigy0}
V_t (\phi ) = \psi^{l}_t \Blr_t + \psi^{b}_t \Bbr_t
+  \sum_{i=1}^d ( \xi^i_t S^i_t  + \psi^{i,b}_t \Bibr_t  )
\ee
where, for every $i=1,2, \dots ,d$ and $t \in [0,T]$, the process  $\psi^{i,b}_t$ satisfies
\be  \lab{biigy2}
 \psi^{i,b}_t =  -(\Bibr_t)^{-1} (\xi^i_t S^i_t)^+ \leq 0.
\ee
Note that since in equation \eqref{biigy0} we use the net position $\xi^i_t$, rather than
$\xi^{i,l}_t$ and $\xi^{i,b}_t$,  the offsetting of long and short positions in every
risky asset $S^i$ is already implicit in this equation. From \eqref{biigy0} and \eqref{biigy2}, we obtain
\bde % \lab{biigy0}
V_t (\phi ) = \psi^{l}_t \Blr_t + \psi^{b}_t \Bbr_t - \sum_{i=1}^d ( \xi^i_t S^i_t )^-.
\ede
Since, as usual, we postulate that $\psi^{l}_t \geq 0$ and $\psi^{b}_t \leq 0$, we obtain
the following equalities
\be \lab{ccbiigy1}
\psi^{l}_t = (\Blr_t)^{-1} \Big( V_t (\phi ) + \sum_{i=1}^d ( \xi^i_t S^i_t )^- \Big)^+, \quad
\psi^{b}_t = - (\Bbr_t)^{-1} \Big( V_t (\phi ) + \sum_{i=1}^d ( \xi^i_t S^i_t )^- \Big)^-.
\ee
Finally, the self-financing condition for the trading strategy $(x, \phi ,A)$ reads
\begin{align*} % \lab{portvf2}
V_t (\phi )  = \, \, & V_0(\phi )+ \sum_{i=1}^d \int_{(0,t]} \xi^i_u \, d(S^i_u + \pA^i_u )
+  \sum_{i=1}^d \int_0^t \psi^{i,b}_u \, d\Bibr_u \\
&+ \int_0^t \psi^{l}_u \, d\Blr_u + \int_0^t \psi^{b}_u \, d\Bbr_u + \pA_t .
\end{align*}
The following result gives the wealth dynamics in the present setup.

\bcor \lab{cortccd}
Assume that $\Bilr = \Blr$ for all $i =1,2, \dots ,d$ and $\psi^{l}_t \geq 0$ and $\psi^{b}_t \leq 0$ for all $t \in [0,T]$.
Then, under assumptions \eqref{biigy0} and \eqref{biigy2}, the dynamics of $V(\phi)=V(x, \phi , A)$ are
\begin{align}  \lab{cb4v4}
dV_t (\phi )  = \, \, & \sum_{i=1}^d \xi^i_t \, (dS^i_t + d\pA^i_t)
- \sum_{i=1}^d ( \xi^i_t S^i_t )^+  (\Bibr_t)^{-1} \, d\Bibr_t + d\pA_t \\
&+  \Big( V_t (\phi ) + \sum_{i=1}^d ( \xi^i_t S^i_t )^- \Big)^+ (\Blr_t)^{-1} \, d\Blr_t
 -  \Big( V_t (\phi ) + \sum_{i=1}^d ( \xi^i_t S^i_t )^- \Big)^- (\Bbr_t)^{-1}\, d\Bbr_t . \nonumber
\end{align}
\ecor

Note that even under an additional assumption that $\Bibr = \Bbr$ for all $i=1,2, \dots , d$, expression \eqref{cb4v4} does not reduce to \eqref{oo44}, since we work here under postulate \eqref{biigy2}, which explicitly states that a long cash position
 in the $i$th risky asset is funded exclusively from the account $\Bibr$.

\bex \lab{ex3b} Under the assumptions of Corollary \ref{cortccd} if, in addition, all account processes $\Bilr$ and $\Bbr$
are absolutely continuous, then \eqref{cb4v4} becomes
\begin{align}  \lab{cbx33}
dV_t(\phi )  = \, \, & \sum_{i=1}^d \xi^i_t \big(dS^i_t + d\pA^i_t  \big)
- \sum_{i=1}^d \ribb_t( \xi^i_t S^i_t )^+  \, dt + d\pA_t
 \\ &+   \rll_t \Big( V_t (\phi ) + \sum_{i=1}^d ( \xi^i_t S^i_t )^- \Big)^+ \, dt
- \rbb_t \Big( V_t (\phi ) + \sum_{i=1}^d ( \xi^i_t S^i_t )^- \Big)^- \, dt \nonumber
\end{align}
and thus the funding costs satisfy
\bde
dF_t(\phi ) =  \rll_t  \Big( V_t (\phi ) + \sum_{i=1}^d ( \xi^i_t S^i_t )^- \Big)^+
 \, dt - \rbb_t  \Big( V_t (\phi ) + \sum_{i=1}^d ( \xi^i_t S^i_t )^- \Big)^- \, dt
 - \sum_{i=1}^d \ribb_t(\xi^i_t S^i_t )^+  \, dt .
\ede
\eex

%%%%%%%%%%%%%%%%%%%%%%%%%%%%%%%%     SECTION 3     %%%%%%%%%%%%%%%%%%%%%%%%%%%%%%%%%%%%%%
%%%%%%%%%%%%%%%%%%%%%%%%%%%%%%%%%%%%%%%%%%%%%%%%%%%%%%%%%%%%%%%%%%%%%%%%%%%%%%%%%%%%%%%%%
%%%%%%%%%%%%%%%%%%%%%%%%%%%%%%%%%%%%%%%%%%%%%%%%%%%%%%%%%%%%%%%%%%%%%%%%%%%%%%%%%%%%%%%%%
\section{Pricing under Funding Costs}\label{Sec2}
%%%%%%%%%%%%%%%%%%%%%%%%%%%%%%%%%%%%%%%%%%%%%%%%%%%%%%%%%%%%%%%%%%%%%%%%%%%%%%%%%%%%%%%%%
%%%%%%%%%%%%%%%%%%%%%%%%%%%%%%%%%%%%%%%%%%%%%%%%%%%%%%%%%%%%%%%%%%%%%%%%%%%%%%%%%%%%%%%%%
%%%%%%%%%%%%%%%%%%%%%%%%%%%%%%%%%%%%%%%%%%%%%%%%%%%%%%%%%%%%%%%%%%%%%%%%%%%%%%%%%%%%%%%%%

Our goal in the preceding section was to analyze the wealth dynamics for self-financing strategies
under alternative assumptions about trading and netting. In the next step, we will
provide sufficient conditions for the no-arbitrage property of a market model under various trading specifications.
It is worth stressing that this issue is apparently overlooked in most papers dealing with funding
costs and collateralization. Instead, most authors work under an ad hoc postulate of the existence of a
very vaguely specified `martingale measure' and they focus on the `risk-neutral valuation' under this probability measure.
Most likely,  a `martingale measure' in these papers should then be interpreted as a `pricing' probability measure, which is obtained from the market data via a model's calibration, rather than a sound theoretical construct.
The main contribution of the existing vast literature in this vein thus lies in a thorough analysis of market conventions regarding margin account and closeout  payoff at default and numerical implementations of sophisticated models for risky assets and default times. By contrast, their authors show relatively little interest in searching for a sound theoretical underpinning of alternative computations of various funding and credit risk adjustments to the so-called `clean' prices.

Obviously, this tentative approach to valuation adjustments hinges on mimicking the classic results for frictionless market models.
However, due to peculiarities in the wealth dynamics under nowadays ubiquitous market frictions, the classic approach should be carefully reexamined, since its straightforward application is manifestly unjustified. To clarify this statement, we will now analyze the applicability of classic paradigms when dealing with trading under funding costs. Specifically, in Sections \ref{tyht} and \ref{tyyht}, respectively, we will address two different, albeit related, questions.

Our first question reads: given a market model and a contract $A$ with an exogenously specified
market price, is it possible for the hedger to produce a risk-free profit by taking a long hedged position in $A$
and simultaneously assuming a short unhedged position in the same contract (note that the level of the market price of $A$
is not relevant for this problem)? If this is the case, the model is manifestly not viable for the hedger, since for any level of the market price for a contract $A$, he would be able to guarantee a risk-free profit for himself. Otherwise, we say that a model is {\it arbitrage-free for the hedger} with respect to a given contract $A$. Intuitively, the level of a model's viability rises when this desirable property holds for a sufficiently large class of contracts that encompasses~$A$.

 The second question is: assuming that the model is arbitrage-free for the hedger with respect to a contract $A$ (or some class of contracts that encompasses $A$), we would like to describe all possible levels of a hedger's price $p$, such that the hedger cannot make a risk-free profit by selling the contract at price $p$ and implementing a smart trading strategy $(\phi ,A)$?  Any number $p$ satisfying this property is referred to as a {\it fair hedger's price} for a contract $A$.

We thus see that the first question deals with a possibility of making a risk-free profit by the hedger through taking back-to-back
offsetting positions in a contract $A$ at an exogenously given market price $p$ for $A$ (and the market price $-p$ for $-A$), whereas the second problem addresses the situation when the hedger is an outright seller of a contract $A$ at price $p$. Let us observe that the issue how to quantify a `risk-free profit' should be carefully analyzed as well, especially when the lending and borrowing rates differ. We will argue that thanks to a judicious specification of the netted wealth process, it is possible to give formal definitions that also enjoy plausible financial interpretations. It should be acknowledged, however, that we do not offer a satisfactory solutions to all problems arising in the context of a non-linear and asymmetric pricing, so several important issues are merely outlined.

%%%%%%%%%%%%%%%%%%%%%%%%%%%%%%%%%%%%%%%%%%%%%%%%%%%%%%%%%%%%%
\ssc{Hedger's Arbitrage under Funding Costs} \lab{tyht} \lab{abop0}
%%%%%%%%%%%%%%%%%%%%%%%%%%%%%%%%%%%%%%%%%%%%%%%%%%%%%%%%%%%%%

The arbitrage-free property of a model under funding costs is a non-trivial concept, even when no margin account (collateral) is involved. However, in some cases it can indeed be dealt with using a judicious description of an arbitrage opportunity and a suitably defined `martingale measure'. Let us stress that the notion of a martingale measure in the present setup is far from obvious and indeed its definition will depend on adopted market conventions. Specifically, for
each particular market convention, an astute choice of a definition is required in order to make this general concept useful for our purposes, namely, for verifying whether a given market model is arbitrage-free and for valuing OTC derivatives.

%%%%%%%%%%%%%%%%%%%%%%%%%%%%%%%%%%%%%%%%%%%%%%%%%%%%%%%%
\sssc{Generic Market Model}\lab{GMM}
%%%%%%%%%%%%%%%%%%%%%%%%%%%%%%%%%%%%%%%%%%%%%%%%%%%%%%%%

By a {\it generic market model}, we mean a general class of models encompassing, but not restricted to, all cases
of trading arrangements considered in the preceding section. We only assume that the concept of the wealth process
$V(x, \phi , A )$ and the discounted wealth  $\wh V(x, \phi , A )$ are well defined, where the choice of a discount
factor is fairly arbitrary and thus it may depend on particular circumstances at hand. Hence all market models introduced in
Section \ref{mm} should now be seen as particular instances of a generic market model.

Since, in principle, the lending and borrowing accounts, $\Blr$ and $\Bbr$ may be different in a generic market model,
the netted wealth is defined by the following natural extension of Definition \ref{nhnhnh}.
For the interpretation of the concept of the netted wealth, see Section \ref{secelem}.

\bd \lab{nhnhnx}
The {\it netted wealth} $\Vnet(x, \phi , \pA )$ of a trading strategy $(x, \phi, \pA)$ is given by the equality
$\Vnet(x, \phi , \pA ) = V(x, \phi , \pA ) + V(0, \wt \phi  , -\pA )$ where $(0,  \wt \phi ,-A)$
is the unique self-financing strategy satisfying the following conditions: \hfill \break
(i) $V_0(0, \wt \phi  , -\pA ) = - A_0 $, \hfill \break
(ii) the equalities $\xi^i_t = \psi^i_t = 0$ hold for every $i=1,2,\dots ,d$ and all $t \in [0,T]$, \hfill \break
(iii) $\wt{\psi}^l_t \geq 0 , \wt{\psi}^b_t \leq 0$ and  $\wt{\psi}^l_t \wt{\psi}^b_t =0$ for all $t \in [0,T]$.
\ed

We note that
\bde
\Vnet_0 (x, \phi , \pA ) = V_0(x, \phi , \pA ) + V_0(0, \wt \phi  , -\pA )= x+ A_0 - A_0=x
\ede
so that the initial netted wealth $\Vnet (x, \phi , \pA )$ is independent of $p$. In view of Assumption \ref{hyp23}, we set $A_0=0$ when using the concept of the netted wealth. According to the financial interpretation, the initial cash flows $A_0$ and $-A_0$ cancel out if the market prices of $A$ and $-A$ satisfy $p^m_0(-A)=-p^m_0(A)$, so that this assumption is reasonable (albeit it reduces slightly the generality of our approach). The following result, which is an extension of Lemma \ref{nhnhn},
is also valid in a model in which some form of netting of positions in risky assets is postulated.

\bl \lab{nhnhnx1}
The following equality holds, for all $t \in [0,T]$, {\rm
\be \lab{xportf2c}
\Vnet_t (x, \phi , \pA ) =  V_t (x, \phi , \pA ) + U_t(A)
\ee }
where the $\gg$-adapted process of finite variation $U(A)$ is the unique solution to the following equation
\be \lab{pyy2}
U_t(A) = \int_0^t (\Blr_u)^{-1} ( U_u(A) )^+\, d\Blr_u - \int_0^t (\Bbr_u)^{-1} ( U_u(A) )^-\, d\Bbr_u - \pA_t .
\ee
\el

\proof
We set $\xi^i_t = \psi^i_t = 0$ in \eqref{vifi} and \eqref{sfvifi}. Then the process $V_t := V_t(0, \wt{\psi}^l, \wt{\psi}^b , -\pA )$ satisfies $V_t = \wt{\psi}^{l}_t \Blr_t + \wt{\psi}^{b}_t \Bbr_t$ and
\bde
V_t = \int_0^t (\Blr_u)^{-1} ( V_u  )^+\, d\Blr_u - \int_0^t (\Bbr_u)^{-1} ( V_u )^- \, d\Bbr_u - \pA_t .
\ede
Hence the assertion of the lemma follows.
\endproof

The next definition is an extension of the classic definition of an arbitrage opportunity,
which is suitable when dealing with the basic model with funding costs. Let us stress that we only
consider here the classic concept of an {\it arbitrage opportunity}. For an exhaustive study of alternative
versions of no-arbitrage conditions, the interested reader may consult the recent paper by Fontana \cite{Fon}.

Let $x$ be an arbitrary real number. We denote by $\VLL (x)$ the wealth process
of a self-financing strategy $(x,\phi^0,0)$ where $\phi^0$ is the portfolio with all components equal to zero,
except for $\psi^0$ (resp. $\psi^l$ and $\psi^b$ if the lending and borrowing rates are different).
It is easy to see that the wealth process $\VLL (x)$ is uniquely specified by $x$ and these
conditions, specifically, it equals $xB $ (resp. $x^+ \Blr - x^- \Bbr$).
For any $t\in (0,T]$, the random variable $\VLL_t (x)$ represents the future value at time $t$ of the hedger's initial endowment $x$. For a given contract $A$, an arbitrage opportunity for the hedger arises if,  through a clever choice of a dynamic portfolio $\phi$, he can generate a higher netted wealth at $T$ than the future value of his initial endowment.
The issue of {\it admissibility} of trading strategy needs to be examined for each model at hand (see, for instance,
Definition \ref{defiadmi}).

\bd \lab{abop0dd}
An admissible trading strategy $(x, \phi ,A)$ is an {\it arbitrage opportunity for the hedger} with respect to $A$ whenever the following conditions are satisfied:
$\P ( \Vnet_T(x, \phi , A ) \geq \VLL_T (x))=1$ and $\P ( \Vnet_T (x, \phi , A ) > \VLL_T (x) ) > 0 $.
\ed

 Definition \ref{abop0dd} states that the hedger with the initial endowment $x$ can produce an arbitrage opportunity by entering into a  contract $A$, if he can find an admissible strategy $(x,\phi , A)$ and such that the netted wealth at the contract's maturity date $T$ is always no less than $ \VLL_T(x)$, and is strictly greater than $ \VLL_T(x)$ with a positive probability.

Let us consider the classic case when $\Blr =\Bbr = \Bilr = \Bibr = B$ for all $i$. Then for any contract $A$,
 due to the additivity of self-financing strategies in the classic setting, for any self-financing strategy $(x,\phi ,A)$,
 we obtain
 \begin{align*}
 \Vnet (x,\phi ,A) - \VLL (x) & = V(x, \phi , \pA ) + V(0, \wt \phi  , -\pA ) - V (x, \phi^0, 0) 
  = V(0, \phi + \wt \phi - \phi^0  , 0 ) = V(0, \wh \phi )
 \end{align*}
where $V(0, \wh{\phi })$ is the wealth process of a trading strategy $\wh{\phi}$, which is self-financing in the usual sense.
Also, if $\wh{\phi}$ is any self-financing trading strategy in the classic sense, then we may set $x=0$ and $A=0$, so that
$V(0, \wh{\phi }) =\Vnet(0, \wh{\phi },0)$.

\brem It is fair to acknowledge that Definition \ref{abop0dd} is only the first step towards a more general view of arbitrage opportunities that might arise in the context of differing funding costs and credit qualities of potential counterparties.
A more sophisticated approach relies on a comparison of two opposite dynamically hedged positions,
so that we would end up with the following condition: an {\it extended arbitrage opportunity} is a pair $(x_1, \phi , A)$ and
$(x_2 , \wt \phi , -A )$ of admissible strategies where $x_1+x_2 = x$ and
\begin{align*}
&\P (V_T(x_1,\phi , A ) + V_T(x_2,\wt \phi , -A) \geq V^0_T(x) ) =1 , \\
&\P ( V_T(x_1, \phi , A ) + V_T(x_2,\wt \phi , -A) > V^0_T(x) ) > 0.
\end{align*}
This more general view means that an arbitrage opportunity can also be created by taking advantage of the presence of two potential counterparties with identical or different creditworthiness. The extended definition requires the possibility of taking back-to-back offsetting positions in OTC deals with identical contractual features, but initiated with different counterparties. Therefore, a minimal trading model now includes the hedger and his two counterparties. For further results in this vein, see Section 3.2 in Nie and Rutkowski \cite{NR2} where the model with partial netting is examined in detail.
\erem

The arguments in favor of Definition \ref{abop0dd} can be summarized as follows: \hfill \break
$\bullet \, $ in specific cases of market models, its implementation is relatively easy, \hfill \break
$\bullet \, $ it yields explicit conditions that make financial sense, and \hfill \break
$\bullet \, $ last but not least, it can be used to clarify and justify the use of the concept
of a {\it martingale measure} in the general setup of a market with funding costs, collateralization and defaults.
%Of course, more work is still needed to demonstrate the last feature.

To sum up, although Definition \ref{abop0dd} could be further refined, it nevertheless seems to be a
sufficient tool to deal with the issue of arbitrage in a non-linear trading environment.
Using Definition \ref{abop0dd}, we may now introduce the notion of an arbitrage-free model either with respect
to all contracts that can be covered by a particular model or by selecting first a particular class ${\cal A}$
of contracts of our interest. Note that, in principle, the arbitrage-free property may depend on the
hedger's initial endowment $x$.

\bd \lab{defarbi}
We say that a generic market model is {\it arbitrage-free} for the hedger with respect to the class ${\cal A}$
of financial contracts whenever no arbitrage opportunity associated with any contract $A$ from the class ${\cal A}$
exists in the class of all trading strategies admissible for the hedger. In other words, a model is arbitrage-free if for any contract $A \in {\cal A}$ and any admissible strategy $(x, \phi , A)$ for the hedger, we have that either $\P \big(  \Vnet_T (x, \phi , A ) = \VLL_T (x) \big) =1$ or $\P \big(  \Vnet_T (x, \phi , A ) < \VLL_T (x) \big) > 0 .$
\ed

Let us stress that if a model is arbitrage-free for the hedger, it is not necessarily true that it is arbitrage-free for the counterparty as well. Observe also that in the classic case when $\Blr =\Bbr = \Bilr = \Bibr = B$ for all $i$,
Definition \ref{defarbi} reduces to the classic definition of an arbitrage-free market model.
Hence, as expected, the methodology developed here agrees with the standard arbitrage pricing theory
if there are no frictions in trading strategies or, at least, when they do not affect the class of
contracts at hand, so that they can be safely ignored.

%%%%%%%%%%%%%%%%%%%%%%%%%%%%%%%%%%%%%%%%%%%%%%%%%%%%%%%%%%%%%
\sssc{Basic Model with Funding Costs}
%%%%%%%%%%%%%%%%%%%%%%%%%%%%%%%%%%%%%%%%%%%%%%%%%%%%%%%%%%%%%

Let us now specify the concepts introduced in the preceding subsection to the basic model with funding costs of Section \ref{secelem} with the cash account $B^0=B$. We now have that $\VLL_T(x) = xB_T$ and thus conditions of Definition \ref{abop0dd} become
\be \lab{7446c}
\P (\Vnet_T (x, \phi ,A) \geq x B_T )=1, \quad \P (\Vnet_T (x,\phi ,A) > xB_T ) > 0
\ee
or, equivalently,
\bde
\P ( \Vnett_T (x, \phi ,A) \geq x )=1, \quad \P (\Vnett_T (x,\phi ,A) > x ) > 0
\ede
where the netted wealth $\Vnet (x, \phi ,A)$ is given by Definition \ref{nhnhnh} or, equivalently, Lemma \ref{nhnhn}.
Recall also that for an arbitrary self-financing trading strategy $(x, \phi ,A)$, equation \eqref{class1} yields
\be  \lab{prtf3b}
\Vnett_t (x,\phi ,A) = x + \sum_{i=1}^d \int_{(0,t]} \xi^i_u \wt B^i_u \, d\wh S^{i,\textrm{cld}}_u  + \sum_{i=1}^d \int_0^t ( \psi^i_u + \xi^i_u \wh S^{i}_u ) \, d\wt B^i_u .
\ee
We thus observe that to examine the arbitrage-free property of the basic model, it suffices to consider trading strategies with
null initial value. In other words, the no-arbitrage property of the basic model does not depend on the hedger's initial endowment.
Note that according to \eqref{prtf3b}, the hedger's trading in risky assets is unrestricted, meaning that each risky asset can be funded from arbitrarily chosen funding accounts. In addition, we make the usual postulate that a strategy $(x,\phi ,A)$ need to satisfy some form of admissibility. In the framework of the basic model with funding costs, we adopt the following definition of the class of admissible strategies; they are usually referred to as {\it tame strategies.}

\bd \lab{defiadmi}
A self-financing trading strategy $(x,\phi ,A)$ is {\it admissible for the hedger} whenever the discounted netted wealth process $\Vnett(x, \phi ,A)$ is bounded from below by a constant.
\ed

The condition that the discounted netted wealth process $\Vnett(x, \phi , A )$ is bounded
from below by a constant is a commonly used requirement of {\it admissibility}, which ensures that, if
the process $\Vnett(x, \phi , A )$ a local martingale under some equivalent probability measure, then it is also
a supermartingale. It is well known that some technical assumption of this kind cannot be avoided even in the classic case of
the Black and Scholes model. Let us stress that the choice of a discount factor was left unspecified in Definition \ref{abop0dd}.
If a constant mentioned in Definition \ref{defiadmi} equals zero, so that the netted wealth of an admissible strategy is bound to
stay non-negative, then it suffices to consider the netted wealth without any discounting and thus the choice of a discount factor in Definition \ref{abop0dd} is manifestly irrelevant. Otherwise, this choice will depend on the problem and model under study (see, for instance, Proposition \ref{prparb1}).

\bl \lab{lm766}
Assume that for any admissible trading strategy $( \phi , A)$  there exists a probability measure $\PT^{\phi ,A} $ on $(\Omega , \G_T)$ such that $\PT^{\phi ,A}$ is equivalent to $\P$ and the process $\Vnett (x, \phi ,A)$ is a $(\PT^{\phi ,A}, \gg)$-local martingale. Then the basic market model with funding costs is arbitrage-free for the hedger.
\el

A probability measure $\PT^{\phi ,A} $  is then called a {\it equivalent local martingale measure} (ELMM) for the process  $\Vnett (x, \phi ,A)$. Of course, the sufficient condition of Lemma \ref{lm766} is very cumbersome to check, in general, and thus it does not seem to be of practical interest. For this reason, we will search for more explicit conditions that will be relatively easy to verify. They will refer to the existence of some universal equivalent local martingale measure for a given trading framework and
for a sufficiently large class of contracts under study.

To this end, we will first re-examine the concepts of an arbitrage opportunity and arbitrage price since,
as we will argue in what follows, the classic definitions do not reflect adequately the present general framework.
In particular, we show that the study of the arbitrage-free property of a market model cannot be separated from
an analysis of hedging strategies for a given class of contracts. This is due to the fact that the presence of
either incoming or outgoing cash flows associated with a contract (that is, external cash flows $A$) may exert a non-additive impact on the dynamics of the wealth process, and thus also on the total gains and/or losses from hedger's trading activities.

Obviously, if there exists $B^k \ne B$, then an arbitrage opportunity arises. Indeed, it is easy to produce it by taking $\xi^1 = \ldots = \xi^d=0$ and $\psi^j =0$ for every $j$, except for $j=k$. Then we obtain
\bde
\Vnett_t (x,\phi ,A) = x + \int_0^t  \psi^k_u \, d\wt B^k_u
\ede
and thus we see that the existence of an equivalent local martingale measure  $\PT^{\phi ,A} $ for the process $\Vnett(x,\phi ,A)$ is by no means ensured, in general. Therefore, additional conditions need to be imposed on the class of trading strategies and/or funding rates to guarantee that the basic model with funding costs is arbitrage-free. In the next result, we thus preclude the occurrence of a mixed funding for any risky asset. Recall that condition \eqref{conee} holds if, for instance, the equality $\psi^i_t B^i_t + \xi^i_t S^{i}_t =0 $ is satisfied for all $t \in [0,T]$. We write $\Q \sim \P $ to denote that the probability measures $\Q $ and $\P $ are equivalent on $(\Omega , \G_T)$.

\bp \lab{proarb1}
Assume that all strategies available to the hedger are admissible and satisfy condition \eqref{conee}.
If there exists a probability measure $\PT$ on $(\Omega , \G_T)$ such that $\PT \sim \P$ and the processes $\wh S^{i,\textrm{cld}},\, i=1,2,\dots, d$ are $(\PT, \gg)$-local martingales then the basic model with funding costs is arbitrage-free for the hedger.
\ep

\proof
It suffices to observe that, under the present assumptions, equation  \eqref{prtf3b}  reduces to
\be  \lab{xxccvv}
\Vnett_t (x, \phi ,A) = x + \sum_{i=1}^d \int_{(0,t]} \xi^i_u \wt B^i_u \, d\wh S^{i,\textrm{cld}}_u
\ee
and to apply the usual argument that any local martingale (or even a sigma-martingale, which may arise
as a stochastic integral in \eqref{xxccvv} when the semimartingales
$S^1,S^2,\dots, S^d$ are not continuous) that is bounded from below by a constant,
is necessarily a supermartingale.
\endproof

%%%%%%%%%%%%%%%%%%%%%%%%%%%%%%%%%%%%%%%%%%%%%%%%%%%%%%%%%%%%%%
\ssc{Hedger's Fair Valuation under Funding Costs} \lab{tyyht}
%%%%%%%%%%%%%%%%%%%%%%%%%%%%%%%%%%%%%%%%%%%%%%%%%%%%%%%%%%%%%%

In the next step, we focus on fair pricing of contracts. The fair pricing of contracts
in a market model that allows for arbitrage opportunities is  obviously not viable, so we henceforth work under the
standing assumption that a model under study is arbitrage-free for the hedger with a given initial endowment $x$
and for a sufficiently large class ${\cal A}$ of contracts, which encompasses a given contract $A$ (see Definition \ref{defarbi}).

Our goal is to propose a realistic definition of a {\it hedger's fair price} and to show how to apply it to some models with funding costs. Let us observe that the definition of an arbitrage-free model is not symmetric, that is, a model in which no arbitrage opportunities for the hedger exist may still allow for arbitrage opportunities for the counterparty. Moreover, even when the market conditions are identical for both parties, they have the same initial endowment and a given model is arbitrage-free for both parties, the cash flows of a contract are obviously asymmetric and thus the range of fair prices computed by the two counterparties may be different. By the usual convention adopted throughout this paper, we will focus the following discussion on one party, which is called the hedger.

%%%%%%%%%%%%%%%%%%%%%%%%%%%%%%%%%%%%%%%%%%%%%%%%%%%%%%%%%%%%%%%%%%%%%
\sssc{Generic Market Model}
%%%%%%%%%%%%%%%%%%%%%%%%%%%%%%%%%%%%%%%%%%%%%%%%%%%%%%%%%%%%%%%%%%%%%

Our next goal is to describe the range of hedger's arbitrage prices of a contract with cash flows $A$.
Let $x$ be the hedger's initial endowment and let $p$ stand for a generic price of a contract at time 0 from the perspective of the hedger. A positive value of $p$ means that the hedger receives at time 0 the cash amount $p$  from the counterparty, whereas a
negative value of $p$ means that he makes the payment $-p$ to the counterparty at time 0. It is clear from the next
definition that a {\it hedger's price} may depend on the hedger's initial  endowment $x$ and it may fail to be unique,
in general.

It is important to stress that the {\it admissibility} of a trading strategy is now defined using the discounted
wealth, as opposed to the discounted netted wealth, as was the case in Section \ref{tyht}. For this reason,
to avoid a possibility of confusion with Definition \ref{defiadmi}, we decided to state explicitly the
admissibility condition  each time when it is relevant. Moreover, the choice of a discount factor depends on a model under consideration,  but it is otherwise fairly arbitrary, so that the {\it discounted wealth process}, formally represented
 by the generic symbol $\wh V(x, \phi , A )$, is not necessarily given as $B^{-1} V(x , \phi , A )$. For instance, in Section \ref{sscsec}, the discounted wealth will be given by $\wh V(x , \phi , A ):= (\Blr)^{-1} V(x, \phi , \pA )$. As a rule of thumb, we suggest that the choice of discounting should be the same when we address either the first or the second question stated at the beginning of this section.

%\bd
%A self-financing trading strategy $(\phi ,A)$ is {\it admissible for the hedger} whenever the discounted wealth process
%$\wt V(\phi ,A)$ is bounded from below by a constant.
%\ed

\bd \lab{deffi7}
We say that a real number $\ppff = A_0$ is a {\it hedger's fair price} for $A$ at time 0 whenever for any self-financing trading strategy $(x, \phi ,A)$ such that the discounted wealth process $\wh V(x , \phi , A )$ is bounded from below by a constant, we have that either
\be \lab{vvy33a}
\P \big( V_T (x, \phi , A ) = \VLL_T (x) \big) = 1
\ee
or
\be \lab{vv33a}
\P \big( V_T (x, \phi , A ) < \VLL_T (x) \big) > 0 .
\ee
\ed

One may observe that the two conditions in Definition \ref{deffi7} are analogous to conditions of Definition \ref{defarbi},
although they are not identical and, indeed, they have quite different financial interpretations. Recall that Definition \ref{defarbi} deals with a possibility of offsetting a dynamically hedged contract $A$ by an unhedged contract $-A$, whereas Definition \ref{deffi7} is concerned with finding a fair price for $A$ from the viewpoint of the hedger as a contract's seller. In the latter case, it is natural to say that a price level $\ppff $ is too high for the hedger, if he can produce an arbitrage opportunity (in the sense that is implicit in Definition \ref{deffi7}) by selling $A$ at price $\ppff$ and devising a suitable hedging strategy for his short position. Once again, the hedger's profits are measured with respect to his idiosyncratic cost of raising cash or, more precisely, with respect to the future value of his current endowment $x$, as represented by the random variable $\VLL_T(x)$. This leads to the following natural definition of a hedger's arbitrage opportunity for $A$ at price $p$. As usual in the arbitrage pricing theory, we need to postulate that trading strategies are admissible.

\bd \lab{deffi7b}
We say that a quadruplet $(p ,x, \phi ,A)$, where $p = A_0$ is a real number and $(x,\phi ,A)$ is an admissible trading strategy such that the discounted wealth process $\wh V(x , \phi , A )$ is bounded from below by a constant, is a {\it hedger's arbitrage opportunity for $A$ at price} $p$ if
\bde
\P \big( V_T ( x , \phi , A ) \geq \VLL_T(x) \big) = 1
\ede
and
\bde
\P \big( V_T ( x , \phi , A ) > \VLL_T(x) \big) > 0.
\ede
\ed

Assume that the hedger has the initial endowment $x$ and he sells the contract $A$ at price $\ppff$. Then $\ppff$ is a  hedger's fair price for $A$, in the sense of Definition \ref{deffi7}, whenever he is not able to find an arbitrage opportunity for $A$ at price $p=\ppff$, in the sense of Definition \ref{deffi7b}.

In practice, the hedger's initial endowment $x<0$ can be interpreted as the amount of cash borrowed by the trading desk from the bank's internal funding unit, which should be repaid with interest $\Bbr_T$ at a given horizon date $T$.  Therefore, an arbitrage opportunity at price $\ppff$ for $A$ means that the price $\ppff$ is high enough to allow the hedger to make a risk-free profit, where the `profits' are assessed in relation to the hedger's idiosyncratic cost of capital, as formally represented by the account $\Bbr$.

%%%%%%%%%%%%%%%%%%%%%%%%%%%%%%%%%%%%%%%%%%%%%%%%%%%%%%%%%%%%%%%%%%%%%
\sssc{Basic Model with Funding Costs } \lab{sscfir}
%%%%%%%%%%%%%%%%%%%%%%%%%%%%%%%%%%%%%%%%%%%%%%%%%%%%%%%%%%%%%%%%%%%%%

In the basic model with funding costs, we obtain the following result, which bears a close resemblance to its classic counterpart, which deals with a market model with a single cash account. Note that we work here under the assumptions of Proposition \ref{proarb1} so that the basic model with funding costs is arbitrage-free for the hedger with respect to any contract $A$.
Recall that here $\Blr= \Bbr = B$, so that the discounted wealth is defined as
$\wt{V}(x,\phi ,A) = B^{-1} V(x, \phi ,A)$,  the admissibility is specified by Definition \ref{defiadmi},
and trading strategies are assumed to satisfy condition \eqref{conee}.

\bp
Under the assumptions of Proposition \ref{proarb1}, a real number $\ppff$ is a hedger's fair price whenever,
for any admissible trading strategy $(x,\phi ,A)$ satisfying condition \eqref{conee}, we have that
either
\bde
\P \bigg( \ppff + \sum_{i=1}^d  \int _{(0,T]} \xi^i_u \wt{B}^i_u \, d\wh S^{i,\textrm{cld}}_u
+ \int_{(0,T]} B_u^{-1} \, d\pA_u = 0 \bigg) = 1
\ede
or
\bde
\P \bigg( \ppff + \sum_{i=1}^d  \int _{(0,T]}  \xi^i_u \wt{B}^i_u \, d\wh S^{i,\textrm{cld}}_u
+ \int_{(0,T]} B_u^{-1} \, d\pA_u < 0 \bigg) > 0 .
\ede
\ep

\proof
It suffices to combine Definition \ref{deffi7} with equation  \eqref{clacss1}.
\endproof

Note that, in this basic framework where $\Blr= \Bbr = B$,  the set of all hedger's fair prices does not depend on the hedger's initial endowment  $x$, although it manifestly depends on funding accounts $B^i,\, i=1,2, \dots ,d$. Also, the real-world probability measure $\P$ can be replaced by an equivalent local martingale measure $\PT$ for processes $\wh S^{i,\textrm{cld}},\, i=1,2, \dots ,d$.

As an example, let us take $A_t = -X \I_{\{t=T\}}$ and let us assume $B^i = B$ for every $i=1, 2, \dots , d$.
Then we obtain the following characterization of a hedger's price $\ppff $: for any admissible trading strategy $(\phi ,A)$, either
\bde
\P \bigg( \ppff + \sum_{i=1}^d  \int _{(0,T]} \xi^i_u \, d\wt S^{i,\textrm{cld}}_u = B^{-1}_T X \bigg) = 1
\ede
or
\bde
\P \bigg( \ppff + \sum_{i=1}^d  \int _{(0,T]} \xi^i_u \, d\wt S^{i,\textrm{cld}}_u < B^{-1}_T X \bigg) > 0.
\ede
We recognize here the classic case, namely, the notion of the hedger's fair price
as an arbitrary level of $\ppff$ that does not allow for creation of a hedger's super-hedging strategy for a European claim $X$.

%%%%%%%%%%%%%%%%%%%%%%%%%%%%%%%%%%%%%%%%%%%%%%%%%%%%%%%%%%%%%%%%%%%%%
\ssc{Illustrative Example} \lab{eexxam}
%%%%%%%%%%%%%%%%%%%%%%%%%%%%%%%%%%%%%%%%%%%%%%%%%%%%%%%%%%%%%%%%%%%%%

As a simple illustration of pricing problems studied in this section, we propose to consider the extension of the Black-Scholes model to the case of different lending and borrowing rates, which satisfy $r^b \geq r^l \geq 0$. It is known that this model is arbitrage-free in the classic sense when one considers self-financing trading strategies with a non-negative wealth (see, for instance, Bergman \cite{Bergman} and Example 1.1 in El Karoui et al. \cite{EKPQ}). We assume that the hedger's initial endowment $x$ satisfies $x>0$ and we complement our model by a simple contract $A$ with only two cash flows after time 0, namely, an outgoing cash flow of $\bar{A}_1 := \cac $ units of cash at time $0<t_0<T$ and an incoming cash flow of $\bar{A}_2 := \cac e^{\wh{r}(T-t_0)}$ units of cash at time $T$.  Hence the contract $A$ can be seen as a loan of $\cac $ units of cash granted at time $t_0$ by the hedger to the counterparty at a continuously compounded interest $\wh{r}$. Note that the existence of the account associated with the rate $\wh{r}$ is not postulated.
According to Definition \ref{defcocc}, the process $A$ is thus given by $A_t = - \bar{A}_1 \I_{[t_0,T]} (t) +  \bar{A}_2 \I_{[T]} (t)$ for every $t \in [0,T]$.

%%%%%%%%%%%%%%%%%%%%%%%%%%%%%%%%%%%%%%%%%%%%%%%%%%%%%
\sssc{Hedger's Arbitrage} \lab{eexxam1}
%%%%%%%%%%%%%%%%%%%%%%%%%%%%%%%%%%%%%%%%%%%%%%%%%%%%%

We first focus on the hedger's arbitrage, in the sense of Definition \ref{abop0dd}, when a contract $A$ is available to the hedger. To this end, we temporarily assume that the initial price $p$ of this contract at time 0 is unspecified.
Let us consider a self-financing strategy $(x,\wh{\phi} ,A)$ in which the initial endowment $x$ is invested in lending and borrowing accounts $\Blr$ and $\Bbr$ only, and a part of this investment is used by the hedger at time $t_0$ to pay $c$ units of cash to the counterparty. Note that it is postulated that no investment in shares of the risky asset is ever made by the hedger. Hence, from   \eqref{rtf1} and \eqref{portf2}, we obtain
$V_t(x, \wh{\phi }, \pA ) = \psi^l_t \Blr_t + \psi^b_t \Bbr_t$ where, by assumption,
$\psi^{l}_t \geq 0,\, \psi^{b}_t \leq 0$ and $\psi^{i,l}_t \psi^{i,b}_t =0$
and
\be  \lab{prortf2}
V_t(x, \wh{\phi }, \pA ) = x  + \int_0^t \psi^l_u \, d\Blr_u +  \int_0^t \psi^b_u \, d\Bbr_u + \pA_t .
\ee
If we assume that $x e^{r^l t_0} \geq \cac $, then the unique strategy $\wh{\phi } = (\psi^l , \psi^b)$ satisfying these assumptions is given as:
$\psi^b_t=0$ for all $t \in [0,T]$, and
\bde
\psi^l_t = x \I_{[0,t_0)} + \frac{\widetilde{x}}{\Blr_{t_0}}  \I_{[t_0,T)} +  \frac{\wh{x}}{\Blr_{T}}  \I_{[T,T]}, \quad t \in [0,T],
\ede
where
\bde
\widetilde{x} = x e^{r^l t_0} - \cac , \quad \wh{x} = x e^{r^l T} - \cac  e^{r^l(T-t_0)} + \cac  e^{\wh{r}(T-t_0)}.
\ede
Hence, in this case, the wealth of the hedger's strategy $(\wh{\phi} ,A)$ at time $T$ equals
\be \lab{5353}
V_T (x, \wh{\phi} , A) = \big( x e^{r^l t_0} - \cac  \big)  e^{r^l (T-t_0)} + \cac e^{\wh{r}(T-t_0)}
=  x e^{r^l T} - \cac  e^{r^l(T-t_0)} + \cac  e^{\wh{r}(T-t_0)} .
\ee
If, on the contrary, the inequality $x e^{r^l t_0}< \cac $ is valid, then the hedger's wealth at $T$ necessarily
satisfies
\bde
V_T (x, \wh{\phi} , A) = \big( x e^{r^l t_0} - \cac  \big)  e^{r^b(T-t_0)} +\cac  e^{\wh{r}(T-t_0)}
=  x e^{r^l t_0} e^{r^b(T-t_0)} - \cac e^{r^b(T-t_0)} + \cac e^{\wh{r}(T-t_0)}
\ede
since now the unique portfolio $\wh{\phi } = (\psi^l , \psi^b)$ available to the hedger involves borrowing of $\cac - x e^{r^l t_0}$ units of cash at time $t_0$ (this is needed to pay $\cac $ units of cash to the counterparty). Similar arguments show that
if we set $x=0$ and consider the contract $-A$, then the wealth at $T$ of the unique portfolio $\wt{\phi } = ( \wt \psi^l , \wt \psi^b)$ available to the hedger equals
\bde
V_T(0, \wt \phi , -\pA ) = \cac e^{r^l (T-t_0)} - \cac e^{\wh{r}(T-t_0)}.
\ede
We thus see that, if $x e^{r^l t_0} \geq \cac $, then the netted wealth equals
\begin{align*}
 \Vnet_T (x,\wh{\phi} , \pA ) & =   V_T(x, \wh{\phi} , \pA ) + V_T(0, \wt{\phi }  , -\pA ) \\
 & =  x e^{r^l T} - \cac  e^{r^l(T-t_0)} + \cac  e^{\wh{r}(T-t_0)} + \cac e^{r^l(T-t_0)} - \cac e^{\wh{r}(T-t_0)}   \\
 &= x e^{r^l T} = \VLL_T(x)
\end{align*}
and for $x e^{r^l t_0} < \cac $ it satisfies
\begin{align*}
 \Vnet_T (x,\wh{\phi} , \pA ) & =   V_T(x, \wh{\phi }, \pA ) + V_T(0, \wt{\phi }  , -\pA ) \\
 & =  x e^{r^l t_0} e^{r^b(T-t_0)} - \cac e^{r^b(T-t_0)} + \cac e^{\wh{r}(T-t_0)}
  + \cac e^{r^l(T-t_0)} - \cac e^{\wh{r}(T-t_0)}   \\
 & =  x e^{r^l T} + \big( \cac  -  x e^{r^l t_0} \big) \big( e^{r^l(T-t_0)} - e^{r^b(T-t_0)} \big) \leq
   x e^{r^l T} = \VLL_T(x)
\end{align*}
where the last inequality is strict whenever $r^b > r^l$. This means that the unique strategy $(\wh{\phi} ,A)$ does not constitute an arbitrage opportunity for the hedger, in the sense of Definition \ref{abop0dd}. Of course, more sophisticated hedger's strategies $(\phi ,A)$ should be examined as well, but it is unlikely that an arbitrage opportunity for the hedger may arise when a possibility of investing in the risky asset is also taken into account.

\sssc{Hedger's Fair Valuation} \lab{eexxam2}
%%%%%%%%%%%%%%%%%%%%%%%%%%%%%%%%%%%%%%%%%%%%%%%%%%%%%

We will now focus on fair valuation of $A$ from the hedger's perspective.
Let us assume, in addition, that  $\wh{r} > r^l$ and $x e^{r^l t_0} \geq \cac $, so that
equation \eqref{5353} with $A_0=0$ yields $V_T (x, \wh{\phi }, A) > x e^{r^l T}$, and thus it is obvious that $p=0$ is not the fair hedger's price for the contract, in the sense of Definition  \ref{deffi7}. We thus expect that a hedger's fair price for $A$ is necessarily a strictly negative number. If we still postulate that the hedger does not invest in the risky asset and the inequalities
\be \lab{qmml}
x+p \geq 0 , \quad (x+p) e^{r^lt_0} > \cac
\ee
are satisfied, then we obtain for $A_0=p$
\bde
V_T (x , \wh{\phi} , A) = \big( (x+p)e^{r^l t_0} - \cac \big)  e^{r^l (T-t_0)} + \cac e^{\wh{r}(T-t_0)}
=  (x+p) e^{r^l T} - \cac e^{r^l(T-t_0)} + \cac e^{\wh{r}(T-t_0)}
\ede
Obviously, the equality  $V_T (x , \wh{\phi} , A)= x e^{r^l T }$ holds whenever
\be \lab{qmmld}
p = \cac e^{- r^l T} \big( e^{r^l(T-t_0)} - e^{\wh{r}(T-t_0)} \big).
\ee
Observe that $p$ given by \eqref{qmmld} is strictly negative, since we assumed that  $\wh{r} > r^l$.
It is thus natural to conjecture that the value given by \eqref{qmmld} is the upper bound for hedger's fair prices for $A$, in the sense of Definition \ref{deffi7}, provided that $p$ given by the formula above is such that conditions \eqref{qmml}
are also met, that is, the absolute value of $p$ given by \eqref{qmmld} is indeed sufficiently small with respect
to $x$. Otherwise, the computations leading to the fair value of $p$ should be modified accordingly and a different result is expected. This example, albeit stylized and not solved completely, shows that the classic arbitrage pricing techniques should indeed be modified when dealing with more realistic models of trading by financial institutions.

%%%%%%%%%%%%%%%%%%%%%%%%%%%%%%%%%%%%%%%%%%%%%%%%%%%%%%%%%%%%%%%%%%%%%%%
\ssc{Model with Funding Costs and Partial Netting}  \lab{sscsec}
%%%%%%%%%%%%%%%%%%%%%%%%%%%%%%%%%%%%%%%%%%%%%%%%%%%%%%%%%%%%%%%%%%%%%%%

To provide a non-trivial illustration of the novel concepts introduced in this section,
let us consider the market model with partial netting of short cash positions, which was introduced in Section \ref{sec2.2.3}.
Recall that, in principle, the choice of a discount factor is
unrestricted, so any particular choice is motivated by convenience for the problem at hand.
Let the hedger's initial endowment be $x \geq 0$.
We will first show that, under mild assumptions, the model is arbitrage-free for the hedger with respect to a contract $A$.
To this end, we define the discounted wealth and the discounted wealth
of $(\phi ,A)$ by setting $\wt V^l_t (x, \phi , \pA ) := (\Blr_t)^{-1} V_t(x, \phi , \pA )$ and
$\Vnettl_t (x , \phi , \pA ) := (\Blr_t)^{-1} \Vnet_t (x, \phi , \pA )$, respectively.
The choice of $\Blr$ for discounting is related here to the assumption that $x \geq 0$; when
$x<0$ it is more natural to take $\Bbr$ instead, since in that case the hedger has a debt at time 0 that has
to be repaid with interest determined by $\Bbr$.

%%%%%%%%%%%%%%%%%%%%%%%%%%%%%%%%%%%%%%%%%%%%%%%%%%%%%
\sssc{Hedger's Arbitrage} \lab{eeoxxam1}
%%%%%%%%%%%%%%%%%%%%%%%%%%%%%%%%%%%%%%%%%%%%%%%%%%%%%

The following result hinges on a plausible assumption that all borrowing rates $\ribb $ are higher than the common lending rate $\rll$.
Note that we assume here that all cash accounts are absolutely continuous.

\bp \lab{prparb1}
Assume that $x \geq 0 , \, \rll_t \leq \rbb_t$ and $\rll_t \leq \ribb_t$ for $i=1,2, \dots , d$.
Let us denote
\be \label{qqww3}
\wt S^{i,l,{\textrm{cld}}}_t =  (\Blr_t)^{-1}S^i_t + \int_{(0,t]} (\Blr_u)^{-1} \, d\pA^i_u .
\ee
If there exists a probability measure $\PT^l \sim \P $ such that the
processes $\wt S^{i,l,\textrm{cld}},\, i=1,2, \dots ,d$ are $(\PT^l , \gg)$-local martingales,
then the market model of Section \ref{sec2.2.3} is arbitrage-free for the hedger with respect to
any contract $A$.
\ep

\proof
From Corollary \ref{cortccd}, we know that the wealth process $V(x, \phi , \pA )$ of a self-financing strategy $(x, \phi ,A)$ satisfies (see equation \eqref{cbx33})
\begin{align*}
dV_t(x,\phi , \pA )  = \, \, & \sum_{i=1}^d \xi^i_t \big(dS^i_t + d\pA^i_t  \big)
- \sum_{i=1}^d \ribb_t( \xi^i_t S^i_t )^+  \, dt + d \pA_t
 \\ &+   \rll_t \Big( V_t (x,\phi , \pA ) + \sum_{i=1}^d ( \xi^i_t S^i_t )^- \Big)^+ \, dt
- \rbb_t \Big( V_t (x,\phi , \pA ) + \sum_{i=1}^d ( \xi^i_t S^i_t )^- \Big)^- \, dt . \nonumber
\end{align*}
Since we assumed that $\rll_t \leq \rbb_t$, we obtain
\begin{align*}
dV_t(x,\phi , \pA )  \leq \, \, & \sum_{i=1}^d \xi^i_t \big(dS^i_t + d\pA^i_t  \big)
- \sum_{i=1}^d \ribb_t( \xi^i_t S^i_t )^+  \, dt + d\pA_t
 \\ &+   \rll_t \Big( V_t (x,\phi , \pA ) + \sum_{i=1}^d ( \xi^i_t S^i_t )^- \Big)^+ \, dt
- \rll_t \Big( V_t (x,\phi , \pA ) + \sum_{i=1}^d ( \xi^i_t S^i_t )^- \Big)^- \, dt \\
= \, \, &\rll_t  V_t (x,\phi , \pA )\, dt + \sum_{i=1}^d \xi^i_t \big(dS^i_t + d\pA^i_t  \big) + d\pA_t
- \sum_{i=1}^d \ribb_t( \xi^i_t S^i_t )^+  \, dt +   \rll_t \sum_{i=1}^d ( \xi^i_t S^i_t )^- \, dt
 \\ \leq \, \, &\rll_t  V_t (x,\phi , \pA )\, dt + \sum_{i=1}^d \xi^i_t \big(dS^i_t - \rll_t S^i_t \, dt + d\pA^i_t \big)
  + d\pA_t
\end{align*}
where the last inequality holds, since it is also postulated that $\rll_t \leq \ribb_t$.
Consequently, the discounted wealth $\wt V^{l}_t (x,\phi , \pA ) = (\Blr_t)^{-1} V_t(x,\phi , \pA )$
satisfies
\bde
d \wt V^{l}_t (x,\phi , \pA ) \leq
\sum_{i=1}^d \xi^i_t (\Blr_t)^{-1} \big( dS^i_t - \rll_t S^i_t \, dt + d\pA^i_t  \big) + (\Blr_t)^{-1} \, d\pA_t
\ede
and thus, in view of \eqref{qqww3}, we obtain
\bde % \label{qqwxw3}
d \wt V^{l}_t (x,\phi , \pA ) \leq  \sum_{i=1}^d \xi^i_t \, d\wt S^{i,l,{\textrm{cld}}}_t + (\Blr_t)^{-1} \, d\pA_t .
\ede
Furthermore, the netted wealth equals $\Vnet_t (x,\phi , \pA ) =  V_t (x,\phi , \pA ) + U_t(A)$  (see Lemma \ref{nhnhnx1})
where the $\gg$-adapted process of finite variation $U(A)$ is the unique solution to the following equation
\be \lab{44tty}
U_t(A) = \int_0^t (\Blr_u)^{-1} ( U_u(A) )^+\, d\Blr_u - \int_0^t (\Bbr_u)^{-1} ( U_u(A) )^- \, d\Bbr_u - \pA_t
\ee
where $U(\pA ) = (U(\pA ))^+ - (U(\pA ))^-$ is the decomposition of the process $U(\pA )$ into its increasing and decreasing components. Hence the netted discounted wealth $\Vnettl_t (x,\phi , \pA ) := (\Blr_t)^{-1} \Vnet_t (x,\phi , \pA )$ satisfies
\begin{align*}
&d\Vnettl_t (x,\phi , \pA ) =  d\wt{V}^l_t (x,\phi , \pA ) +  d(  (\Blr_t)^{-1} U_t(A) ) \\
&\leq   \sum_{i=1}^d \xi^i_t \, d\wt S^{i,l,{\textrm{cld}}}_t
+ (\Blr_t)^{-2} ( U_t(A) )^+\, d\Blr_t - (\Blr_t)^{-1} (\Bbr_t)^{-1} ( U_t(A) )^- \, d\Bbr_t + U_t(A) \, d(\Blr_t)^{-1} \\
&=  \sum_{i=1}^d \xi^i_t \, d\wt S^{i,l,{\textrm{cld}}}_t
+ r^l_t (\Blr_t)^{-1} ( U_t(A) )^+\, dt -  r^b_t (\Blr_t)^{-1} ( U_t(A) )^- \, dt - r^l_t (\Blr_t)^{-1} U_t(A) \, dt\\
& = \sum_{i=1}^d \xi^i_t \, d\wt S^{i,l,{\textrm{cld}}}_t + (r^l_t - r^b_t )(\Blr_t)^{-1} ( U_t(A) )^- \, dt
\end{align*}
and thus
\be \label{vgyh}
\Vnettl_t (x,\phi , \pA ) - \Vnettl_0 (x,\phi , \pA ) \leq \sum_{i=1}^d \int_{(0,t]} \xi^i_u \, d\wt S^{i,l,{\textrm{cld}}}_u .
\ee
The arbitrage-free property of the model for the hedger can  now be established using the standard arguments.
First, from \eqref{vgyh} and the assumption that the process $\Vnettl (x,\phi , \pA )$ is bounded from below by a constant, we deduce that the right-hand side in \eqref{vgyh} is a $(\PT^l,\ff)$-supermartingale, which is null at $t=0$. Next, since the initial endowment  $x$ is non-negative, we have that $\VLL_T(x) = \Blr_T x$. From inequality \eqref{vgyh}, we obtain
\bde
 (\Blr_T)^{-1} \big( \Vnet_T (x,\phi , A ) - \VLL_T (x) \big) \leq \sum_{i=1}^d \int_0^T \xi^i_t \, d\wt S^{i,l,{\textrm{cld}}}_t .
\ede
Since $\PT^l $ is equivalent to $\P$, we conclude that either the equality $\Vnet_T (x,\phi , A ) = \VLL_T(x)$
holds or the inequality $\P ( \Vnet_T (x,\phi , A ) < \VLL_T(x))>0$ is satisfied.
This means that arbitrage opportunities are indeed precluded and thus the market model with partial netting is arbitrage-free
for the hedger in respect of any contract $A$.
\endproof

\brem  \lab{remzz1}
We claim that assertion of Proposition \ref{prparb1} is also true for $x\leq 0$ under the stronger assumption that $r^b \leq r^{i,b}$ for all $i$ provided that the processes $\wt S^{i,l,\textrm{cld}},\, i=1,2, \dots ,d$ are replaced by $\wt S^{i,b,\textrm{cld}},\, i=1,2, \dots ,d$, where the process $\wt S^{i,b,\textrm{cld}}$ is obtained by replacing $\Blr$ by $\Bbr$ in the right-hand side of \eqref{qqww3}.  Since $\rll \leq \rbb \leq \ribb $, we now obtain
\begin{align*}
dV_t(x,\phi , \pA )  \leq \, \, & \sum_{i=1}^d \xi^i_t \big(dS^i_t + d\pA^i_t  \big)
- \sum_{i=1}^d \ribb_t( \xi^i_t S^i_t )^+  \, dt + d\pA_t
 \\ &+   \rbb_t \Big( V_t (x,\phi , \pA ) + \sum_{i=1}^d ( \xi^i_t S^i_t )^- \Big)^+ \, dt
- \rbb_t \Big( V_t (x,\phi , \pA ) + \sum_{i=1}^d ( \xi^i_t S^i_t )^- \Big)^- \, dt \\
= \, \, &\rbb_t  V_t (x,\phi , \pA )\, dt + \sum_{i=1}^d \xi^i_t \big(dS^i_t + d\pA^i_t  \big) + d\pA_t
- \sum_{i=1}^d \ribb_t( \xi^i_t S^i_t )^+  \, dt +   \rbb_t \sum_{i=1}^d ( \xi^i_t S^i_t )^- \, dt
 \\ \leq \, \, &\rbb_t  V_t (x,\phi , \pA )\, dt + \sum_{i=1}^d \xi^i_t \big(dS^i_t - \rbb_t S^i_t \, dt + d\pA^i_t \big)
  + d\pA_t .
\end{align*}
Therefore,  the discounted wealth $\wt V^{b}_t (x,\phi , \pA ) = (\Bbr_t)^{-1} V_t(x,\phi , \pA )$ satisfies
\bde
d \wt V^{b}_t (x,\phi , \pA ) \leq  \sum_{i=1}^d \xi^i_t \, d\wt S^{i,b,{\textrm{cld}}}_t + (\Bbr_t)^{-1} \, d\pA_t
\ede
and for the netted discounted wealth $\Vnettb_t (x,\phi , \pA ) := (\Bbr_t)^{-1} \Vnet_t (x,\phi , \pA )$, we obtain
from \eqref{44tty}
%\bde
%d\Vnettb_t (x,\phi , \pA ) =  d\wt{V}^b_t (x,\phi , \pA ) +  d(  (\Bbr_t)^{-1} U_t(A) ) \leq 0
%\ede
\begin{align*}
&d\Vnettb_t (x,\phi , \pA ) =  d\wt{V}^b_t (x,\phi , \pA ) +  d(  (\Bbr_t)^{-1} U_t(A) ) \\
&\leq   \sum_{i=1}^d \xi^i_t \, d\wt S^{i,b,{\textrm{cld}}}_t
+ (\Bbr_t)^{-1} (\Blr_t)^{-1} ( U_t(A) )^+\, d\Blr_t - (\Bbr_t)^{-2}(U_t(A) )^- \, d\Bbr_t + U_t(A) \, d(\Bbr_t)^{-1} \\
&=  \sum_{i=1}^d \xi^i_t \, d\wt S^{i,b,{\textrm{cld}}}_t
+ r^l_t (\Bbr_t)^{-1} ( U_t(A) )^+\, dt -  r^b_t (\Bbr_t)^{-1} ( U_t(A) )^- \, dt - r^b_t (\Bbr_t)^{-1} U_t(A) \, dt .
\end{align*}
Since $r^l \leq r^b$, this yields
\bde
 (\Bbr_T)^{-1} \big( \Vnet_T (x,\phi , A ) - \VLL_T (x) \big) \leq \sum_{i=1}^d \int_0^T \xi^i_t \, d\wt S^{i,b,{\textrm{cld}}}_t
\ede
where $\VLL_T(x) = \Bbr_T x$ since $ x \leq 0$. Hence the conclusion follows if there exists a probability measure $\PT^b \sim \P $ such that the processes $\wt S^{i,b,\textrm{cld}},\, i=1,2, \dots ,d$ are $(\PT^b , \gg)$-local martingales.
\erem

%%%%%%%%%%%%%%%%%%%%%%%%%%%%%%%%%%%%%%%%%%%%%%%%%%%%%
\sssc{Hedger's Fair Valuation} \lab{eeoxxam2}
%%%%%%%%%%%%%%%%%%%%%%%%%%%%%%%%%%%%%%%%%%%%%%%%%%%%%

We now address the issue of the hedger's fair valuation of a contract $A$.
In the present setup, Definition \ref{deffi7} is applied to the discounted wealth $\wh V(x , \phi , A )$
with $A_0 = \ppff $, which is given by the following equation
\bde
\wh V_t(x , \phi , A )= \wt{V}^l_t (x, \phi , \pA )= (\Blr_t)^{-1} V_t(x, \phi , \pA ),
\ede
that is, the admissibility of a trading strategy $(x,\phi ,A)$ is defined using the discounted wealth
$\wt{V}^l_t (x, \phi , \pA )$. In view of Corollary \ref{cortccd} (see also equation \eqref{cbx33}), the set of hedger's fair prices $\ppff$ in the model with partial netting can be characterized as follows: for any admissible strategy $(x, \phi ,A)$, we have that either
\begin{align*}
\P \bigg(& x + \ppff + \sum_{i=1}^d \int _{(0,T]} \xi^i_t \big(dS^i_t + d\pA^i_t  \big)
- \sum_{i=1}^d \int_0^T \ribb_t( \xi^i_t S^i_t )^+  \, dt + \int_{(0,T]} dA_t
 \\ &+  \int_0^T \rll_t \Big( V_t (x, \phi , \pA ) + \sum_{i=1}^d ( \xi^i_t S^i_t )^- \Big)^+ \, dt
-\int_0^T \rbb_t \Big( V_t (x, \phi , \pA ) + \sum_{i=1}^d ( \xi^i_t S^i_t )^- \Big)^- \, dt
 < \VLL_T(x) \bigg) > 0  \nonumber
\end{align*}
or
\begin{align*}
\P \bigg(& x + \ppff + \sum_{i=1}^d \int _{(0,T]} \xi^i_t \big(dS^i_t + d\pA^i_t  \big)
- \sum_{i=1}^d \int_0^T \ribb_t( \xi^i_t S^i_t )^+  \, dt + \int_{(0,T]} dA_t
 \\ &+  \int_0^T \rll_t \Big( V_t (x, \phi , \pA ) + \sum_{i=1}^d ( \xi^i_t S^i_t )^- \Big)^+ \, dt
-\int_0^T \rbb_t \Big( V_t (x, \phi , \pA ) + \sum_{i=1}^d ( \xi^i_t S^i_t )^- \Big)^- \, dt
 = \VLL_T(x) \bigg) = 1 .  \nonumber
\end{align*}
 It is clear that that $\int_{(0,T]} dA_u = A_T-A_0$. However, the terms $\ppff$ and $-A_0$ do not cancel out in the formulae above, since $\ppff $ is the yet unknown initial fair price of the contract, whereas the random variable $A_T - A_0$ represents all contract's cash flows on $(0,T]$, and thus it is explicitly specified through the contract's covenants. Of course, this formal characterization of a fair price $\ppff$ does not offer any tangible computational algorithm. Hence, in the next step, one needs to develop more explicit methods for finding fair prices (for instance, via a suitable extension of the BSDE approach, which is examined in Section \ref{BSDErr}).

%%%%%%%%%%%%%%%%%%%%%%%%%%%%%%%%     SECTION 4     %%%%%%%%%%%%%%%%%%%%%%%%%%%%%%%%%%%%%%
%%%%%%%%%%%%%%%%%%%%%%%%%%%%%%%%%%%%%%%%%%%%%%%%%%%%%%%%%%%%%%%%%%%%%%%%%%%%%%%%%%%%%%%%%
%%%%%%%%%%%%%%%%%%%%%%%%%%%%%%%%%%%%%%%%%%%%%%%%%%%%%%%%%%%%%%%%%%%%%%%%%%%%%%%%%%%%%%%%%
\section{Trading under Funding Costs and Collateralization} \lab{seccoll}
%%%%%%%%%%%%%%%%%%%%%%%%%%%%%%%%%%%%%%%%%%%%%%%%%%%%%%%%%%%%%%%%%%%%%%%%%%%%%%%%%%%%%%%%%
%%%%%%%%%%%%%%%%%%%%%%%%%%%%%%%%%%%%%%%%%%%%%%%%%%%%%%%%%%%%%%%%%%%%%%%%%%%%%%%%%%%%%%%%%
%%%%%%%%%%%%%%%%%%%%%%%%%%%%%%%%%%%%%%%%%%%%%%%%%%%%%%%%%%%%%%%%%%%%%%%%%%%%%%%%%%%%%%%%%

In this section, we will examine the situation when the hedger enters a contract with cash flows $A$ and either receives or posts collateral with the value formally represented by a stochastic process~$\pC $. The process $C$ is called the {\it margin account} or the {\it collateral amount} and the mechanism of either posting or receiving a collateral is referred to as {\it margining}. Let
\be \lab{collss}
\pC_t = \pC_t \I_{\{ \pC_t \geq 0\}} +  \pC_t \I_{\{ \pC_t < 0\}} = \pC^+_t - \pC^-_t
\ee
be the usual decomposition of the random variable $\pC_t$ into the positive and negative components. By convention, $\pC^+_t$ is
the cash value of collateral received by the hedger, whereas $\pC^-_t$ represents the cash value of collateral posted by him.

For simplicity of presentation, it is postulated throughout that only shares of particular {\it collateral assets}, henceforth denoted by $S^{d+1}$ (resp. $S^{d+2}$) may be delivered (resp. received) by the hedger as a collateral. In principle, this assumption can be relaxed to cover the case where a collateral asset is not predetermined, but it may be chosen from a larger class of assets. However, the notation and computations would become heavier, so we decided to consider a simple case only.
Unless explicitly stated otherwise, we work under the following standing assumptions: \hfill \break
(a)  lending and borrowing cash rates $\Blr$ and $\Bbr$ are equal, so that $\Blr = \Bbr = B$,  \hfill \break
(b) long and short funding rates for each risky asset $S^i$ are identical, that is, $\Bilr=\Bibr=B^i$ for
$i=1, 2, \dots, d$.

We make the following standing assumption regarding the behavior of the margin account at the contract's
maturity date.

\bhyp \lab{cxzcxz}
We postulate that the $\gg$-adapted collateral amount process satisfies $C_T=0$. Hence any particular specification of the
collateral amount $C_t$ discussed in that follows will only be valid for $0 \leq t <T$ and, invariably, we set $C_T=0$.
\ehyp

The postulated equality $C_T=0$ is a convenient way of ensuring that any collateral amount posted is returned in full to its owner when a contract matures, provided that the default event does not occur at $T$. Of course, if the default event is also modeled,
then one needs to specify the closeout payoff. Let us mention that the case of an exogenously given collateral was studied in
\cite{NR2,NR4}, whereas the case of an endogenous collateral (as given, for instance by equation \eqref{ty6v} below) was examined in \cite{NR5}.

%%%%%%%%%%%%%%%%%%%%%%%%%%%%%%%%%%%%%%%%%%%%%%%%%%%%
\ssc{Collateral Conventions}
%%%%%%%%%%%%%%%%%%%%%%%%%%%%%%%%%%%%%%%%%%%%%%%%%%%%

In the market practice, the complexity of the issue of collateralization is enormous and obviously beyond the scope of this work,
in which we will only focus on the impact of collateralization on the dynamics of the hedger's portfolio and thus on
the valuation from the perspective of the hedger. Let us first make some comments regarding the crucial features of the margin accounts that underpin our stylized approach to the costs of margining. As usual, we take the perspective of the hedger.
\begin{itemize}
\item
The current financial practice typically requires the collateral amounts to be held in {\it segregated} margin accounts,
so that the hedger, when he is a collateral taker, cannot make use of the collateral amount for trading.
Therefore, under segregation the hedger's wealth dynamics do not depend
on whether the collateral amount was posted by the counterparty in cash or shares of a risky asset $S^{d+2}$. By contrast,
the character of delivered assets always matters to him when the hedger is a collateral giver.
\item Another collateral convention encountered  in practice is {\it rehypothecation}, which refers to the situation where a bank is allowed to reuse the collateral pledged by its counterparties as collateral for its own borrowing.
In our approach to rehypothecation, we will distinguish between the case when
the collateral amount was delivered to the hedger in the form of shares of a risky asset (and thus it can only be reused as a collateral) and the case of cash collateral where it can be used for an outright trading.
\item If the hedger is a collateral giver, then a particular convention regarding segregation or rehypothecation is immaterial for the wealth dynamics of his portfolio.  Of course, the distinction between segregation and rehypothecation becomes important when the closeout payoff at default of either the hedger or the counterparty is evaluated. However,
the latter issue, as well as a rather complex mechanism of updating the margin account, are left aside, since they were already thoroughly studied in the literature.
\end{itemize}

We first introduce the general notation, which will be used when analyzing various conventions regarding collateralization.
Let us make clear that we set here to introduce an abstract setup, which is flexible enough to cover various collateral convention.
By contrast, we do not pretend that any particular convention should be seen as a prevailing or desirable one.

\bd \lab{tsx2}
A {\it collateralized hedger's trading strategy} is a quadruplet $(x,\phi , \pA , \pC )$ where a portfolio $\phi $, given by
\be \lab{vty}
\phi = \big( \xi^1,\dots , \xi^{d+1},\psi^0 ,\dots ,\psi^{d+1}, \etab, \etal , \eta^{d+2}\big)
\ee
is composed of the {\it risky assets} $S^i,\, i=1,2,\ldots,d+1$, the {\it unsecured cash account} $B^0=B$, the {\it funding accounts} $B^i,\, i=1,2,\ldots,d+1$, the {\it borrowing account} $B^{d+1}$  for the posted cash collateral,
the {\it collateral accounts} $B^{\pCc,b}$ and $B^{\pCc,l}$, and the {\it lending account} $B^{d+2}$ associated with the
received collateral asset $S^{d+2}$.
\ed

\brem \lab{vgt}
The collateral account $B^{\pCc,b}$ (resp. $B^{\pCc,l}$) plays the following role: if the hedger receives (resp. posts) cash or risky collateral with the equivalent cash value $C^+$ (resp. $C^-$), then he  pays (resp. receives) interest on this nominal amount, as specified by the process $B^{\pCc,b}$ (resp. $B^{\pCc,l}$).
\erem

All funding and collateral accounts are assumed to be continuous processes of finite variation, whereas the price of the
collateral asset $S^{d+1}$ is assumed to be a c\`adl\`ag semimartingale. Let us formulate
two definitions, which clarify the distinction between the conventions of the {\it risky asset collateral} and the {\it cash collateral} and introduce the notation, which will be used in what follows.

\bd \lab{rrcc1}
The {\it risky collateral} is described by the following postulates: \begin{itemize}
\item If the hedger receives at time $t$ the number $\xi^{d+2}_t >0$ of shares of the risky asset $S^{d+2}$ as collateral, then he pays to the counterparty interest determined by the amount $\pC^+_t = \xi^{d+2}_t S^{d+2}_t$ and the collateral account $B^{\pCc,b}$. Formally, there is no reason to postulate that the process $\xi^{d+2}$ is a component of the hedger's trading strategy.
    However, in cases where the collateral amount is related to the hedging strategy, this process is explicitly given in terms of the wealth of
     of the hedger's portfolio.
     Under {\it segregation}, the hedger also receives (possibly null) interest determined by the amount $\pC^+_t$ and the account $B^{d+2,s}$,
     whereas under {\it rehypothecation}, he also receives interest determined by the amount $\pC^+_t$ and the secured funding account $B^{d+2,h}$.
\item If the hedger posts a collateral at time $t$, then he delivers $\xi^{d+1}_t >0$ of shares of the risky asset $S^{d+1}$ funded from the (unsecured) funding account $B^{d+1}$ and he receives interest determined by the amount $\pC^-_t =
    \xi^{d+1}_t S^{d+1}_t$ and the collateral account $B^{\pCc,l}$. Formally, we thus postulate that
\be \lab{5656}
\xi^{d+1}_t S^{d+1}_t = C^-_t, \quad \xi^{d+1}_t S^{d+1}_t + \psi^{d+1}_t B^{d+1}_t = 0 .
\ee
This implies, in particular, that the equality $\psi^{d+1}_t B^{d+1}_t = - C^-_t $ holds for all $t$.
\end{itemize}
\ed

Note that the lending account $B^{d+2}$ is equal to  $B^{d+2,s}$ or $B^{d+2,h}$, depending on the adopted collateral convention.
In practice, deliverable collateral assets should have low credit risk and should be uncorrelated with the underlying trading portfolio. For this reason, it is assumed  in Definition \ref{rrcc1} that, even under rehypothecation, the received risky asset
$S^{d+2}$ cannot be used for hedging purposes, but it is assumed instead to yield interest, denoted by $B^{d+2,h}$, by being pledged as collateral in a repo contract and thus raising an equivalent amount $\pC^+$ of cash.
Note that the hedger's advantages when $S^{d+2}$ is delivered as collateral in his another contract are
not examined here; to this end, we would need to consider a portfolio of hedger's contracts, rather than to focus on a single contract in isolation. Under segregation, the account $B^{d+2,s}$ reflects a (perhaps unlikely in practice) possibility that the hedger receives interest from the collateral custodian, whenever he maintains a positive equivalent cash amount $\pC^+$ in the segregated account. We now move on to the case when all collateral amounts
are delivered in cash.

\bd   \lab{irrcc1}
The {\it cash collateral} is described by the following postulates:
\begin{itemize}
\item  If the hedger receives at time $t$ the amount $\pC^+_t$ as collateral in cash, then he pays
   to the counterparty interest determined by the amount $\pC^+_t$ and the account $B^{\pCc,b}$.
   Under segregation, he receives interest determined by the amount $\pC^+_t$ and the account $B^{d+2,s}$. When rehypothecation is considered,  the hedger may temporarily (that is, before the contract's maturity date or the default time, whichever comes first) utilize the cash amount $\pC^+_t$ for his trading purposes.
\item If the hedger posts a cash collateral at time $t$, then the collateral amount is borrowed from
 the dedicated {\it collateral borrowing account} $B^{d+1}$ (which, of course, may coincide with $B$). He receives interest determined by the amount $\pC^-_t$ and the collateral account $B^{\pCc,l}$. Instead of \eqref{5656}, we now postulate that
\be \lab{565656}
\xi^{d+1}_t =0, \quad  \psi^{d+1}_t B^{d+1}_t = - C^-_t .
\ee
\end{itemize}
\ed

In the context of a collateralized contract, we find it convenient to introduce the following three processes:
\hfill \break $\bullet $ the process $V_t(x,\phi , \pA ,\pC)$ representing the hedger's wealth at time $t$,
\hfill \break $\bullet $ the process $V^p(x,\phi , \pA ,\pC)$ representing the value of hedger's portfolio at time $t$,
\hfill \break $\bullet $ the adjustment process $\VCc_t(x,\phi , \pA ,\pC) := V_t(x,\phi , \pA ,\pC) - V^p_t(x,\phi , \pA ,\pC)$,
 which measures the impact of the margin account.

An explicit specification of the process $\VCc(x,\phi , \pA ,\pC)$ depends on the adopted collateral convention, however,
we always have that $\VCc(x,\phi , \pA ,\pC)=0$ when $\pC$ vanishes, so that the adjustment is not needed.
Let us consider, for example, the case where $\VCc_t(x,\phi , \pA ,\pC)=-C_t$. Then the portfolio's value satisfies $V^p_t(x,\phi , \pA ,\pC) = V_t(x,\phi , \pA ,\pC)+C_t$, meaning that the hedger also invests in his portfolio of traded assets the collateral amount $C^+_t$
received at time $t$, but when he posts collateral at time $t$, then to compute the portfolio's value, we need to subtract $C^-_t$ from the hedger's wealth. In particular, for the hedger with an initial endowment $x$, at time 0 we have $V_0(x,\phi ,0,0) = x,\,  V_0(x,\phi , \pA ,\pC) = x +A_0 $ and $V^p_0(x,\phi , \pA ,\pC) = x + A_0 + C_0$, where the first two equalities are always true and, in general, the
last one reads $V^p_0(x,\phi , \pA ,\pC) = x + A_0 - V_0^c(x,\phi , \pA ,\pC)$.

We are now in a position to formally define the processes $V(x,\phi , \pA ,\pC),\, V^p_t(x,\phi , \pA ,\pC)$
and $\VCc_t(x,\phi , \pA ,\pC)$ in our framework. For alternative explicit specifications of the process $\eta^{d+2}$, we refer to Propositions \ref{collssp}, \ref{xcollssp} and \ref{ttmmi2}. Similarly, in the next definition, we formally identify $B^{d+2,h}$ with $B^{d+2,s}$ and we denote them generically as $B^{d+2}$. This is possible, since these two accounts will play a similar role in our further computations, although their financial interpretation is different and thus in practice they are not necessarily equal.

\bd \lab{ts2x}
The hedger's {\it portfolio's value} $V^p(x,\phi , \pA ,\pC)$ is given by
\be \lab{poutf1}
V^p_t (x,\phi , \pA ,\pC) =  \sum_{i=1}^{d+1} \xi^i_t S^i_t + \sum_{j=0}^{d+1} \psi^j_t B^j_t .
\ee
The hedger's {\it wealth} $V(x,\phi , \pA ,\pC)$ equals
\be \lab{portf1}
V_t (x,\phi , \pA ,\pC) =  \sum_{i=1}^{d+1} \xi^i_tS^i_t + \sum_{j=0}^{d+1} \psi^j_tB^j_t
+  \etab_t B^{\pCc,b}_t+ \etal_t B^{\pCc,l}_t +  \eta^{d+2}_t B^{d+2}_t  .
\ee
The {\it adjustment process} $\VCc(x,\phi , \pA ,\pC)$ satisfies
\be \lab{portf1b}
 \VCc_t (x,\phi , \pA ,\pC) = \etab_t B^{\pCc,b}_t+ \etal_t B^{\pCc,l}_t +  \eta^{d+2}_t B^{d+2}_t
 = - C_t +  \eta^{d+2}_t B^{d+2}_t
\ee
where $\etab_t =-  (B^{\pCc,b}_t)^{-1}\pC_t^+$ and $\etal_t =(B^{\pCc,l}_t)^{-1} \pC_t^-$ (see Remark \ref{vgt}).
\ed

Various specifications of the adjustment process $\VCc(\phi )$ are now encoded in the process denoted generically as $\eta^{d+2}$,
 which will be sometimes complemented by superscripts $s$ or $h$, so that it can also be denoted as $\eta^{d+2,s}$ or $\eta^{d+2,h}$.
 For explicit specifications of these processes, we refer to Propositions \ref{collssp}, \ref{xcollssp} and \ref{ttmmi2}.

 The self-financing property of the hedger's strategy is defined in terms of the dynamics of the value process
 of his portfolio. This definition is a natural extension of Definition \ref{ts1} the the case of collateralized contracts.
 Note that we use here the process $V^p(x,\phi , \pA ,\pC)$, and not $V(x,\phi , \pA ,\pC)$ as was done in Definition \ref{ts1},
 to emphasize the role of $V^p(x,\phi , \pA ,\pC)$ as the value of the hedger's portfolio of traded assets
 (recall also that  $V^p(x,\phi , \pA ,\pC) = V(x,\phi , \pA ,\pC)$ when the process $\pC$ vanishes).

\bd \lab{ts2}
A collateralized hedger's trading strategy $(x,\phi , \pA , \pC )$ with $\phi $ given by \eqref{vty} is {\it self-financing} whenever
the {\it portfolio's value} $V^p(x,\phi , \pA ,\pC)$, which is given by \eqref{poutf1}, satisfies, for every $t \in [0,T]$,
\begin{align} \lab{porfx}
V^p_t (x,\phi , \pA ,\pC)  = \, \, &x + \sum_{i=1}^{d+1} \int_{(0,t]} \xi^i_u \, d(S^i_u + \pA^i_u )
+  \sum_{j=0}^{d+1} \int_0^t \psi^j_u \, dB^j_u  + \pA_t \\
&+ \int_0^t \etab_u\, dB^{\pCc,b}_u + \int_0^t \etal_u \, dB^{\pCc,l}_u + \int_0^t \eta^{d+2}_u\, dB^{d+2}_u
  - \VCc_t (x,\phi , \pA ,\pC). \nonumber
\end{align}
\ed

It is clear that the terms $\int_0^t \etab_u\, dB^{\pCc,b}_u$, $ \int_0^t \etal_u \, dB^{\pCc,l}_u$ and $\int_0^t \eta^{d+2}_u\, dB^{d+2}_u$ represent the cumulative interest due to the presence of the margin account. The first two processes are given explicitly in terms
of $C$ since  $\etab_t =-  (B^{\pCc,b}_t)^{-1}\pC_t^+$ and $\etal_t =(B^{\pCc,l}_t)^{-1} \pC_t^-$, whereas the last one depends on the collateral convention.

\brem
As was already mentioned, the process $\VCc (x,\phi , \pA ,\pC)$ is aimed to measure the impact of the margin account on the part of hedger's wealth that can be used for trading in primary traded assets. Typically, it is given as $\VCc_t (\phi ) = g(C_t(\phi ))$ for some real function $g$ (typically, $g(x)=-x$ or $g(x)=x^-$). Hence, in view of Assumption \ref{cxzcxz}, the equality  $V_T(x,\phi , \pA ,\pC) = V^p_T(x,\phi , \pA ,\pC)$ is always satisfied provided that $g(0)=0$. In the remainder of this work (with one exception, namely, Section \ref{sxsxsx}), we have that either
$\VCc(x,\phi ,A,C)= -C$  or $\VCc (x,\phi ,A,C )= C^-$ for an exogenously give process $C$.
Then equations \eqref{poutf1} and \eqref{porfx} are autonomous, so that they uniquely specify the
portfolio's value $V^p(\phi )$, meaning that we do not use \eqref{portf1} for this purpose.
One can observe that we formally deal here with an example of a self-financing strategy with the wealth $V^p_t (x,\phi , \pA ,\pC)$,
in the sense of Definition \ref{ts1}, but where the process $A$ is substituted with $A- \VCc (x, \phi ,A,C)$.
\erem

%Let us make further comments on the interpretation of equation \eqref{portf1}. It provides a decomposition of the total
%hedger's wealth at time $t$ into the component $V^p_t(\phi )$, which represents the wealth currently invested in risky assets, cash account and funding accounts, and the component $\VCc_t(\phi )$ representing the margin account under segregation or rehypothecation. The process $\VCc_t(\phi )$ depends on a convention regarding the margin account and thus its specification will be crucial for each of results presented  this section. In particular, the corresponding terms in the second line of \eqref{porfx}, which represent the positive and negative cash flows from the margin account, should reflect the covenants described in Definitions \ref{rrcc1} and \ref{irrcc1}. Assuming that the processes $\etab , \etal $ and $\eta^{d+2}$ are known, equalities \eqref{poutf1} and \eqref{porfx} define a self-financing strategy $ (\phi , {\cal A}^c)$ where the process ${\cal A}^c$ is given by the last five terms in the right-hand side in \eqref{porfx}, that is,
%\[
%{\cal A}^c_t = \pA_t + \int_0^t \etab_u\, dB^{\pCc,b}_u + \int_0^t \etal_u \, dB^{\pCc,l}_u + \int_0^t \eta^{d+2}_u\, dB^{d+2}_u
%  - \VCc_t (\phi ).
%\]

 Although these are obviously very important practical issues, neither an explicit specification of the process $C$, nor the rules governing the way in which the margin account is adjusted, are studied in detail here. Let us only remark that the collateral amount is typically tied to the regularly updated {\it marked-to-market value} of a contract, whose level at time $t$ is henceforth denoted as $M_t$.  In that case, the process $\pC$ can be specified as follows
\be \lab{ty6}
\pC_t = (1 +\delta^1_t) M_t \I_{\{ M_t > 0\}} + (1+ \delta^2_t) M_t \I_{\{ M_t < 0\}}
 = (1+\delta^1_t) M^+_t - (1+\delta^2_t) M^-_t
\ee
for some non-negative {\it haircut processes} $\delta^1$ and $\delta^2$. In our theoretical framework, the goal is to develop valuation of a contract based on hedging, so that it is natural to relate the marked-to-market value to the (so far unspecified) hedger's value of a contract. To be more specific, since the wealth process $V(\phi )$ of the hedger is aimed to cover his future liabilities, it is natural to postulate that the stylized `market value' of a contract, as seen by the hedger, coincides with the negative of his wealth. Consequently, we formally identify the marked-to-market value $M$ with the negative of the wealth of the hedger's portfolio. More precisely, one may set
$M_t = V^0_t (x) - V (x,\phi ,A,\pC)$ (see also Definition \ref{defe:replicate} of the ex-dividend price of a contract
$(A,C)$ for a justification of this postulate). Then formula \eqref{ty6} becomes
\be  \lab{ty6v}
\pC_t = \pC_t (\phi ) := (1+\delta^1_t) ( V^0_t(x) - V_t (x,\phi ,A,\pC ))^+
-  (1+ \delta^2_t) ( V^0_t(x)- V_t (x,\phi ,A,\pC))^- .
\ee
The case of a {\it fully collateralized contract} is obtained by setting $\delta^1_t = \delta^2_t=0$ for all $t$ in \eqref{ty6v},
which implies that the process $\pC (\phi )$ is implicitly given by the equation $\pC (\phi )= V^0_t(x) - V(x,\phi ,A, \pC)$. Of course, an analogous analysis can be done for the counterparty. However, since the market conditions will typically be different for the two parties, it is unlikely that their computations of the contract's value (hence the collateral amount) will yield the same value. Obviously, specification \eqref{ty6v} of collateral amount with the wealth $V(\phi )$ computed by the hedger makes practical sense only when it is bilaterally accepted in the contract's CSA ({\it Credit Support Annex}).

The remaining part of this section is organized as follows. First, in Proposition  \ref{collssp}, which covers both segregation and rehypothecation, we derive more explicit representation for the dynamics of the hedger's wealth in the case of a risky collateral. Subsequently, in Propositions  \ref{xcollssp} and  \ref{ttmmi2}, we examine the case of cash collateral under segregation and rehypothecation, respectively.

%%%%%%%%%%%%%%%%%%%%%%%%%%%%%%%%%%%%%%%%%%%%%%%%%%%%%%%%%%%%%%%%%%%%%%%%%%%%%%%
\ssc{Risky Collateral}
%%%%%%%%%%%%%%%%%%%%%%%%%%%%%%%%%%%%%%%%%%%%%%%%%%%%%%%%%%%%%%%%%%%%%%%%%%%%%%%

In this subsection, we work under the assumption that the collateral amount is delivered by the
hedger in the form of shares of the risky asset $S^{d+1}$ and we follow the conventions described in Definition \ref{rrcc1}. In particular,  from \eqref{5656}, we see that the net wealth invested in risky asset $S^{d+1}$ and the account $B^{d+1}$ is null. We denote by $\pFCh$ the process given by
\be \lab{funcos}
\pFCh_t := F^c_t +  \int_0^t\pC^+_u (B^{d+2,h}_u)^{-1} \, dB^{d+2,h}_u
\ee
where $F^c$ is the {\it cumulative interest of margin account}
\be  \lab{hfuncos}
F^c_t := \int_0^t \pC^-_u  (B^{\pCc,l}_u)^{-1} \, dB^{\pCc,l}_u -\int_0^t \pC^+_u (B^{\pCc,b}_u)^{-1} \, dB^{\pCc,b}_u.
\ee
We will show that the process $\pFCh$ represents all positive and negative cash flows from the margin account under rehypothecation, as specified by Definition \ref{rrcc1}. Note that if segregation of the delivered asset $S^{d+2}$ is postulated, then all statements in Proposition \ref{collssp} remain valid if $B^{d+2,h}$ is substituted with $B^{d+2,s}$ and thus this result also covers the case of a segregated risky collateral. In the latter case, the symbol $\pFCh$ will be replaced by $\pFCs$.

\bp \lab{collssp}
Consider the case of the segregated margin account when the collateral is posted in shares of a risky asset
$S^{d+1}$ and received in any form. Assume that a trading strategy $(x,\phi , \pA , \pC )$ is self-financing
and the following equalities hold, for all $t \in [0,T]$,
\begin{align} \label{qqpp1}
\xi^{d+1}_t  =( S^{d+1}_t)^{-1} C^-_t ,\quad \psi^{d+1}_t = -(B^{d+1}_t)^{-1}\pC^-_t  ,
\quad \eta^{{d+2}}_t = (B^{{d+2,h}}_t)^{-1}\pC^+_t .
\end{align}
Then the hedger's wealth $V(\phi ) = V(x, \phi , \pA , \pC )$ equals, for every $t \in [0,T]$,
\be  \lab{portf1v}
V_t (\phi ) = V^p_t (\phi ) + C^-_t = \sum_{i=1}^{d} \xi^i_t S^i_t + \sum_{j=0}^{d} \psi^j_t B^j_t  + C^-_t
\ee
and the dynamics of the portfolio's value $V^p(\phi )= V^p(x, \phi , \pA , \pC )$ are
\be \lab{pof3d1x}
dV^p_t (\phi)  =  \wt V^p_t (\phi) \, dB_t + \sum_{i=1}^{d} \xi^i_t \, dK^i_t
+ (S^{d+1}_t)^{-1}C^-_t\, dK^{d+1}_t + \sum_{i=1}^{d}  \zeta^i_t (\wt B^i_t)^{-1}  \, d\wt B^i_t + d\pACh_t - dC^-_t
\ee
where $\wt{V}^p_t (\phi ) := (B_t)^{-1}V^p_t(\phi )$ and $\pACh := \pA + \pFCh$.
In particular, under assumption \eqref{conee}, we obtain
\be  \lab{xx2i}
dV_t (\phi) =  \wt V_t (\phi) \, dB_t + \sum_{i=1}^{d} \xi^i_t B^i_t \, d\wh S^{i,\textrm{cld}}_t
+ (S^{d+1}_t)^{-1}C^-_t B^{d+1}_t \, d\wh S^{d+1,\textrm{cld}}_t + d\pFChb_t + d\pA_t
\ee
where
\be \lab{fuuncos}
\pFChb_t :=  F^c_t +  \int_0^t \pC^+_u (B^{{d+2,h}}_u)^{-1} \, dB^{{d+2,h}}_u - \int_0^t \pC^-_u (B_u)^{-1} \, dB_u .
\ee
The hedger's wealth admits the following decomposition
\be \lab{portfx}
V_t (\phi ) = x + G_t(\phi ) + F_t(\phi ) + \pFCh_t + \pA_t
\ee
where $G_t(\phi)$ is given by \eqref{gg66} with $d$ replaced by $d+1$ and $F_t(\phi )$ satisfies \eqref{gg666}
with $d$ replaced by $d+1$.
\ep

\begin{proof}
Equality \eqref{portf1v} is an immediate consequence of the specification of $\phi $ and assumptions \eqref{qqpp1}.
We now focus on dynamics of the process $V^p(\phi)$. First, we observe that, in view of \eqref{qqpp1}, we have $ \zeta^{d+1}_t := \xi^{d+1}_t S^{d+1}_t + \psi^{d+1}_t B^{d+1}_t = 0$. Second, the term $\pFCh$, which is deduced from \eqref{porfx} and \eqref{qqpp1}, may be combined with $\pA $ to yield $\pACh = \pA + \pFCh $. We are now in a position to apply Corollary \ref{correx} to the process $V^p(\phi )$ satisfying \eqref{poutf1}--\eqref{porfx}. This yields equality \eqref{pof3d1x}, which in turn after simple computations becomes \eqref{xx2i} when $\zeta^i=0$ for all $i=1,2, \dots ,d$. Finally, decomposition is immediate from  \eqref{portf1v} and \eqref{porfx}.
\end{proof}

\brem  If the assumption that $\Blr = \Bbr = B$ is relaxed, then the dynamics of the portfolio's value (hence also
the hedger's wealth) should be adjusted along the same lines as in Section \ref{sec2.2.1}.
Specifically, if $\zeta^i=0$ for all $i$, then we obtain the following equality, which combines formulae \eqref{sfvifi} and \eqref{pof3d1x},
\begin{align}   \lab{sfpvifi}
dV^p_t (\phi )=  \psi^{l}_t\, d\Blr_t + \psi^{b}_t \, d\Bbr_t + \sum_{i=1}^{d} \xi^i_t \, dK^i_t
+ (S^{d+1}_t)^{-1}C^-_t\, dK^{d+1}_t  + d\pACh_t - dC^-_t
\end{align}
where the processes $\psi^{l}_t$ and  $\psi^{b}_t$ satisfy
\bde
\psi^{l}_t = (\Blr_t)^{-1} \Big( V^p_t (\phi ) -  \sum_{i=1}^d \xi^i_t S^i_t -\sum_{i=1}^d \psi^i_t B^i_t - C^-_t \Big)^+
\ede
and
\bde
\psi^{b}_t = - (\Bbr_t)^{-1} \Big( V^p_t (\phi ) -  \sum_{i=1}^d \xi^i_t S^i_t- \sum_{i=1}^d \psi^i_t B^i_t - C^-_t \Big)^-.
\ede
Formula \eqref{sfpvifi} leads to a suitable extension of Proposition \ref{collssp}.
Similar extensions can be derived for the case of cash collateral; since they are rather straightforward,
they are rather to the reader.
\erem

%%%%%%%%%%%%%%%%%%%%%%%%%%%%%%%%%%%%%%%%%%%%%%%%%%%%%%%%%%%%%%%%%%%%%%%%%%%%%%%
\ssc{Cash Collateral} \lab{casd}
%%%%%%%%%%%%%%%%%%%%%%%%%%%%%%%%%%%%%%%%%%%%%%%%%%%%%%%%%%%%%%%%%%%%%%%%%%%%%%%

In this section, we work under the conventions of cash collateral, as specified in Definition \ref{irrcc1}.
Of course, the risky assets $S^{d+1}$ plays no role in this subsection and thus its existence can be safely ignored.
Formally, we will postulate that $\xi^{d+1}_t  =  0$ for all $t$.

%%%%%%%%%%%%%%%%%%%%%%%%%%%%%%%%%%%%%%%%%%%%%%%%%%%%%%%%%%%%%%%%%%%%%%%%%%%%%%%
\sssc{Margin Account under Segregation}
%%%%%%%%%%%%%%%%%%%%%%%%%%%%%%%%%%%%%%%%%%%%%%%%%%%%%%%%%%%%%%%%%%%%%%%%%%%%%%%

Assume first that the cash amount received by the hedger as collateral cannot be used for trading. Then only the interest
on $C^+$, denoted as $B^{d+2,s}$, matters and the fact that the collateral is received in cash is immaterial here. The cash amount $C^-$ posted by the hedger is borrowed from the account $B^{d+1}$ and it yields interest paid by the counterparty, as determined by the process $B^{\pCc,l}$.  These features of the margin account are reflected through equalities \eqref{qqpp} in the statement of the next result. Recall that the process $\pFCs$ is given by \eqref{funcos} with the superscript $h$ substituted with $s$.

\bp  \lab{xcollssp}
Consider the case of the segregated margin account when the collateral is posted by the hedger in cash borrowed from the
account $B^{d+1}$ and it is received in any form. Assume that a trading strategy $(\phi , \pA , \pC )$ is self-financing
and the following equalities hold, for all $t \in [0,T]$,
\begin{align} \lab{qqpp}
\xi^{d+1}_t  =  0  ,\quad \psi^{d+1}_t = -(B^{d+1}_t)^{-1}\pC^-_t  , \quad \eta^{{d+2,s}}_t = (B^{{d+2,s}}_t)^{-1}\pC^+_t .
\end{align}
Then the hedger's wealth $V(\phi ) = V(x,\phi , \pA , \pC )$ equals, for every $t \in [0,T]$,
\be  \lab{pkrtf1v}
V_t (\phi ) = V^p_t (\phi ) + C^-_t = \sum_{i=1}^{d} \xi^i_t S^i_t + \sum_{j=0}^{d+1} \psi^j_t B^j_t + C^-_t
\ee
and the dynamics of the portfolio's value $V^p(\phi )= V^p(x, \phi , \pA , \pC )$ are
\be \lab{pof3x}
dV^p_t (\phi)  =  \wt V^p_t (\phi) \, dB_t + \sum_{i=1}^{d} \xi^i_t \, dK^i_t
 + \sum_{i=1}^{d}  \zeta^i_t (\wt B^i_t)^{-1}  \, d\wt B^i_t - C^-_t (B^{d+1}_t)^{-1}  \, dB^{d+1}_t + d\pACs_t
 - dC^-_t
\ee
where $\wt{V}^p_t (\phi ) := (B_t)^{-1}V^p_t(\phi )$ and $\pACs := \pA + \pFCs$.
In particular, under assumption \eqref{conee} we obtain
\be  \lab{xxo2i}
dV_t (\phi) =  \wt V_t (\phi) \, dB_t + \sum_{i=1}^{d} \xi^i_t B^i_t \, d\wh S^{i,\textrm{cld}}_t
- \pC^-_t (B^{d+1}_t)^{-1}  \, dB^{d+1}_t + d\pFCs_t + d\pA_t
\ee
or, equivalently,
\be  \lab{uuyy}
dV_t (\phi) =  \wt V_t (\phi) \, dB_t + \sum_{i=1}^{d} \xi^i_t B^i_t \, d\wh S^{i,\textrm{cld}}_t + d\whpFCs_t + d\pA_t
\ee
where
\bde
\whpFCs_t :=  F^c_t + \int_0^t \pC^+_u  (B^{{d+2,s}}_u)^{-1} \, dB^{{d+2,s}}_u -  \int_0^t \pC^-_u (B^{d+1}_u)^{-1} \, dB^{d+1}_u  . \nonumber
\ede
The hedger's wealth admits the following decomposition
\be \lab{poptfx}
V_t (\phi ) = x + G_t(\phi ) + F_t(\phi ) + \whpFCs_t + \pA_t
\ee
where $G_t(\phi)$ is given by \eqref{gg66} and $F_t(\phi )$ satisfies \eqref{gg666}.
\ep

\begin{proof}
We use the arguments similar to those in the proof of Proposition \ref{collssp}.
We start by noting that \eqref{pkrtf1v} yields
\bde
V^p_t (\phi ) = \sum_{i=1}^{d} \xi^i_t S^i_t +  \psi^0_t B_t + \sum_{i=1}^{d+1} \psi^i_t B^i_t .
\ede
Hence, in view of \eqref{porfx}, we may observe that we deal here with a self-financing strategy
$(\phi , A)$ introduced in Definition \ref{ts2} with $d$ replaced by $d+1$ and $\pACs = \pA + \pFCs$
such that $\zeta^{d+1}_t = \psi^{d+1}_tB^{d+1}_t  = - \pC^-_t $. An application of Corollary \ref{correx}
gives \eqref{pof3x}. The equivalence of \eqref{xxo2i} and \eqref{uuyy} follows by direct computations using the
equality  $V(\phi ) = V^p(\phi ) + C^-$ and the assumptions that $B$ and $B^{d+1}$ are continuous processes of finite variation.
\end{proof}

%%%%%%%%%%%%%%%%%%%%%%%%%%%%%%%%%%%%%%%%%%%%%%%%%%%%%%%%%%%%%%%%%%%%%%%%%%%%%%%
\sssc{Margin Account under Rehypothecation}
%%%%%%%%%%%%%%%%%%%%%%%%%%%%%%%%%%%%%%%%%%%%%%%%%%%%%%%%%%%%%%%%%%%%%%%%%%%%%%%

In the case of cash collateral under rehypothecation, we assume that the hedger, when he is a collateral taker, is granted an unrestricted use of the full collateral amount $\pC^+$. As usual, we postulate that the hedger then pays interest to the counterparty determined by the collateral amount $\pC^+$ and $B^{\pCc,b}$. Furthermore, we assume that when the hedger is a collateral giver, then collateral is delivered in cash and he receives interest specified by $C^-$ and $B^{\pCc,l}$. We maintain the assumption that the hedger borrows cash for collateral delivered to the counterparty from the dedicated account $B^{d+1}$. Of course, the case when $B^{d+1}=B$ is not excluded, but we decided to use a different symbol for the dedicated account to facilitate
identification of each cash flow. Recall that the process $\pFCh$, which is now given by expression \eqref{funcos} with $B^{d+2,h}= B$, is aimed to
%\be \lab{hfuncos}
%\pFCh_t :=  \int_{(0,t]} \pC^-_u  (B^{\pCc,l}_u)^{-1} \, dB^{\pCc,l}_u
%- \int_{(0,t]} \pC^+_u (B^{\pCc,b}_u)^{-1} \, dB^{\pCc,b}_u,
%\ee
represent the cash flows from the margin account under rehypothecation, as specified by Definition \ref{irrcc1}.
The proof of the next result is also based on Corollary \ref{correx} and thus it is omitted.

\bp \lab{ttmmi2}
Consider the case of a rehypothecated margin account when the cash collateral is posted and received by the hedger.
We assume that a trading strategy $(\phi , \pA , \pC )$ is self-financing and the following equalities hold, for all $t \in [0,T]$,
\begin{align} \lab{lqqpp}
\xi^{d+1}_t  =  0  ,\quad \psi^{d+1}_t = -(B^{d+1}_t)^{-1}\pC^-_t  , \quad \eta^{{d+2,h}}_t = 0 .
\end{align}
Then the hedger's wealth $V(\phi ) = V(x, \phi , \pA , \pC )$ equals, for every $t \in [0,T]$,
\be  \lab{lxtf1v}
V_t (\phi ) = V^p_t (\phi ) - C_t = \sum_{i=1}^{d} \xi^i_t S^i_t + \sum_{j=0}^{d+1} \psi^j_t B^j_t - C_t
\ee
and the dynamics of the portfolio's value $V^p(\phi )= V^p(x, \phi , \pA , \pC )$ are
\be \lab{lpof3d1x}
dV^p_t (\phi)  =  \wt V^p_t (\phi) \, dB_t + \sum_{i=1}^{d} \xi^i_t \, dK^i_t
 + \sum_{i=1}^{d}  \zeta^i_t (\wt B^i_t)^{-1}  \, d\wt B^i_t - C^-_t (B^{d+1}_t)^{-1}  \, dB^{d+1}_t + d\pACh_t
  + dC_t
\ee
where $\wt{V}^p_t (\phi ) := (B_t)^{-1}V^p_t(\phi )$ and $\pACh := \pA + \pFCh$.
In particular, under assumption \eqref{conee} we obtain
\be  \lab{lxx2i}
dV_t (\phi) =  \wt V_t (\phi) \, dB_t + \sum_{i=1}^{d} \xi^i_t B^i_t \, d\wh S^{i,\textrm{cld}}_t
 - \pC^-_t (B^{d+1}_t)^{-1}  \, dB^{d+1}_t + d\pFCh_t + d\pA_t
\ee
or, equivalently,
\be \lab{lxxhh2i}
dV_t (\phi)=\wt V_t (\phi)\, dB_t+\sum_{i=1}^{d} \xi^i_t B^i_t \, d\wh S^{i,\textrm{cld}}_t + d\whpFCh_t + d\pA_t
\ee
where
\be \lab{vvfbbcos}
\whpFCh_t:=F^c_t+\int_0^t \pC^+_u (B_u)^{-1} \, dB_u -\int_0^t \pC^-_u(B^{d+1}_u)^{-1} \, dB^{d+1}_u.
\ee
The hedger's wealth admits the following decomposition
\be \lab{lportfx}
V_t (\phi ) = x + G_t(\phi ) + F_t(\phi ) + \whpFCh_t + \pA_t
\ee
where $G_t(\phi)$ is given by \eqref{gg66} and $F_t(\phi )$ satisfies \eqref{gg666}.
\ep

%%%%%%%%%%%%%%%%%%%%%%%%%%%%%%%%%%%%%%%%%%%%%%%%%%%%%%%%%%%%%%%%%%%%%%%%%%%%%%%%%%%%%
\ssc{Trading Strategies with Benefit or Loss at Default} \lab{secben}
%%%%%%%%%%%%%%%%%%%%%%%%%%%%%%%%%%%%%%%%%%%%%%%%%%%%%%%%%%%%%%%%%%%%%%%%%%%%%%%%%%%%%

Let us now assume that the hedger may default on his contractual obligations before or on the maturity date $T$ of a contract under consideration. In particular, in the case of his default, he will fail to make a full repayment on his unsecured debt, which is represented by a negative position in the unsecured cash account $\Bbr$. Let $\theta $ be a random time of hedger's default and let $R \in [0,1]$ stand for his recovery rate process, which is assumed to be $\gg$-adapted.
It is now natural to assume that all trading activities of the hedger will stop at the random time horizon $\theta \wedge T$.
To account for the hedger's benefit at the moment $\theta$ of his own default, we propose to introduce the
{\it default-adjusted borrowing account} $\bar \Bbr$ by setting $\bar \Bbr_0=1$ and
\be
d\bar \Bbr_t = d\Bbr_t - \Bbr_t (1 - R_t ) \, dH_t
\ee
where we denote $H_t = \I_{\{t \geq \theta \}}$. It is clear before default, that is,
on the event $\{ \theta > t\}$, the equality $\bar \Bbr_t = \Bbr_t$ holds.
Note also that the size of the jump of $\bar \Bbr$ at time $\theta$ equals
$\Delta \bar \Bbr_{\theta } = - (1- R_{\theta } ) \Bbr_{\theta }$.
We also replace $\psi_t$ by  $\psi_{t-}$ in dynamics \eqref{portf2}
in order to ensure that this process is $\gg$-predictable. Then a non-negative jump of the wealth process $V(\phi )$, which
is triggered by the jump of the process $\bar \Bbr$ at the random time $\theta $, is given by the following
expression $\psi_{\theta -} \, \Delta \bar \Bbr_{\theta } = - (1- R_{\theta } ) \psi_{\theta - } \Bbr_{\theta }$.
The financial interpretation of this jump is the hedger's benefit at his own default, due to the fact that his
debt to the external lender is not repaid in full.

The last step is to evaluate the loss at the moment of default of either party.
In case of a default of either one of the counterparties prior to or maturity of the contract, the contract is terminated and closeout payoffs are transferred. Since alternative specifications for the closeout payoff (and thus also the hedger's loss of default) were presented and discussed in numerous papers (see, e.g., \cite{BCJZ10}, \cite{Durand} or the recent monograph \cite{CBB}), we are not going into details here. Theoretical issues related to the specification of defaults of counterparties, closeout  payoffs, and the impact of benefits at defaults on pricing results will be examined in the second part of this work.

%%%%%%%%%%%%%%%%%%%%%%%%%%%%%%%%     SECTION 5     %%%%%%%%%%%%%%%%%%%%%%%%%%%%%%%%%%%%%%
%%%%%%%%%%%%%%%%%%%%%%%%%%%%%%%%%%%%%%%%%%%%%%%%%%%%%%%%%%%%%%%%%%%%%%%%%%%%%%%%%%%%%%%%%
%%%%%%%%%%%%%%%%%%%%%%%%%%%%%%%%%%%%%%%%%%%%%%%%%%%%%%%%%%%%%%%%%%%%%%%%%%%%%%%%%%%%%%%%%
\section{Pricing under Funding Costs and Collateralization}\label{Sec21}
%%%%%%%%%%%%%%%%%%%%%%%%%%%%%%%%%%%%%%%%%%%%%%%%%%%%%%%%%%%%%%%%%%%%%%%%%%%%%%%%%%%%%%%%%
%%%%%%%%%%%%%%%%%%%%%%%%%%%%%%%%%%%%%%%%%%%%%%%%%%%%%%%%%%%%%%%%%%%%%%%%%%%%%%%%%%%%%%%%%
%%%%%%%%%%%%%%%%%%%%%%%%%%%%%%%%%%%%%%%%%%%%%%%%%%%%%%%%%%%%%%%%%%%%%%%%%%%%%%%%%%%%%%%%%

We will now focus on valuation of a collateralized contract that can be replicated by the hedger with the initial endowment $x$ at time 0.  We consider throughout the hedger's self-financing trading strategies $(x, \phi , \pA , \pC )$, as specified by Definition \ref{ts2} and, unless explicitly stated otherwise, we postulate that condition \eqref{conee} is met.
It will be implicitly assumed that all trading strategies considered in what follows are {\it admissible}, in a suitable sense.

As usual, the price of a contract will be defined from the perspective of a hedger. We assume that $p_0=A_0$ is an unknown real number, which should be found through contract's replication, whereas the {\it cumulative dividend stream} $A-A_0$ of a contract $A$ is predetermined. Therefore, by pricing of $A$, we mean in fact valuation of the cumulative dividend stream $A-A_0$ (or $A-A_t$ if we search for the price of $A$ at time $t$), which is supplemented by the collateral process $C$.

\bd \lab{def:replicate}
For a fixed $t \in [0,T]$, a self-financing trading strategy $(\VLL_{t}(x)+p_t, \phi , A- A_t , \pC )$,
where $p_t$ is a ${\cal G}_{t}$-measurable random variable, is said to {\it replicate the collateralized
contract} $(A,C)$ on $[t,T]$ whenever $V_T(\VLL_{t}(x)+p_t , \phi , A-A_t ,C) = \VLL_T (x)$.
\ed

In the next definition, we consider the situation when the hedger with the initial wealth $x$ at time 0
enters the contract $A$ at time $t$.

\bd
Any ${\cal G}_{t}$-measurable random variable $p_t$ for which a replicating strategy for $(A,C)$ over $[t,T]$ exists is called the {\it ex-dividend price at time $t$ of the contract $\pA$ associated with} $\phi $ and it is denoted by $S_t(x,\phi ,A,C)$.
\ed

It is worth noting that for $t=0$ we always have that $p_0=A_0$ and thus, for any portfolio $\phi $, the strategies $(x+p_0, \phi , A- A_0 , \pC )$ and $(x, \phi , A , \pC )$ are in fact identical. Therefore, we may simply say that a self-financing trading strategy $(x, \phi , A , \pC )$ {\it replicates} $(A,C)$ on $[0,T]$ whenever the equality $V_T(x , \phi , A ,C) = \VLL_T (x)$ holds. This equality is in fact consistent with equation \eqref{vvy33a} in Definition \ref{deffi7} of a hedger's fair price, so we conclude that any ex-dividend price $p_0$ of $A$ at time 0 is also a hedger's fair price $\ppff$ for $A$ at time 0 (though the converse does not hold {in general}).

\brem
In general, the ex-dividend price $S_t(x,\phi ,A,C)$ depends on $x$, so that the knowledge of the hedger's initial endowment
is essential for our (non-linear) pricing rule. In the special case when $x=0$, the price $p_t$ at
time $t$ corresponds to the existence of a trading strategy $\phi $ such that $V_T( p_t , \phi , A-A_t ,C) = 0$. In particular,
when $x=0, \, C=0$ and the process $A-A_t$ is given as a single cash flow $X$ at time $T$, then $p_t$ is the initial wealth
of a self-financing strategy $\phi $ with the wealth equal to $-X$ just prior to $T$, more precisely,
the wealth satisfying the equality $V_{T-}(\phi )=- X$ (since here $\Delta A_T = A_T-A_{T-} = X$).
In a frictionless market model, we thus obtain the classic definition of the replicating price of a European claim $X$.
\erem

%\bd
%A self-financing trading strategy $(x, \phi , A , \pC )$ {\it replicates} the cumulative dividend stream $A-A_0$ on $[0,T]$ whenever $V_T(x, \phi , A ,C) = \VLL_T (x)$ where $V_0(x, \phi , A , C ) = x+p_0$ for some real number $p_0$. Any number $p_0$ for which a replicating strategy for $A$ exists is called an {\it ex-dividend price of $\pA$ associated with} $\phi $ and it is denoted by $S_0(\phi )$.
%\ed

It is not difficult to check that necessarily $S_T (x,\phi ,A,C) = 0$ for any contract $A$. By contrast, it is not clear a priori whether {$S_t(x,\phi ,A,C)$}   for some $t<T$ depends on the initial endowment $x$ and a portfolio $\phi $
(recall also that $C = C(\phi )$, in general). If model's where the uniqueness of $S_t (x,\phi ,A,C)$ fails to hold, it would be natural to search for the least expensive way of replication for a given initial endowment~$x$. One could also address the issue of finding the least expensive way of super-hedging a contract $A$ by imposing the weaker condition that $V_T(x , \phi , A ,C) \geq \VLL_T (x)$ instead of insisting on the equality $V_T(x , \phi , A ,C) = \VLL_T (x)$.

If we assume that the hedger can replicate the contract $A$ on $[0,T]$ using a trading strategy initiated at time 0,
then it is not necessarily true that, starting with the initial endowment $\VLL_{t}(x)$ at some date $0<t<T$, he can also replicate the cumulative dividend stream $A-A_{t}$ representing the contract $A$ restricted to the interval $[t,T]$.
Let us thus consider the situation when a contract $(A,C)$ can be replicated on $[0,T]$. Then we may propose
an alternative definition of an ex-dividend price at time $t$. In fact, Definition \ref{defe:replicate}
mimics the classic definition of arbitrage price obtained through replication of a contingent claim  when $x=0$.
Of course, in the classic case, we may assume, without loss of generality that $x=0$, since arbitrage prices
obtained through replication is independent of the hedger's initial endowment.

\bd \lab{defe:replicate}
Assume that a self-financing trading strategy $(x, \phi , A , \pC )$ replicates $(A,C)$ on $[0,T]$. Then the process  $\wh{p}_t := V_t(x , \phi , A ,C) - \VLL_t(x)$ is called the {\it valuation ex-dividend price of $\pA$ associated with} $\phi $ and it is denoted by $\wh{S}_t(x,\phi ,A,C)$.
\ed

We note that the equality $\wh{S}_T(x,\phi ,A,C)=0$ is always satisfied. Furthermore, when $x=0$, Definition \ref{defe:replicate} states that  the reduced ex-dividend price of $\pA$ associated with $\phi $ is simply the wealth  $V(0, \phi , A ,C)$ of a replicating strategy. Observe also that a replicating strategy for the hedger with null initial endowment starts from the initial wealth $p_0$ at time 0 and terminates with null wealth at time $T$. We will argue that Definitions \ref{def:replicate} and \ref{defe:replicate} of ex-dividend prices are equivalent in the basic model with funding costs (where indeed the ex-dividend prices will be shown to be independent of
$x$ and $\phi $, provided that the collateral process $C$ is exogenously given), but the two prices do not necessarily
coincide in a generic market model with different borrowing and lending rates and/or other restrictions on trading.
The latter observation and the aim to cover all sorts of market restrictions, not necessarily exemplified in what follows,
motivated us to introduce a more general Definition \ref{def:replicate}, which is subsequently used in Definition \ref{defe:replicate},
which is sufficient in simpler models.

%%%%%%%%%%%%%%%%%%%%%%%%%%%%%%%%%%%%%%%%%%%%%%%%%%%%%%%%%%
\ssc{Basic Model with Funding Costs and Collateralization} \lab{xswsx}
%%%%%%%%%%%%%%%%%%%%%%%%%%%%%%%%%%%%%%%%%%%%%%%%%%%%%%%%%%

We consider the basic model with funding costs  introduced in  Section \ref{sscfir} and we postulate that: \\
(i) the assumptions of Proposition \ref{proarb1} are met, so that the model is arbitrage-free for the hedger, \\
(ii) the collateral process $\pC $ is exogenously given, that is, it is independent of a hedger's portfolio $\phi $.

We assume that the random variables whose conditional expectations are evaluated are integrable and
we write $\wt \E_t(\cdot ):=\E_{\wt \P}(\, \cdot \, |\, {\cal G}_t)$ where $\PT$ is any martingale measure for the processes
$\wh{S}^{i,\textrm{cld}},\, i=1, 2, \dots ,d$ (for the existence of $\PT$, see Proposition \ref{proarb1}).
We use a generic symbol $\wh{A}^c$ to denote either of the processes $A + \pFChb ,\, A + \whpFCs$ or $A + \whpFCh$,
depending on the adopted convention for the margin account, and we assume that the process $\wh{A}^c$ is bounded.
Also, we postulate that the cash account process $B$ is increasing.

We first show that, under mild technical assumptions, the price can be computed using the conditional expectation under $\PT$.
It is worth noting that the impact of collateralization is relatively easy to handle in the present setting by
quantifying additional gains or losses generated by the margin account, as explicitly given by either of processes
$\pFChb ,\, \whpFCs$ or $\whpFCh$, and aggregating them with the cumulative cash flows $A$.
We write here $S(A , \pC )$, rather than $S(x, \phi , A , \pC )$, in order to emphasize that, under the present
assumptions, the price does not depend on $(x,\phi)$.

\bp \lab{xxss}
Under assumptions (i)--(ii), if the collateralized contract $(A,C)$ can be replicated by an admissible trading strategy $(x,\phi, \pA, \pC )$
on $[0,T]$ and the stochastic integrals with respect to $\wh{S}^{i,\textrm{cld}},\, i=1, 2, \dots  ,d$ in \eqref{ooiip}
$($or with respect to $\wh{S}^{i,\textrm{cld}},\, i=1, 2, \dots  ,d+1$ in \eqref{oxxoiip}$)$ are $\PT$-martingales,
then its ex-dividend price process $S (x,\phi ,A,C)$ is independent of $(x,\phi )$ and equals, for all $t \in [0,T]$,
\be \lab{price}
S_t (A,C) =- B_t \, \wt \E_t\bigg(\int_{(t,T]} B_u^{-1} \, d\wh{A}^{\pCc}_u \bigg).
\ee
\ep

\proof
Assume that a strategy $(x, \phi , \pA , \pC )$ replicates the collateralized contract $(A,C)$ on $[t,T]$. By applying \eqref{xx2i},
we obtain
\be \lab{uxx2i}
d\wt V_t (x,\phi, A,C) = \sum_{i=1}^{d} \xi^i_t \wt{B}^i_t \, d\wh S^{i,\textrm{cld}}_t
+ B_t^{-1} (S^{d+1}_t)^{-1}C^-_t B^{d+1}_t \, d\wh S^{d+1,\textrm{cld}}_t + B_t^{-1} d\wh{A}^{\pCc}_t
\ee
whereas   \eqref{uuyy} and \eqref{lxxhh2i} yield
\be \lab{ccct2i}
d\wt V_t (x,\phi ,A,C) =  \sum_{i=1}^d \xi^i_t \wt{B}^i_t  \, d \wh{S}^{i,\textrm{cld}}_t + B_t^{-1} d\wh{A}^{\pCc}_t
\ee
where the specification of the process $\wh{A}^{\pCc}$ depends on the convention regarding the margin account.
Using equation \eqref{clacss1} in Corollary \ref{correx} with $\zeta^i=0$ for all $i$ and a suitable choice of $A$,
we deduce that the trading strategy given by \eqref{ccct2i} is self-financing, in the sense of Definition
\ref{ts1} and, obviously, it satisfies condition \eqref{conee}.

For a fixed $0 \leq t <T$, equality $V_T(\VLL_{t}(x)+p_t , \phi , A-A_t ,C) = \VLL_T (x)$ where $\VLL_t (x)= xB_t$,
combined with equation \eqref{ccct2i}, yields
\be \lab{ooiip}
- B^{-1}_t p_t =  \sum_{i=1}^d \int_{(t,T]} \xi^i_u \wt{B}^i_u  \, d\wh{S}^{i,\textrm{cld}}_u+ \int_{(t,T]} B_u^{-1} \,  d\wh{A}^{\pCc}_u .
\ee
By the definition of $\PT$, the processes $\wh{S}^{i,\textrm{cld}},\, i=1, 2, \dots  ,d$ are $\PT$-local martingales.
Consequently, equality \eqref{price} follows provided that the integrals with respect to $\wh{S}^{i,\textrm{cld}},\, i=1, 2, \dots  ,d$ are martingales under $\PT$, rather than merely local (or sigma) martingales.
The arguments used in the case of a risky collateral, as described by \eqref{uxx2i}, are analogous.
We now obtain
\be \lab{oxxoiip}
- B^{-1}_t p_t =  \sum_{i=1}^d \int_{(t,T]} \xi^i_u \wt{B}^i_u  \, d\wh{S}^{i,\textrm{cld}}_u+
\int_{(t,T]} B_u^{-1} (S^{d+1}_u)^{-1}C^-_u B^{d+1}_u \, d\wh S^{d+1,\textrm{cld}}_u
+ \int_{(t,T]} B_u^{-1} \,  d\wh{A}^{\pCc}_u
\ee
and we postulate that all integrals with respect to $\wh{S}^{i,\textrm{cld}},\, i=1, 2, \dots  ,d+1$ in equation \eqref{oxxoiip} are martingales under $\PT$. The last postulate is indeed justified, since we assumed, in particular, that a replicating strategy is admissible.
\endproof

Note that the minus sign in equation \eqref{price} % and \eqref{cprice}
is due to the fact that all cash flows and prices are considered from the viewpoint of the hedger.
For instance, a negative payoff $X$ at $T$, which represents the hedger's liability at time $T$ to
his counterparty, is compensated by a positive price collected by the hedger at time 0.

The existence of a replicating strategy can be ensured by postulating that the local martingales
$\wh{S}^{i,\textrm{cld}},\, i=1, 2, \dots  ,d$ have the predictable representation property with respect
to $\gg$ under $\PT$. Moreover, since the ex-dividend price is independent of $x$ and $\phi $, it is easy to verify that
the equality ${S}_t(A,C) = \wh{S}_t(A,C)$ holds for every $t \in [0,T]$. In essence, we deal here with only a minor modification of the standard linear pricing rule, which is very well understood in a market model with a single cash account.
A more complex situation where the pricing mechanism is non-linear is a subject of the next subsection.

\brem \lab{zwiazek}
Proposition \ref{xxss} sheds some light on the connection between arbitrage-free property of the model, in the sense of Definition \ref{defarbi}, and existence and representation of the hedger's fair price, in the sense of Definition \ref{deffi7}.
\erem

%%%%%%%%%%%%%%%%%%%%%%%%%%%%%%%%%%%%%%%%%%%%%%%%%%%%%%%%%
\ssc{Model with Partial Netting and Collateralization} \lab{BSDErr}
%%%%%%%%%%%%%%%%%%%%%%%%%%%%%%%%%%%%%%%%%%%%%%%%%%%%%%%%%

We now consider the market model from Section \ref{sscsec}, and we work under the assumptions
of Proposition \ref{prparb1}. Specifically, we assume that $x \geq 0,\, 0 \leq \rll \leq \rbb$ and $\rll \leq \ribb$ for $i=1,2, \dots , d$, and we postulate the existence of a probability measure $\PT^l $ equivalent to $\P $ and such that the
processes $\wt S^{i,l,\textrm{cld}},\, i=1,2, \dots ,d$ are $(\PT^l , \gg)$-local martingales, where
\bde
\wt S^{i,l,{\textrm{cld}}}_t =  (\Blr_t)^{-1}S^i_t + \int_{(0,t]} (\Blr_u)^{-1} \, d\pA^i_u .
\ede
For a collateralized contract $(A,C)$, we search for the value process of a replicating strategy (of course, we will also need to show that such a strategy exists). We consider here the special case of an exogenous margin account with rehypothecated cash collateral $C$.
%which is either borrowed from (resp. lent to) the account $\Bbr$ (resp., $\Blr$) and, to simplify the presentation, we set %$r^{c,l}=r^{c,b}=0$. This restriction is not crucial,
%since otherwise it would suffice to include in the dynamics of the process $V^p_t (x,\phi ,A^c)$ the cumulative interest
%of margin account, that is, the process $F^c$ given by equation \eqref{hfuncos}.

%%%%%%%%%%%%%%%%%%%%%%%%%%%%%%%%%%%%%%%%%%%%%%%%%%%%%%%%%%%
\sssc{Dynamics of Discounted Portfolio's Wealth}
%%%%%%%%%%%%%%%%%%%%%%%%%%%%%%%%%%%%%%%%%%%%%%%%%%%%%%%%%%%

By applying a slight extension of Definition \ref{ts2} (see also Proposition \ref{ttmmi2}) to the case of different lending and borrowing rates, one notes that a hedger's trading strategy $(x, \phi , \pA , \pC )$ is self-financing whenever the hedger's wealth, which is given by the equality
\bde
V(x, \phi , \pA ,\pC) =  \sum_{i=1}^{d} \xi^i_tS^i_t + \psi^l_t \Blr_t + \psi^b_t\Bbr_t + \sum_{j=1}^{d} \psi^j_tB^j_t - C_t
 = V^p_t (x, \phi ,A^c) - C_t
\ede
where $\psi^l_t \geq 0, \, \psi^b_t \leq 0$ and $\psi^l_t \psi^b_t =0$ for all $t \in [0,T]$, is such that
the portfolio's value satisfies
\begin{align*}
V^p_t (x ,\phi , A^c ) = x + \sum_{i=1}^{d} \int_{(0,t]} \xi^i_u \, d(S^i_u + \pA^i_u )
+ \int_0^t\psi^l_t \, d\Blr_t +\int_0^t \psi^b_t\, d\Bbr_t + \sum_{j=1}^{d} \int_0^t \psi^j_u \, dB^j_u  + A^c_t
\end{align*}
where we set $A^c = A  + \pC + F^c$ and the process $F^c$ is given by equation \eqref{hfuncos}.

 Observe that here $V^p_t (x, \phi ,A^c)= V_t (x,\phi ,A,C) + C_t$ for every $t \in [0,T)$ and $V^p_T (x ,\phi, A^c)= V_T (x,\phi ,A,C )$ since, by Assumption \ref{cxzcxz}, the equality $C_T=0$ holds.
The following lemma shows, in particular, that one could also write $\wt V^{p,l}_t(x,\phi , A^c ) = \wt V^{p,l}_t(x, \xi , A^c )$
in order to emphasize that within the present framework the process $\xi$ uniquely determines the trading strategy $\phi$,
as can be seen from Corollary \ref{cortccd} and equations \eqref{biigy2}--\eqref{ccbiigy1} in Section \ref{sec2.2.3}.

\bl \lab{zaq1}
The discounted wealth  $Y_t := \wt V^{p,l}_t(\phi , A^c) = (\Blr_t)^{-1} V^p_t (x, \phi , A^c)$ satisfies
\be \lab{ttrr}
dY_t  =\sum_{i=1}^d \xi^i_t \, d \wt S^{i,l,{\textrm{cld}}}_t
+ \wt{f}_l(t, Y_t ,\xi_t )\, dt +(\Blr_t)^{-1} \, dA^c_t
\ee
where
\be \lab{wtf00}
\wt{f}_l( t, Y_t ,\xi_t ): = (\Blr_t)^{-1} f_l(t,\Blr_t Y_t ,\xi_t ) -  \rll_t Y_t
\ee
where in turn for any process $X$ (note that $f_l$ depends on $(t,\omega)$ through $\rll_t, \rbb_t, \ribb_t $ and $S^i_t$)
\bde
f_l(t , X_t,\xi_t )  :=   \sum_{i=1}^d \rll_t \xi^i_t S^i_t
- \sum_{i=1}^d \ribb_t( \xi^i_t S^i_t )^+
+   \rll_t \Big( X_t  + \sum_{i=1}^d ( \xi^i_t S^i_t )^- \Big)^+
 - \rbb_t \Big( X_t + \sum_{i=1}^d ( \xi^i_t S^i_t )^- \Big)^-  .
\ede
\el

%\bl \lab{zaq1}
%The discounted wealth  $\wt V^{p,l}_t(\phi , A^c) := (\Blr_t)^{-1} V^p_t (x, \phi , A^c)$ satisfies
%\be \lab{ttrr}
%d\wt V^{p,l}_t(\phi , A^c )  =\sum_{i=1}^d \xi^i_t \, d \wt S^{i,l,{\textrm{cld}}}_t
%+ \wt{f} \big(t, \wt{V}^{p,l}_t(\phi , A^c ),\xi_t  \big)\, dt +(\Blr_t)^{-1} \, dA^c_t
%\ee
%where
%\be \lab{wtf00}
%\wt{f}\big( t,\wt{V}^{p,l}_t (\phi , A^c),\xi_t \big): = (\Blr_t)^{-1}
%f\big(t,\Blr_t \wt{V}^{p,l}_t (\phi , A^c),\xi_t  \big) -  \rll_t \wt V^{p,l}_t(\phi , A^c )
%\ee
%where in turn for any process $X$ (note that $f$ depends on $(t,\omega)$ through $\rll_t, \rbb_t, \ribb_t $ and $S^i_t$)
%\bde
%f(t , X_t,\xi_t )  :=   \sum_{i=1}^d \rll_t \xi^i_t S^i_t
%- \sum_{i=1}^d \ribb_t( \xi^i_t S^i_t )^+
%+   \rll_t \Big( X_t  + \sum_{i=1}^d ( \xi^i_t S^i_t )^- \Big)^+
% - \rbb_t \Big( X_t + \sum_{i=1}^d ( \xi^i_t S^i_t )^- \Big)^-  .
%\ede
%\el

\proof
We note that the processes $\pC$ and $F^c$, which represent additional cash flows due to the presence
of the margin account, do not depend on $\phi $. It thus follows from \eqref{cbx33} that
the portfolio's value $V^p(\phi ,A^c)$ satisfies
\begin{align*}
dV^p_t(\phi , A^c )  = \, \, & \sum_{i=1}^d \xi^i_t \big(dS^i_t + d\pA^i_t  \big)
- \sum_{i=1}^d \ribb_t( \xi^i_t S^i_t )^+  \, dt + d A^c_t
 \\ &+   \rll_t \Big( V^p_t (\phi , A^c ) + \sum_{i=1}^d ( \xi^i_t S^i_t )^- \Big)^+ \, dt
- \rbb_t \Big( V^p_t (\phi , A^c ) + \sum_{i=1}^d ( \xi^i_t S^i_t )^- \Big)^- \, dt  \nonumber
\end{align*}
so that we may represent the dynamics of  $V^p(\phi , A^c )$ as follows
\bde
dV^p_t(\phi , A^c )  =  \sum_{i=1}^d \xi^i_t \big(dS^i_t - \rll_t S^i_t \, dt + d\pA^i_t  \big)
+ f_l \big(t,\xi_t , V^p_t(\phi ,A^c ) \big)\, dt + dA^c_t .
\ede
Consequently, the discounted wealth  $\wt V^{p,l}_t(\phi , A^c) = (\Blr_t)^{-1} V^p_t (\phi , A^c)$ is governed by
\bde
d\wt V^{p,l}_t(\phi , A^c )  =\sum_{i=1}^d \xi^i_t \, d \wt S^{i,l,{\textrm{cld}}}_t
-  \rll_t \wt V^{p,l}_t(\phi , A^c )\, dt
+ (\Blr_t)^{-1} f_l \big(t , \Blr_t \wt V^{p,l}_t(\phi , A^c ) ,\xi_t \big) \, dt +(\Blr_t)^{-1} \, dA^c_t ,
\ede
which means that
\bde
d\wt V^{p,l}_t(\phi , A^c )  =\sum_{i=1}^d \xi^i_t \, d \wt S^{i,l,{\textrm{cld}}}_t
+ \wt{f}_l \big(t , \wt{V}^{p,l}_t(\phi , A^c ),\xi_t \big)\, dt +(\Blr_t)^{-1} \, dA^c_t
\ede
where the mapping $\wt{f}_l$ is given by \eqref{wtf00}.
\endproof

\brem \lab{remzz2}
Using analogous arguments, it is possible to show that the discounted wealth process  $\wh{Y}_t := \wt V^{p,b}_t(\phi , A^c) = (\Bbr_t)^{-1} V^p_t (x, \phi , A^c)$ satisfies
\be \lab{ccttrr}
d\wh{Y}_t  =\sum_{i=1}^d \xi^i_t \, d \wt S^{i,b,{\textrm{cld}}}_t
+ \wt{f}_b(t, \wh{Y}_t ,\xi_t )\, dt +(\Bbr_t)^{-1} \, dA^c_t
\ee
where the mapping $\wt{f}_b$ is given by
\be \lab{ccwtf00}
\wt{f}_b( t, \wh{Y}_t ,\xi_t ): = (\Bbr_t)^{-1} f_b(t,\Bbr_t \wh{Y}_t ,\xi_t ) -  \rbb_t \wh{Y}_t
\ee
where in turn $f_b$ is given by, for any process $X$,
\bde
f_b(t , X_t,\xi_t )  :=   \sum_{i=1}^d \rbb_t \xi^i_t S^i_t
- \sum_{i=1}^d \ribb_t( \xi^i_t S^i_t )^+
+   \rll_t \Big( X_t  + \sum_{i=1}^d ( \xi^i_t S^i_t )^- \Big)^+
 - \rbb_t \Big( X_t + \sum_{i=1}^d ( \xi^i_t S^i_t )^- \Big)^-  .
\ede
\erem

%%%%%%%%%%%%%%%%%%%%%%%%%%%%%%%%%%%%%%%%%%%%%%%%%%%%%%%%%%%
\sssc{An Auxiliary BSDE}
%%%%%%%%%%%%%%%%%%%%%%%%%%%%%%%%%%%%%%%%%%%%%%%%%%%%%%%%%%%

 We focus here on the case where $x \geq 0$; an analogous
analysis can be done for the case where $x <0$ examined in Remark \ref{remzz2}.
Assume that the processes $M^i,\, i=1,2,\dots , d$ are continuous local martingales on the filtered probability
space $(\Omega, {\cal G}, \gg , \PT^l )$. To proceed further, we need to address the problem of existence and uniqueness
of a solution $(Y,Z)$ to the following BSDE
\bde
dY_t  = \sum_{i=1}^d Z^i_t \, dM^i_t + \wt f_l(t, Y_t , Z_t )\, dt + dU_t
\ede
with a terminal value $Y_T= \eta $ and a given process $U$. Equivalently,
\be  \lab{bsde1x}
Y_t = \eta - \int_t^T \sum_{i=1}^d Z^i_u \, dM^i_u - \int_t^T \wt f_l(u, Y_u , Z_u )\, du - (U_T-U_t).
\ee
If we set $\wh Y_t = Y_t - U_t$,  then equation \eqref{bsde1x} can be written as
\be \lab{bsde1x-new}
\wh Y_t  = \wh{Y}_T - \int_t^T \sum_{i=1}^d Z^i_u \, dM^i_u  - \int_t^T \wh f_l(t, \wh Y_u , Z_u )\, du
\ee
where the terminal value satisfies $\wh Y_T = \eta - U_T$ and where the driver $\wh f_l$ satisfies
\be\lab{fhat}
\wh f_l(t, \wh Y_t , Z_t ) : =\wt f_l(t,\wh Y_t + U _t , Z_t ).
\ee
Equation \eqref{bsde1x-new} is a special case of general BSDE studied in El Karoui and Huang \cite{EKH}
(see also Carbone et al. \cite{CFS}). Note, that if a pair $(\wh Y,Z)$ is a solution to \eqref{bsde1x-new} with terminal condition $\wh Y_T=\eta - U_T$, then the pair $(Y,Z)$ with $Y:=\wh Y + U$ is a solution to \eqref{bsde1x} with terminal condition $Y_T= \eta .$

Under the assumption that the processes $\rll , \rbb$ and $\ribb ,\, i=1,2, \dots , d$ are non-negative and bounded, and the prices of risky assets are bounded, it is easy to check that the mapping $\wt f_l: [0,T] \times \rr \times \rr^d \times \Omega  \to \rr$  given by \eqref{wtf00} is a standard {\it driver} (in the terminology of El Karoui and Huang \cite{EKH}). Consequently, under mild integrability assumptions imposed on the process $U$, the mapping $\wh f_l: [0,T] \times \rr \times \rr^d \times \Omega  \to \rr$ given by \eqref{fhat} is a standard driver as well. Therefore, the existence and uniqueness of a solution $(\wh Y,Z)$ to BSDE  \eqref{bsde1x-new} in a suitable space of stochastic processes holds, provided that the $\rr^k$-valued local martingale $M =(M^1,\dots , M^d)$ is continuous and has the predictable representation property with respect to the filtration $\gg$ under $\PT^l$ (see, for instance, El Karoui and Huang \cite{EKH}) and the terminal condition $\eta - Y_T$ satisfies a suitable integrability condition. We conclude that the existence and uniqueness of a solution $(Y, Z )$ to BSDE  \eqref{bsde1x} holds under mild technical assumptions. For technical details, the reader is referred to Nie and Rutkowski \cite{NR2,NR4}.
We also observe that if BSDE  \eqref{bsde1x} has a solution $(Y,Z)$, then the process
\bde
\bar M_t := \sum_{i=1}^d \int_0^t Z^i_u \, dM^i_u
\ede
is a $\PT^l$-martingale, since the property that $\bar M$ is a square-integrable martingale is a part
of the definition of a solution to BSDE \eqref{bsde1x} . Consequently, the process $Y$ admits the following recursive representation
\bde
Y_t =- \, \wt \E^l_t \bigg(\int_t^T \wt f_l(u, Y_u , Z_u )\, du + U_T - U_t \bigg)
\ede
where we denote $\wt \E^l_t(\cdot ):=\E_{\PT^l}(\, \cdot \, |\, {\cal G}_t)$.

%%%%%%%%%%%%%%%%%%%%%%%%%%%%%%%%%%%%%%%%%%%%%%%%%%%%%%%%%%%
\sssc{Pricing and Hedging Result}
%%%%%%%%%%%%%%%%%%%%%%%%%%%%%%%%%%%%%%%%%%%%%%%%%%%%%%%%%%%

Assume that the processes $\wt S^{i,l,{\textrm{cld}}},\, i=1,2,\dots , d$ are continuous.
In the next result, we assume that the $d$-dimensional continuous local martingale $\wt S^{l,{\textrm{cld}}}$ has the predictable
representation property with respect to the filtration $\gg$ under $\PT^l$, meaning that any square-integrable $(\PT^l ,\gg)$-martingale $N$ admits the following integral representation for some process $(\eta^1, \dots , \eta^d)$
\bde
N_t = N_0 + \sum_{i=1}^d \int_0^t \eta^i_u \, d \wt S^{i,l,{\textrm{cld}}}_u.
\ede
For the sake of concreteness, one may assume, for instance, that under $\PT^l$ the processes $\wt S^{i,l,{\textrm{cld}}}$ satisfy, for every $i=1,2,\dots ,d$ and $t \in [0,T]$,
\bde
d \wt S^{i,l,\textrm{cld}}_t =  \sum_{j=1}^d \wt S^{i,l,\textrm{cld}}_t \sigma^{ij}_t \, d\wt{W}^j_t
\ede
where $(\wt{W}^1, \dots , \wt{W}^d)$ is the $d$-dimensional standard Brownian motion generating the filtration $\gg$
and the matrix-valued process $\sigma = [\sigma^{ij}]$ is non-singular.

We are in a position to establish the following pricing result in which we write $S_t (x,A,C)$ instead of $S_t (x,\phi ,A,C)$, in order to emphasize that, for any fixed $x \geq 0$, the replicating strategy $\xi^x$ for the collateralized contract $(A,C)$ is unique. Note also that the price $S_t (x,A,C)$ manifestly depends on the hedger's initial endowment $x$ through the terminal condition in BSDE \eqref{uuttrr}. For the existence and uniqueness of a solution to \eqref{uuttrr} and further properties of
the price $S_t (x,A,C)$, the reader is referred to the follow-up papers by Nie and Rutkowski \cite{NR2,NR3}.

\bp \lab{xxsss}
Let the random variables
\bde
U_T := \int_{(0,T]} (\Blr_t)^{-1} dA^c_t \quad \mbox{and}\quad \int_0^T (U_t )^{2}\, dt
\ede
be square-integrable under $\PT^l$. Then, for any fixed real number $x \geq 0$, the  unique replicating strategy $\xi^x$ equals $Z^x$ and the ex-dividend price satisfies, for every $t \in [0,T)$,
$$
S_t (x,A,C) =  \Blr_t (Y^x_t - x)  - C_t
$$
where the pair $(Y^x,Z^x)$ is the unique solution to the BSDE
\be \lab{uuttrr}
Y^x_t = x - \int_t^T \sum_{i=1}^d Z^{x,i}_u \, d \wt S^{i,l,{\textrm{cld}}}_u
- \int_t^T \wt{f}_l \big(u, Y^x_u , Z^x_u \big)\, du - \int_{(t,T]} (\Blr_u)^{-1} \, dA^c_u .
\ee
Consequently, the following representation is valid
\be \label{bsde22}
S_t (x,A,C) =- \Blr_t \, \wt \E^l_t \bigg(\int_t^T \wt f_l (u , Y^x_u , \xi^x_u )\, du +
\int_{(t,T]} (\Blr_u)^{-1} dA^c_u \bigg) - C_t.
\ee
\ep

\proof
 For a fixed $0 \leq t < T$, we consider replication on the interval $[t,T]$ and valuation at time $t$. Recall that $V^p_t (x, \phi ,A^c)= V_t (x,\phi ,A,C) + C_t$ for every $t \in [0,T)$ and $V^p_T (x ,\phi, A^c)= V_T (x,\phi ,A,C )$. On the one hand, the definition of replication on the interval $[t,T]$ requires that
\bde
V_T(\VLL_{t}(x)+p_t , \phi , A-A_t ,C) = \VLL_T (x)
\ede
where $\VLL(x) = x \Blr$ (recall that we work here under the assumption that $x \geq 0$), so that
\bde
\wt{V}^l_T(\VLL_{t}(x)+p_t , \phi , A-A_t ,C) - \wt{V}^l_t(\VLL_{t}(x)+p_t , \phi , A-A_t ,C)
= x - (\Blr_t)^{-1} (p_t + x \Blr_t) = - (\Blr_t)^{-1} p_t .
\ede
On the other hand, $V(x,\phi ,A,C) = V^p(x, \phi ,A^c)- C$ and thus, since $C_T=0$,
\begin{align*}
\wt{V}^l_T(\VLL_{t}(x)+p_t , \phi , A-A_t ,C) - \wt{V}^l_t(\VLL_{t}(x)+p_t , \phi , A-A_t ,C) = \wt{V}^{p,l,x}_T - \wt{V}^{p,l,x}_t + (\Blr_t)^{-1}C_t
\end{align*}
where the dynamics of the process $\wt{V}^{p,l,x}$ are given by \eqref{ttrr} with
the terminal condition
\bde
\wt{V}^{p,l,x}_T = (\Blr_T)^{-1} (V_T(\VLL_{t}(x)+p_t , \phi , A-A_t ,C) + C_T) = (\Blr_T)^{-1}\VLL_T (x) = x
\ede
where the last equality is obvious, since $\VLL_T (x) = \Blr_T x$ for every $x \geq 0$. Therefore, the ex-dividend price $p_t = S_t(x, \phi ,A,C)$ satisfies
\bde
 - (\Blr_t)^{-1}  S_t(x, \phi ,A,C) = \wt{V}^{p,l,x}_T - \wt{V}^{p,l,x}_t + (\Blr_t)^{-1}C_t .
\ede
This in turn implies that $ S_t(x, \phi ,A,C)$ equals
\bde
 S_t(x, \phi ,A,C) =  \Blr_t \wt{V}^{p,l,x}_t - C_t - x \Blr_t = \Blr_t (Y^x_t - x)  - C_t
\ede
where the pair $(Y^x,Z^x)$ solves the BSDE \eqref{uuttrr} with the terminal condition $Y^x_T =x$.
This in turn yields equality \eqref{bsde22}.
\endproof

For any fixed $t \in [0,T)$, equation \eqref{bsde22} can also be rewritten as follows
\begin{align} \label{oobsde22}
S_t(x,A,C) &=- \Blr_t \, \wt \E^l_t \bigg( \int_t^T \wt f_l \big( u , Y^x_u ,\xi_u \big)\, du
 + \int_{(t,T]} (\Blr_u)^{-1} (dA_u + dF^c_u)    \bigg)
 \\ &\quad - \Blr_t \, \wt \E^l_t \bigg( \int_{[t,T]} (\Blr_u)^{-1} dC^t_u    \bigg) \nonumber
\end{align}
where $C^t_u=C_u$ for $u\in [t,T]$ and $C^t_u=0$ for $u\in [0,t).$ Equation \eqref{oobsde22} follows easily from \eqref{bsde22} and
the fact that, for any fixed $t$, the process $C^t$ in equation \eqref{oobsde22} has the jump at time $t$
equal to $\Delta C^t_t = C^t_t - C^t_{t-}=C^t_t=C_t$. Note that the last integral in this equation is taken over $[t,T]$, whereas
the penultimate one over $(t,T]$. This discrepancy is due to markedly different financial interpretations of the cumulative cash flows process $A$ and the collateral process $C$. Alternative collateral conventions can also be covered through a suitable modification of BSDE \eqref{bsde22}. Although we do not offer here any general result in this vein, some special cases are presented in Section~\ref{secbsdee}.

\brem
In contrast to the linear case studied in Section \ref{xswsx}, we no longer claim here that the ex-dividend price
$S(x,\phi,A,C)$ and the valuation ex-dividend price $\wh{S}(x,\phi,A,C)$ necessarily coincide in the present non-linear setting.
\erem

\brem \lab{remzz3}
In view of Remarks \ref{remzz1} and \ref{remzz2}, it is easy to check that if $x \leq 0$, then the ex-dividend price
$S_t(x,A,C)$ satisfies, for every $t \in [0,T)$,
$$
S_t (x,A,C) =  \Bbr_t (Y^x_t - x)  - C_t
$$
where $(Y^x,Z^x)$ is the unique solution to the following BSDE under $\PT^b$
\bde % \lab{yyttrr}
Y^x_t = x - \int_t^T \sum_{i=1}^d Z^{x,i}_u \, d \wt S^{i,b,{\textrm{cld}}}_u
- \int_t^T \wt{f}_b \big(u, Y^x_u , Z^x_u \big)\, du - \int_{(t,T]} (\Bbr_u)^{-1} \, dA^c_u
\ede
where the mapping $\wt{f}_b$ is given by equation \eqref{ccwtf00}. It can be checked that for $x=0$ the pricing
algorithm of Proposition \ref{xxsss} and the one outlined in this remark coincide, as was expected.
\erem

\sssc{Illustrative Example}

As a sanity check for pricing equation \eqref{bsde22}, let us consider a toy model where $S^i=0$ for all $i$. so that $A^c=A $. We assume that the interest rates $r^l$ and $r^b$ are constant and, for simplicity, we set $r^{c,l} = r^l$ and $r^{c,b} = r^b$.
We fix $0\leq t_0 < T$, and we first consider the contract $(A,C)$ where
\bde
A_t = \I_{[t_0,T]}(t) - e^{r^l(T-t_0)}\I_{[T]}(t)
\ede
and for some constant $0 \leq \alpha <1$
\bde
C_t = - \alpha  e^{r^l(t-t_0)}\I_{[t_0,T)}(t).
\ede
Let us assume that $x=0$. We claim that the contract is fair, in the sense that the hedger's price
at time $t_0$ is null. To this end, we observe that the hedger may easily replicate his net liability
at time $T$ by investing $1- \alpha $ units of cash received from the counterparty at time $t_0$
in the lending account $\Blr$. When the collateral amount  $\alpha  e^{r^l(T-t_0)}$ is returned
to him at time $T$, then the hedger will have the right amount $e^{r^l(T-t_0)}$ units of cash to deliver
to the counterparty.

We thus expect that the price $S_t(0, A,C)$ equals zero for every $t<t_0$,  Under the present assumptions,
equation \eqref{ttrr} reduces to
\be \lab{cgyty}
d\wt V^{p,l}_t(\phi ,A^c)  = (r^l- r^b) \big( \wt{V}^{p,l}_t(\phi,A^c) \big)^{-} dt +(\Blr_t)^{-1} \, dA^c_t
\ee
where $A^c = A+C+F^c$ where $F^c_t = -\int_0^t  r^l C_u \, du $ (note that $C= -C^-$). For $x=0$, the portfolio's wealth $V^{p}(\phi,A^c)$ is always non-negative, so that
$ d\wt V^{p,l}_t(\phi ,A^c)  = (\Blr_t)^{-1} \, dA^c_t$. Using \eqref{oobsde22} with $\wt{f}=0$, we obtain,
for every $t< t_0$,
\begin{align*}
(\Blr_t)^{-1} S_t(0,A,C) & = - \int_{(t,T]} (\Blr_u)^{-1} d(A_u + F^c_u) - \int_{[t,T]} (\Blr_u)^{-1} dC_u \\
&= - (\Blr_{t_0})^{-1} + (\Blr_{T})^{-1} e^{r^l(T-t_0)} - \alpha \int_{t_0}^T (\Blr_u)^{-1} r^l   e^{r^l(u-t_0)} \, du
+ \alpha (\Blr_{t_0})^{-1} \\
& +\alpha  \int_{t_0}^T (\Blr_u)^{-1} d \big( e^{r^l(u-t_0)}\big)
-   \alpha (\Blr_{T})^{-1} e^{r^l(T-t_0)} =0.
\end{align*}
If we take instead the process
$$
A_t = - \I_{[t_0,T]}(t) + e^{r^b(T-t_0)}\I_{[T]}(t),
$$
then the hedger pays one unit of cash at time $t_0$ and thus if $C=0$ then his wealth will be negative, specifically, $V^p_t(\phi ,A,0) = - e^{r^b(t-t_0)}$ for $t \in [t_0,T)$. Hence \eqref{cgyty} and \eqref{oobsde22} with $\wt{f}(t,Y_t) =(r^l- r^b)(Y_t)^{-} $ now yield, for $t< t_0$,
\begin{align*}
(\Blr_t)^{-1} S_t(0,A,0) &= - \int_{t_0}^T (\Blr_u)^{-1} (r^l- r^b) e^{r^b(u-t_0)}\,  du
- \int_{[t_0,T]} (\Blr_u)^{-1} dA_u  \\ &
= e^{-r^b t_0} \big( e^{(r^b-r^l)T} - e^{(r^b-r^l)t_0} \big) + e^{-r^lt_0} -  e^{-r^lT} e^{r^b(T-t_0)} =0.
\end{align*}
Once again, this was expected since if the hedger borrows one unit of cash at time $t_0$ then his debt at time $T$
will match the cash amount, which he receives from the counterparty at this date.

%%%%%%%%%%%%%%%%%%%%%%%%%%%%%%%%%%%%%%%%%%%%%%%%%%%%%%%%%%%%%%%%%%%%%%%%
\ssc{Funding and Counterparty Risk Adjustments}
%%%%%%%%%%%%%%%%%%%%%%%%%%%%%%%%%%%%%%%%%%%%%%%%%%%%%%%%%%%%%%%%%%%%%%%%

Suppose that a self-financing trading strategy $(x, \phi ,A)$ is the hedger's {\it replicating} strategy for the contract $A$ on $[0,T]$, in the sense of Definition \ref{def:replicate}. In addition, let us consider another trading strategy, say $(\wh \xxx, \wh \phi , \wh{A} )$, which also invests in securities $S^1,S^2,\ldots ,S^d$, but uses for funding the accounts denoted as $B^0,\wh B^1, \dots , \wh B^{\wh d}$, and assume that it replicates a related contract, which is denoted as $\wh{A}$.  In Definition \ref{def:FD}, we deal with a single market model in which two different collections of funding accounts are simultaneously defined, namely, $(B^0,B^1 \dots ,B^d)$ for one hedger and $(B^0,\wh B^1, \dots , \wh B^{\wh d})$ for another hedger, whereas the prices of all risky assets are identical for both hedgers.

\bd \label{def:FD}
The {\it total funding adjustment} between a replicating strategy $(\xxx,\phi,A;B^0, B^1,\dots , B^d)$ and a
replicating strategy $(\wh \xxx,\wh \phi, \wh{A};B^0,\wh B^1 , \dots ,\wh  B^{\wh d})$ is defined as
\be \lab{ff66-FD}
\dTFA^{A, \wh A}_t (\xxx,\phi ; \wh \xxx,\wh \phi ) = F_t (\xxx,\phi,A; B^0, B^1, \dots , B^d)- F_t(\wh \xxx,\wh \phi , \wh{A}; B^0,\wh B^1 ,\dots ,\wh  B^{\wh d}).
\ee
\ed

In the next definition, we specialize Definition \ref{def:FD} to the case when the strategies $\phi $ and $\wh{\phi }$
are assumed to replicate the same contract, denoted by $A$. The pure funding adjustment is aimed to reflect the comparative costs for two hedgers with different creditworthiness.

\bd \label{def:FA}
The {\it pure funding adjustment} for a replicating strategy $(\xxx, \phi,A;B^0, B^1,\dots , B^d)$ relative to a
 replicating strategy $(\wh \xxx ,\wh \phi ,A;B^0,\wh B^1 , \dots ,\wh  B^{\wh d})$ is given by
\be \lab{ff66-FA}
\dPFA^A_t (\xxx,\phi ; \wh \xxx,\wh \phi )=  F_t (\xxx,\phi,A; B^0, B^1, \dots , B^d)- F_t(\wh \xxx, \wh \phi ,A; B^0,\wh B^1, \dots ,\wh  B^{\wh d}).
\ee
\ed

A special case of the pure funding adjustment is obtained when we assume that the equalities $B^0 = \wh B^1 = \dots = \wh  B^{\wh d}$ hold, so that \eqref{ff66-FA} can be represented as follows
\be \lab{ff66-FA-bis}
\dPFA^A_t ( \xxx,\phi ; \wh \xxx,\wh \phi ) = F_t (\xxx, \phi,A; B^0, B^1, \dots , B^d)- F_t(\wh \xxx , \wh \phi ,A; B^0).
\ee
Equality \eqref{ff66-FA-bis} corresponds to the `funding adjustment' typically considered in the literature,
where the hedging costs based on actual funding sources of the hedger are compared with the benchmark
case where funding through a single cash account is available for all risky assets.
The total funding adjustment, as specified by Definition \ref{def:FD}, can be computed for any two self-financing trading strategies $(\xxx ,\phi,A)$ and $(\wh \xxx , \wh \phi,\wh A)$, regardless whether they replicate the same financial contract or not. In practical applications, the contracts $A$ and $\wh A$ will be financially related. For instance, a contract $A$ may represent the clean CDS, whereas $\wh A$ may represent the corresponding counterparty risky CDS. This example motivates the following definition in which all traded assets, including identical funding accounts, are common for both hedgers. However, it is reasonable to assume that only some of risky assets will be used by a hedger who ignores the counterparty risk, that is, when a replicating strategy denoted
as $(\xxx ,\phi ,A)$ is used.

\bd \label{def:CA}
Assume that $\wh{A}$ is a counterparty risky contract and $A$ stands for the corresponding contract with
no counterparty risk, that is, when both parties are assumed to be default-free.  Then the {\it counterparty risk funding adjustment} for a replicating strategy $(\xxx,\phi,A;B^0,\wh B^1, \dots ,\wh  B^{\wh d})$ relative to a replicating strategy $(
\wh \xxx ,\wh \phi ,\wh{A};\wh B^1, \dots ,\wh  B^{\wh d})$  is given by
\be \lab{ff66-CA}
\dCFA^{A, \wh A}_t (\xxx,\phi ; \wh \xxx,\wh \phi ) = F_t (\xxx,\phi,A; B^0,\wh B^1, \dots ,\wh  B^{\wh d})- F_t(\wh \xxx , \wh \phi , \wh A; B^0,\wh B^1, \dots ,\wh  B^{\wh d}).
\ee
\ed

It is tempting to decompose the total funding adjustment as follows
\begin{align}
& \dTFA^{A, \wh A}_t (\xxx,\phi; \wh \xxx,\wh \phi ) =F_t (\xxx,\phi,A; B^0, B^1, \dots , B^d)- F_t(\wh \xxx, \wh \phi , A; B^0,\wh B^1 ,\dots ,\wh  B^{\wh d}) \nonumber \\ & + F_t(\wh \xxx , \wh \phi ,A; B^0,\wh B^1 ,\dots ,\wh  B^{\wh d}) - F_t(\wh \xxx , \wh \phi , \wh{A}; B^0,\wh B^1 ,\dots ,\wh  B^{\wh d}) \lab{ssdad} \\ & =\dPFA^A_t (\xxx,\phi ; \wh \xxx,\wh \phi ) +  \dCFA^{A, \wh A}_t (\wh \xxx, \wh \phi ; \wh \xxx,\wh \phi ). \nonumber
\end{align}
However, this algebraic decomposition suffers from a major flaw: the assumption made in Definition \ref{def:FD}
that the replicating strategies $(\xxx,\phi,A; B^0, B^1, \dots , B^d)$ and $(\wh \xxx, \wh \phi , \wh{A}; B^0,\wh B^1 ,\dots ,\wh  B^{\wh d})$ are self-financing strategies does not imply, in general, that the strategy $(\wh \xxx , \wh \phi , A; B^0,\wh B^1 ,\dots ,\wh  B^{\wh d})$ is self-financing as well. In other words, the two kinds of funding adjustments cannot be easily separated since, typically, the two effects are intertwined and thus the additivity property \eqref{ssdad} fails to hold.

Admittedly, representations  \eqref{ff66-FA} and \eqref{ff66-CA} are fairly abstract and thus they need to be further specialized to effectively handle a contract at hand. Our goal was to emphasize an essential conceptual difference between the funding adjustment and the counterparty risk funding adjustment from the perspective of hedging.  Let us finally mention that another important kind of adjustment to hedging strategies (hence also to funding costs) arises in the presence of a margin account -- this issue is studied in some detail in Section \ref{seccoll}.

\sssc{Pure Funding Adjustment in the Basic Model}
%%%%%%%%%%%%%%%%%%%%%%%%%%%%%%%%%%%%%%%%%%%%%%%%%%%%%%%%%%%%%%%%%%%%%%%%%%%%%%%

In the basic model with funding costs, the pure funding adjustment for a replicating strategy $(x,\phi,A;B^0,B^1,\dots , B^d)$ on $[0,T]$ relative to a replicating strategy $(x, \wh \phi ,A;B)$ where $B=B^0$ takes the following form
(note that condition \eqref{conee} is not assumed here)
\begin{align*}
&\dPFA^A_t (x, \phi ; x , \wh \phi ) = \int_0^t (\wt V_u(x,\phi ,A)-\wt V_u(x,\wh \phi ,A))\, dB_u + \sum_{i=1}^d \int_0^t  \zeta^i_u (\wt B^i_u)^{-1}  \, d\wt B^i_u \\& \quad - \sum_{i=1}^d  \int_0^t \xi^i_u \wh S^i_u \, dB^i_u +  \sum_{i=1}^d  \int_0^t \wh{\xi}^i_u \wt S^i_u \, dB_u \nonumber \\ &=\int_0^t (\wt V_u(\phi ,A )-\wt V_u(\wh \phi ,A ))\, dB_u + \sum_{i=1}^d \int_0^t  \zeta^i_u (\wt B^i_u)^{-1}  \, d\wt B^i_u - \sum_{i=1}^d  \int_0^t S^i_u \bigg(\xi^i_u  \, \frac{dB^i_u}{B^i_u} - \wh{\xi}^i_u \, \frac{dB_u}{B_u}\bigg).
\end{align*}
In the special case when condition  \eqref{conee} is met (for instance when processes $\zeta^i$ given by \eqref{portf3a} vanish),
the formula above becomes
\bde
\dPFA^A_t (x,\phi,x, \wh \phi ) = \int_0^t (\wt V_u(x,\phi ,A)-\wt V_u(x,\wh \phi ,A))\, dB_u- \sum_{i=1}^d  \int_0^t S^i_u
\bigg(\xi^i_u  \, \frac{dB^i_u}{B^i_u} - \wh{\xi}^i_u \, \frac{dB_u}{B_u}\bigg).
\ede
Consequently, in the classic case where  $B^i=B$ for all $i$, that is, when we consider two replicating strategies funded
from the common cash account $B$, the pure funding adjustment reduces to
\begin{align}
\dPFA^A_t (x, \phi; x,\wh \phi) &= \int_0^t (\wt V_u(x,\phi ,A)-\wt V_u(x,\wh \phi ,A))\, dB_u- \sum_{i=1}^d  \int_0^t S^i_u
\big(\xi^i_u  - \wh{\xi}^i_u \big)\, \frac{dB_u}{B_u} \nonumber \\
&=   \int_0^t \Big( \sum_{i=1}^d  \wh{\xi}^i_u S^i_u - \sum_{i=1}^d \xi^i_u S^i_u  \Big)\, \frac{dB_u}{B_u} =
\int_0^t \big(  \psi^0_u - \wh{\psi}^0_u   \big)\, dB_u \lab{j667}
\end{align}
where the last two equalities hold in any arbitrage-free model, since then the wealth processes $V(x,\phi ,A)$ and
$V(x, \wh \phi ,A)$ necessarily coincide for any two admissible trading strategies $(x,\phi ,A)$ and $(x,\wh{\phi},A)$ that replicate $A$,
so that
\bde
V_t(x,\phi ,A) =  \sum_{i=1}^d \xi^i_t S^i_t + \psi^0_t B_t=  \sum_{i=1}^d  \wh{\xi}^i_t S^i_t + \wh{\psi}^0_t B_t =  V_t(x,\wh \phi ,A).
\ede
Observe that the pure funding adjustment given by \eqref{j667} does not vanish if the processes
$\sum_{i=1}^d \xi^i S^i$  and $\sum_{i=1}^d \wh{\xi}^i S^i$ (or, equivalently, the processes
$\psi^0$ and $\wh{\psi}^0$, which specify the cash components of strategies $\phi $ and $\wh{\phi}$, respectively)
are not identical.  Interestingly, this may indeed be the case, even in an arbitrage-free and complete market model,
as will be shown by means of an example.

\sssc{Illustrative Example} \lab{efxx1}

Let us consider, for concreteness, the market model in which risky assets $S^1, \dots , S^d$ satisfy
\bde
dS^i_t = S^i_t \Big( \mu^i_t \, dt + \sum_{l=1}^k \sigma^{il}_t \, dW^l_t \Big)
\ede
where $W=(W^1,\dots ,W^k)$ is a standard Brownian motion on $(\Omega, \G, \gg , \P)$ where the
filtration $\gg = \ff^W$ is generated by $W$. We postulate that $dB_t = r_t B_t \, dt$ for some $\gg$-adapted, bounded process $r$
(hence condition \eqref{conee} is trivially satisfied).
Then for any two self-financing strategies $\phi $ and $\wh{\phi}$ the equality $V(x,\phi ,0)= V(x,\wh{\phi},0)$ holds
whenever, for every $l=1,2, \dots , k$,
\be \lab{edc44a}
\int_0^t\sum_{i=1}^d \big(\xi^i_u  - \wh{\xi}^i_u \big) S^i_u \sigma^{il}_u \, dW^l_u = 0
\ee
and
\be \lab{edc44b}
\int_0^t\sum_{i=1}^d \big(\xi^i_u  - \wh{\xi}^i_u \big) S^i_u ( \mu^i_u - r_u) \, du = 0 .
\ee
Condition \eqref{edc44a} is equivalent to
\be \lab{edct44}
\int_0^t \sum_{l=1}^k \Big( \sum_{i=1}^d (\xi^i_u - \wh{\xi}^i_u) S^i_u \sigma^{il}_u \Big)^2 \, du= 0.
\ee
If the model is arbitrage-free then there exists an $\rr^k$-valued, $\gg$-adapted process $\lambda$ such that
the equality $\sigma \lambda = r- \mu $ holds. Then \eqref{edc44b} is a consequence of \eqref{edct44}.
If the model is complete and $d=k$ then a replicating strategy for any contingent claim is unique
and thus the uniqueness of funding costs is obvious. If, however, the model is complete but $d>k$, so that redundancies appear, then the uniqueness of funding costs is not longer valid, in general, as the following counter-example shows.
Let us consider the case when $d=3$ and $k=2$. Specifically, we set, for $i=1,2$,
\bde
dS^i_t = S^i_t \big( \mu^i_t \, dt +  dW^i_t \big)
\ede
for some $\gg$-adapted, bounded processes $\mu^1$  and $\mu^2$, and
\bde
dS^3_t = S^3_t \big( \mu^3_t \, dt +  dW^1_t + dW^2_t \big).
\ede
Our goal is to produce an example of a self-financing trading strategy with null wealth process and non-vanishing funding costs.
We start by noting that model is complete and arbitrage-free whenever $\mu^3_t = \mu^1_t + \mu^2_t - r_t$ for all $t \in [0,T]$.
Under the unique martingale measure $\PT$ we have, for $i=1,2$,
\bde
dS^i_t = S^i_t \big( r_t \, dt +  d\wt{W}^i_t \big)
\ede
and
\bde
dS^3_t = S^3_t \big( r_t \, dt +  d\wt{W}^1_t + d\wt{W}^2_t \big)
\ede
where $\wt{W}=(\wt{W}^1,\wt{W}^2)$ is a standard Brownian motion on $(\Omega, \G, \gg , \PT )$.
We consider the self-financing trading strategy $(x,\phi ,0)$ where the portfolio $\phi  = (\xi , \psi^0)$ is such that $\xi_t = ( (S^1_t)^{-1}, (S^2_t)^{-1}, -(S^3_t)^{-1} )$. Then $\sum_{i=1}^d \xi^i_t S^i_t = 1$ and $\sum_{i=1}^d \xi^i_t \, dS^i_t = r_t \, dt$ for all $t \in [0,T]$, so that $G_t(x, \phi ,0) = \int_0^t r_u \, du$.
Moreover,
\bde
d \wt V_t(x, \phi , 0)= \sum_{i=1}^d \xi^i_t \, d\wt{S}^i_t = 0
\ede
or, equivalently, $V_t(x, \phi ,0)= V_0(\phi ) B_t = xB_t$.
In particular, $x =0$ then $V_t(x, \phi ,0)= 0$ for all $t \in [0,T]$,
and thus $\psi^0_t = - (B_t)^{-1}$ for all $t \in [0,T]$.
Hence the wealth $V(0, \phi ,0)=0$ admits  also the following decomposition
$
V_t (0, \phi ,0) = G_t (0,\phi ,0) + F_t (0, \phi ,0)
$ where $G_t (0,\phi ,0)= \int_0^t r_u \, du$ and $F_t (0,\phi ,0)=- \int_0^t r_u \, du$.
If we now take any replicating strategy for any contract $A$, then by adding the strategy produced
above we obtain another replicating strategy, but with manifestly different funding costs.
We conclude from this example that one should focus on (complete or incomplete) arbitrage-free models in which
redundancies among risky assets are precluded.

\newpage

%%%%%%%%%%%%%%%%%%%%%%%%%%%%%%%%%%%%%%%%%%%%%%%%%%%%%%%%%%%%%%%%%%%%%%%%%%%%%%%%%%%%%%%%%
\ssc{Diffusion-Type Market Models} \lab{secbsdee}
%%%%%%%%%%%%%%%%%%%%%%%%%%%%%%%%%%%%%%%%%%%%%%%%%%%%%%%%%%%%%%%%%%%%%%%%%%%%%%%%%%%%%%%%%

To give an illustration of the general hedging and pricing methodology developed in preceding sections,
we will now present a detailed study of the valuation problem under various conventions regarding collateralization.
A special case of this model was previously examined by Piterbarg \cite{PV10}. We assume that the processes $B^j,\, j=0,1,\dots ,d$ are absolutely continuous, so that they can be represented as $dB^j_t = r^j_t B^j_t \, dt $ for some $\gg$-adapted processes $r^j,\, j=0,1,\dots ,d+1$ (see Example \ref{ex1}). It is also postulated in this section that the lending and borrowing rates
are identical, that is, $r^l = r^b = r$ for some non-negative $\gg$-adapted process $r$. For this reason,
we can mimic (but also slightly extend) here the approach developed in Section \ref{xswsx}.

We postulate the existence of $d+2$ traded risky assets $S^i,\, i=1, 2, \dots  ,d+2$, where the asset $S^{d+1}$ (resp. $S^{d+2}$) can be posted by the hedger (resp. the counterparty) as collateral. Of course, the situation where $S^{d+1}=S^{d+2}$ is not excluded. However, if the risky assets $S^{d+1}$ and $S^{d+2}$ are distinct, then we do not need to model the dynamics of $S^{d+2}$; it suffices to know the identity of this asset or, more precisely, the corresponding repo rate $r^{d+2,h}$. By contrast, an explicit specification of the dynamics of $S^{d+2}$ (but not of $S^{d+1}$) would be needed if the valuation problem were solved from the perspective of the counterparty. Unless explicitly stated otherwise, we postulate in this section that condition \eqref{conee} is satisfied for $i=1,2, \dots , d+1$.

%%%%%%%%%%%%%%%%%%%%%%%%%%%%%%%%%%%%%%%%%%%%%%%%%%%%%%%%%%%%%%%%%%%%%%%%%%%%%%%
\sssc{Martingale Measure}
%%%%%%%%%%%%%%%%%%%%%%%%%%%%%%%%%%%%%%%%%%%%%%%%%%%%%%%%%%%%%%%%%%%%%%%%%%%%%%%

We assume that each risky asset $S^i,\, i=1, 2, \dots  ,d+1$ pays continuously dividends at stochastic rate $\kappa^i $ and has the (ex-dividend) price dynamics under the real-world probability $\P$
\bde % \lab{ff1}
dS^i_t = S^i_t \big( \mu^i_t \, dt+\sigma^i_t \, dW^i_t \big), \quad S^i_0>0,
\ede
where $W^1, W^2, \dots, W^d$ are correlated Brownian motions and the volatility processes $\sigma^1, \sigma^2, \dots, \sigma^d$ are positive and bounded away from zero. The corresponding dividend processes are given by
\bde
\pA^i_t = \int_0^t \kappa^i_u S^i_u \, du .
\ede
As usual, we write $\wh S^i_t = (B^i_t)^{-1}S^i_t$ and $\wh S_t^{i,\textrm{cld}}=(B^i_t)^{-1} S_t^{i,\textrm{cld}}$.
Recall that we denote by $\PT$ a martingale measure for the basic model with funding costs (see Proposition \ref{proarb1}).

\bl
The price process $S^i$ satisfies under $\PT$
\bde % \lab{bbpri2}
dS^i_t = S^i_t \big( (r^i_t- \kappa^i_t) \, dt+\sigma^i_t \, d\wt W^i_t \big)
\ede
where $\wt W^i$ is a Brownian motion under $\PT$. Equivalently, the process $\wh S^{i,\textrm{cld}}$ satisfies
\be \lab{xdx}
d\wh S^{i,\textrm{cld}}_t = \wh S^{i,\textrm{cld}}_t \sigma^i_t \, d\wt W^i_t .
\ee
The process $K^i$ given by \eqref{portf2a} satisfies
\be \lab{kk1}
dK^i_t = dS^i_t - r^i_tS^i_t \, dt + \kappa^i_t S^i_t \, dt = S^i_t\sigma^i_t \, d\wt W^i_t
\ee
and thus it is a (local) martingale under $\wt \P$.
\el

\proof By the definition of a martingale measure $\PT$, the discounted cumulative-dividend price  $\wh S^{i,\textrm{cld}}$
is a (local) martingale under $\PT$. Recall that the process $\wh S^{i,\textrm{cld}}$ is given by
\bde  % \lab{pri2o}
\wh S^{i,\textrm{cld}}_t = \wh S^i_t+\int_{(0,t]} (B^i_u)^{-1} \, d\pA^i_u ,\quad t\in [0,T].
\ede
Consequently,
\bde
\wh S^{i,\textrm{cld}}_t = \wh S^i_t+\int_0^t \kappa^i_u (B^i_u)^{-1} S^i_u \, du
=  \wh S^i_t+\int_0^t \kappa^i_u \wt S^i_u \, du .
\ede
Since
\be\lab{S1x}
d\wh S^i_t = \wh S^i_t\big( (\mu^i_t - r^i_t) \, dt+\sigma^i_t \, dW^i_t\big),
\ee
we obtain
\bde
d \wh S^{i,\textrm{cld}}_t = d \wh S^i_t + \kappa^i_t \wh S^i_t \, dt
= \wh S^i_t\big( (\mu^i_t + \kappa^i_t - r^i_t) \, dt + \sigma^i_t \, dW^i_t\big).
\ede
Hence $\wh S^{i,\textrm{cld}}$ is a (local) martingale under $\PT$ provided that the process
\be \lab{ff2}
d\wt W^i_t =   dW^i_t + (\sigma_t^i)^{-1}(\mu^i_t + \kappa^i_t - r^i_t) \, dt
\ee
is a Brownian motion under $\PT$. By combining \eqref{S1x} with \eqref{ff2} we obtain expression \eqref{xdx}.
Other asserted formulae now follow easily.
\endproof

%%%%%%%%%%%%%%%%%%%%%%%%%%%%%%%%%%%%%%%%%%%%%%%%%%%%%%%%%%%%%%%%%%%%
\sssc{Wealth Dynamics for Collateralized Contracts}
%%%%%%%%%%%%%%%%%%%%%%%%%%%%%%%%%%%%%%%%%%%%%%%%%%%%%%%%%%%%%%%%%%%%%

We postulate, in addition, that the processes $B^{\pCc,b}, B^{\pCc,l}, B^{{d+2,s}}$ and $B^{{d+2,s}}$ are absolutely
continuous as well, so that
\begin{align*}
&dB^{\pCc,b}_t = r^{\pCc,b}_t B^{\pCc,b}_t \, dt, \quad dB^{\pCc,l}_t = r^{\pCc,l}_t B^{\pCc,l}_t \, dt,
\\ & dB^{{d+2,s}}_t = r^{{d+2,s}}_t B^{{d+2,s}}_t \, dt  , \quad  dB^{{d+2,h}}_t = r^{{d+2,h}}_t B^{{d+2,h}}_t \, dt  ,
\end{align*}
for some processes $r^{\pCc,b},r^{\pCc,l}, r^{{d+2,s}}$ and $r^{{d+2,h}}$, which are assumed to be non-negative.

%%%%%%%%%%%%%%%%%%%%%%%%%%%%%%%%%%%%%%%%%%%%%%%%%%
% \sssc{Risky Collateral}
%%%%%%%%%%%%%%%%%%%%%%%%%%%%%%%%%%%%%%%%%%%%%%%%%%%
%%%%%%%%%%%%%%%%%%%%%%%%%%%%%%%%%%%%%%%%%%%%%
\vskip 5 pt \noindent $\bullet $ {\bf Risky collateral.$\ $}
%%%%%%%%%%%%%%%%%%%%%%%%%%%%%%%%%%%%%%%%%%%%%
We first consider the case of risky collateral under the assumptions of  Proposition  \ref{collssp}.
Formally, the cases of rehypothecation and segregation differ only in the choice of
either $r^{{d+2,s}}$ or $r^{{d+2,h}}$ as the hedger's interest on the collateral amount posted by the counterparty.
In practice, it is clear that the repo rate $r^{{d+2,h}}$ is positive, whereas the conventional rate $r^{{d+2,s}}$
is likely to be zero. In the case of  rehypothecation, $\pFChb_t$ is given here by the following expression
(see \eqref{fuuncos})
\bde % \lab{actgmm}
\pFChb_t =  \int_{0}^t  \big( r^{{d+2,h}}_u - r^{\pCc,b}_u \big)\pC^+_u  \, du
 - \int_{0}^t \big( r^{d+1}_u - r_u \big)\pC^-_u  \, du
\ede
and thus, as expected, the term $\pFChb_t$ vanishes when the equalities $r^{{d+2,h}}= r^{\pCc,b}$ and $r^{d+1}=r$ hold, since then
the negative and positive cash flows related to the margin account cancel out.
From equation \eqref{uxx2i}, we obtain the dynamics of the hedger's wealth $V(\phi ) = V(x,\phi,A,C)$
\begin{align*}
d V_t(\phi ) &= r_t V_t (\phi) \, dt + \sum_{i=1}^d \xi^i_t \,  \big(dS^i_t- r^i_t S^i_t \, dt + d\pA^i_t  \big)
\\ &+ (S^{d+1}_t)^{-1}C^-_t \,  \big(dS^{d+1}_t- r^{d+1}_t S^{d+1}_t \, dt + d\pA^{d+1}_t  \big) + d\pFChb_t + dA_t .
\end{align*}
If the collateral $C$ is predetermined, then the sum of the last three terms in the formula above defines a single process $\pDTbh$, which represents all cash flows associated with a collateralized contract except for the gains or losses from trading in risky assets $S^1,S^2, \dots , S^d$. Then we may rewrite the last equation as follows
\be \lab{exx1}
d V_t(\phi ) = r_t V_t (\phi) \, dt + \sum_{i=1}^d \xi^i_t S^i_t\sigma^i_t \, d\wt W^i_t + d\pDTbh_t .
\ee
We note that the process $\pDTbh$ depends also on the dynamics of the risky asset $S^{d+1}$.
As was already mentioned, the dynamics of the asset $S^{d+2}$ are irrelevant, so they are left unspecified.

%%%%%%%%%%%%%%%%%%%%%%%%%%%%%%%%%%%%%%%%%%%%%%%%%%%
% \sssc{Cash Collateral under Segregation}
%%%%%%%%%%%%%%%%%%%%%%%%%%%%%%%%%%%%%%%%%%%%%%%%%%%
%%%%%%%%%%%%%%%%%%%%%%%%%%%%%%%%%%%%%%%%%%%%%
\vskip 5 pt \noindent $\bullet $ {\bf Cash collateral under segregation.$\ $}
%%%%%%%%%%%%%%%%%%%%%%%%%%%%%%%%%%%%%%%%%%%%%
We now consider the case of cash collateral under segregation and we place ourselves within the setup
of Proposition  \ref{xcollssp}. Under the present assumptions, the expression for $\whpFCs$ reduces to
\bde  % \lab{acgmm}
\whpFCs_t =  \int_{0}^t  \big( r^{{d+2,s}}_u - r^{\pCc,b}_u \big)\pC^+_u  \, du
 - \int_{0}^t \big( r^{d+1}_u - r^{\pCc,l}_u \big)\pC^-_u  \, du .
\ede
Formula \eqref{uuyy} yields
\bde  % \lab{ggbb}
dV_t (\phi) = r_t V_t (\phi) \, dt + \sum_{i=1}^d \xi^i_t \,  \big(dS^i_t- r^i_t S^i_t \, dt + d\pA^i_t  \big)
+ d\whpFCs_t + dA_t ,
\ede
so, if we denote the sum of the last three terms by $\pDTsh$, then we obtain
\be \lab{exx2}
d V_t(\phi ) = r_t V_t (\phi) \, dt + \sum_{i=1}^d \xi^i_t S^i_t\sigma^i_t \, d\wt W^i_t + d\pDTsh_t
\ee
where the process $\pDTsh$ does not depend on the dynamics of the risky asset $S^{d+1}$.

%%%%%%%%%%%%%%%%%%%%%%%%%%%%%%%%%%%%%%%%%%%%%%%%%%%
% \sssc{Cash Collateral under  Rehypothecation}
%%%%%%%%%%%%%%%%%%%%%%%%%%%%%%%%%%%%%%%%%%%%%%%%%%%
%%%%%%%%%%%%%%%%%%%%%%%%%%%%%%%%%%%%%%%%%%%%%
\vskip 5 pt \noindent $\bullet $ {\bf Cash collateral under rehypothecation.$\ $}
%%%%%%%%%%%%%%%%%%%%%%%%%%%%%%%%%%%%%%%%%%%%%
Recall that the case of cash collateral under rehypothecation was examined in Proposition \ref{ttmmi2}.
Under the present assumptions, we deduce from \eqref{vvfbbcos} that
\be \lab{funicos}
\whpFCh_t =  \int_0^t \pC^+_u \,  \big( r_u - r^{\pCc,b}_u  \big) du
 -  \int_0^t \pC^-_u \,  \big( r^{d+1}_u - r^{\pCc,l}_u \big) du
\ee
and thus \eqref{lxxhh2i} becomes
\be \lab{portf3d}
dV_t(\phi )= r_t V_t (\phi) \, dt + \sum_{i=1}^d \xi^i_t \,  \big(dS^i_t- r^i_t S^i_t \, dt + d\pA^i_t  \big)
+ d\whpFCh_t + dA_t .
\ee
If we denote the sum of the last three terms by $\pDThh$, then, using also \eqref{kk1}, we obtain
\be \lab{exx3}
d V_t(\phi ) = r_t V_t (\phi) \, dt + \sum_{i=1}^d \xi^i_t S^i_t\sigma^i_t \, d\wt W^i_t + d\pDThh_t
\ee
where, once again, the process $\pDThh$ does not depend on the dynamics of $S^{d+1}$.

%%%%%%%%%%%%%%%%%%%%%%%%%%%%%%%%%%%%%%%%%%%%%%%%%%%%%%%%%
\sssc{Pricing with an Exogenous Collateral}
%%%%%%%%%%%%%%%%%%%%%%%%%%%%%%%%%%%%%%%%%%%%%%%%%%%%%%%%%

Our goal is to value and hedge a collateralized contract within the framework of a diffusion-type
model. We postulate that the process $A$ is adapted to the filtration $\mathbb{F}^S$ generated by risky
assets $S^1,S^2, \dots,S^d$. We first assume that a collateral process $\pC$ is predetermined, so it does not
depend on the hedger's trading strategy. We use the generic symbol $A^c$ to denote either of the processes $\pDTbh, \pDThh , \pDTsh $ introduced in the preceding subsection. Assume that all short-term rates and the processes $A$ and $C$ are bounded, so that
the process $A^c$ is bounded as well. In fact, it would be enough to postulate that the conditional
expectation in \eqref{xxK3} is well defined for all $t \in [0,T]$. The following result can be seen
as a corollary to Proposition \ref{xxss}.

\bp \lab{prox4}
A collateralized contract $(A,C)$ with the predetermined collateral process $C$ can be replicated by
an admissible trading strategy. The ex-dividend price $S(A,C)$ satisfies, for
every $t \in [0,T)$,
\be \lab{xxK3}
S_t (A,C) =  - B_t \, \wt \E_t \bigg( \int_{(t,T]}  B_u^{-1} \, dA^c_u \bigg).
\ee
\ep

\proof
We formally consider $A^c$ as the total cash flow process associated with the contract $A$.
Hence it suffices to check that the assumptions of Proposition \ref{xxss} are met. For the existence
of an admissible replicating strategy under the cash collateral convention, we note that the processes
$C$ and $A$ are adapted with respect to filtration generated by risky assets $S^1,S^2, \dots,S^d$
and thus the predictable representation property of the Brownian filtration entails that
\bde
\int_{(0,T]} B_u^{-1} \, dA^c_u = S_0 + \sum_{i=1}^d \int_0^T  \xi^i_u S^i_u \sigma^i_u \, d\wt W^i_u.
\ede
In the case of the risky collateral, the trading strategy is complemented by $\xi^{d+1}= (S^{d+1}_t)^{-1} C^-_t$, so that we now
use the following representation
\be \lab{trtry}
\int_{(0,T]}  B_u^{-1} \, (d\pFChb_t + dA_u )= \wt S_0 + \sum_{i=1}^{d+1} \int_0^T  \xi^i_u S^i_u\sigma^i_u \, d\wt W^i_u
\ee
where, by assumption, the process $A$ is adapted to the filtration $\mathbb{F}^S$ generated by the risky
assets $S^1,S^2, \dots,S^d$, and thus the right-hand side in \eqref{trtry} defines a bounded ${\cal F}^S_T$-measurable
random variable. Hence the existence of an admissible replicating strategy satisfying condition \eqref{conee} follows.
\endproof

For the sake of concreteness, let us consider a particular instance of a collateralized contract, specifically,
the valuation of a single cash flow $X$ at maturity date $T$ under the convention of cash collateral with rehypothecation.
We assume that $X$ is a bounded random variable, which is measurable with respect to the $\sigma$-field
${\cal F}^S_T$. It is natural to assume that $r^{d+1}=r$, meaning that the cash for collateral posted is borrowed from the risk-free account. We first obtain the non-linear pricing formula \eqref{K3}.
Under an additional assumption of symmetry, $r^{\pCc,b}=r^{\pCc,l}=r^{\pCc}$, we denote by $B^c$ the process satisfying
$dB^{\pCc}_t = r^{\pCc}_t B^{\pCc}_t\, dt$, and we obtain the linear pricing formula  \eqref{vvK3}.

\bcor \lab{ccprox4}
A collateralized contract with the cumulative dividend $A_t = p \, \I_{[0,T]}(t)+  X \I_{[T]}(t)$ and the predetermined collateral process $\pC$ can be replicated by an admissible trading strategy. The ex-dividend price $S(A,C)$ satisfies, for
every $t \in [0,T)$,
\be\lab{K3}
S_t (A,C) =  - B_t \, \wt \E_t \bigg( B_T^{-1} X  +
\int_t^T B^{-1}_u \pC^+_u \,  \big( r_u - r^{\pCc,b}_u  \big) du
 -  \int_t^T B^{-1}_u \pC^-_u \,  \big( r^{d+1}_u - r^{\pCc,l}_u \big) du \bigg).
\ee
In particular, if $r^{d+1}=r$ and $ r^{\pCc,b}= r^{\pCc,l} =r^{\pCc}$, then
\be \lab{vvK3}
S_t (A,C) =  - B_t \, \wt \E_t \bigg( B_T^{-1} X  + \int_t^T B_u^{-1} ( r_u - r^{\pCc}_u )\pC_u \, du \bigg).
\ee
\ecor

\proof
Equality \eqref{K3} is an immediate consequence of  \eqref{funicos} and \eqref{xxK3}.
To obtain \eqref{vvK3}, it suffices to observe that equalities $r^{d+1}=r$ and $ r^{\pCc,b}= r^{\pCc,l} =r^{\pCc}$
imply that
\bde
\whpFCh_t =  \int_0^t \pC_u \,  \big( r_u - r^{\pCc}_u  \big) du
\ede
and thus  \eqref{vvK3}  is an immediate consequence of \eqref{K3}.
\endproof

\brem Piterbarg \cite{PV10} examined a diffusion-type market model with three cash accounts
\bde
B_t=e^{\int_0^t r_u \, du},\quad B^1_t=e^{\int_0^t r^1_u \, du},\quad B^{\pCc}_t=e^{\int_0^t r^{\pCc}_u \, du},
\ede
where the spreads $r^1-r^{\pCc},\, r^1 -r ,\, r^{\pCc}-r$ represent the \textit{bases}
between the funding rates, that is, the {\it funding bases}. According to our classification and notation,
he dealt with cash collateral under rehypothecation with $r^2=r$ and $ r^{\pCc,b}= r^{\pCc,l} =r^{\pCc}$.
Our formulae agree with those derived by  Piterbarg \cite{PV10}, although our convention for the collateral amount is
slightly different than the one adopted in \cite{PV10}, specifically, our collateral process $C$ corresponds to the process $-C$ in \cite{PV10}.
\erem

\brem
Observe that the equivalence of formulae \eqref{vvK3} and \eqref{mmK3} shows that the choice of a particular discount factor can be rather arbitrary, as long as the (cumulative) cash flow process of a security under valuation is appropriately adjusted. In the case of formula \eqref{vvK3}, the discount factor is chosen as the price process $B$ representing a traded asset, whereas in the case of formula \eqref{mmK3}, we deal with the process $B^{\pCc}$, which does not even represent the price process of a traded asset in the present setup.

Suppose, for instance, that $d=1$ and the dividend rate $\kappa^1 =0$. Then none of the two above mentioned choices of the discount factor correspond to the usual martingale measure for the stock price which corresponds to the choice of $B^1$ as the discount factor. In Section \ref{P}, we offer a more extensive discussion of this peculiar feature in the context of the pricing approach
recently proposed by Pallavicini et al.~\cite{PPB12}.
\erem

%%%%%%%%%%%%%%%%%%%%%%%%%%%%%%%%%%%%%%%%%%%%%%%%%%%%%%%%%%%%%%%%%%
\sssc{Pricing with Hedger's Collateral}  \lab{sxsxsx}
%%%%%%%%%%%%%%%%%%%%%%%%%%%%%%%%%%%%%%%%%%%%%%%%%%%%%%%%%%%%%%%%%%

As already mentioned in Section \ref{seccoll}, the collateral amount $\pC$ can be specified in terms of the marked-to-market value of a contract and thus, at least in theory, it can be given in terms of the wealth process $V(\phi )$ of the hedger's strategy.
To this end, we introduce the process $\wh{V}(\phi ) := V(\phi ) - xB$; for the interpretation of the process $\wh{V}(\phi )$, see Definition \ref{defe:replicate}.
Then, for instance, the process $\pC$ may be given as follows (see \eqref{ty6v})
\be \lab{cc3c}
\pC_t (\phi) = (1+ \delta^1_t) \wh{V}^-_t (\phi )  - (1+ \delta^2_t)  \wh{V}^+_t (\phi )
= \bar \delta^1_t \wh{V}^-_t (\phi )  - \bar \delta^2_t \wh{V}^+_t (\phi )
\ee
for some bounded, $\ff^S$-adapted processes $\delta^1$ and $\delta^2$, where for brevity we set $\bar \delta^i_t = 1+ \delta^i_t$.
Hence the generic process $A^c$, which as before is aimed to represent either of the processes $\pDTbh, \pDThh , \pDTsh $,
depends here in a non-linear manner on the hedger's wealth when he implements a replicating portfolio. Consequently, the conditional expectation in equation \eqref{xxK3} can now be informally interpreted as a BSDE with the shorthand notation
\be \lab{xixK3}
V_t (\phi ) =  - B_t \, \wt \E_t \bigg( \int_{(t,T]}  B_u^{-1} \, dA^c_u ( \wh{V} (\phi )) \bigg)
\ee
where the notation $A^c( \wh{V} (\phi ))$ is used to emphasize that the process $A^c$ depends on $\wh{V} (\phi )$.
A more explicit form of BSDE \eqref{xixK3} can be derived as soon as a particular convention for the margin account
is adopted. Let us consider, for instance, the special case of cash collateral with rehypothecation (recall that this was also our choice in Section \ref{BSDErr}). To simplify expressions, we also assume that $r^{d+1}=r$ and $r^{\pCc,b}=r^{\pCc,l}=r^{\pCc}$, so that the process $F^c$ satisfies $F^c_t = \int_0^t r^{\pCc}_u C_u \, du$ for all $t \in [0,T]$. Then, from equations \eqref{funicos} and \eqref{portf3d}, the wealth process of a self-financing strategy $\phi$ satisfies
\be \lab{mexx3}
d V_t(\phi ) = r_t V_t (\phi) \, dt + \sum_{i=1}^d \xi^i_t S^i_t\sigma^i_t \, d\wt W^i_t +
(r_t - r^{\pCc}_t )( \bar \delta^1_t \wh{V}^-_t (\phi )  - \bar \delta^2_t  \wh{V}^+_t (\phi ) ) \, dt + d A_t .
\ee
In the next pricing result, we once again focus on a collateralized contract $(A,C)$ where
$A_t = p \, \I_{[0,T]}(t)+  X \I_{[T]}(t)$. For any fixed $t \in [0,T]$, we search for a ${\cal G}_{t}$-measurable random variable $p_t$ such that
$$
V_T(\VLL_{t}(x)+p_t , \phi , A-A_t ,C) = \VLL_T (x)
$$
for some admissible trading strategy $\phi $. Obviously, for any fixed $t \in [0,T)$, we
have $A_u - A_t = X \I_{[T]}(u)$ for all $u \in [t,T]$.

It is worth noting that in Proposition \ref{probbg} we obtain a non-linear pricing rule, although we work
there under the assumption that the lending and borrowing rates are identical. Due to this postulate, the price process $S(x,A,C)$ is in fact independent of the hedger's initial endowment -- this property can be easily deduced from equation \eqref{pcxxdm}.
The non-linearity of the pricing rule is now due to specification \eqref{cc3c} of the
collateral amount $C$ and thus the non-linear BSDEs  \eqref{uuttrr} and \eqref{pcxxdm} do not have the same shape.
For a detailed study of pricing BSDEs and fair prices for both parties when $C$ is given by \eqref{cc3c},
the interested reader is referred to Nie and Rutkowski \cite{NR5}.

\bp \lab{probbg}
Let $X$ be an ${\cal F}^S_T$-measurable, bounded random variable. The BSDE
\be  \lab{pcxxdm}
dY^x_t = r Y^x_t \, dt + \sum_{i=1}^d Z^{x,i}_t S^i_t\sigma^i_t \, d\wt W^i_t + (r_t - r^{\pCc}_t )
( \bar \delta^1_t (Y^x_t - xB_t)^-  - \bar \delta^2_t  (Y^x_t - x B_t)^+ ) \, dt
\ee
with the terminal condition $xB_T - X$ has a unique solution $(Y^x,Z^x)$.
For any fixed $x$ and $t \in [0,T)$, the contract $A_t = p \,\I_{[0,T]}(t)+  X \I_{[T]}(t)$ with the collateral process $\pC$ given by \eqref{cc3c} can be replicated  on $[t,T]$ by an admissible trading strategy $\xi^x = Z^x$ and the ex-dividend price
satisfies $S_t (x,A,C)= Y^x_t - x B_t $. Furthermore, the price $S_t (x,A,C)$  admits the following representation, for
every $t \in [0,T)$,
\be \lab{K3cc}
S_t (x,A,C)= - B_t \, \wt \E_t \bigg( B_T^{-1} X  + \int_t^T B_u^{-1} ( r_u - r^{\pCc}_u )
\big( \bar \delta^1_u (Y^x_u - xB_u)^-   - \bar  \delta^2_u  (Y^x_u - xB_u)^+ \big) \, du \bigg).
\ee
Equivalently, the price $S_t (x,A,C) = Y_t$ for every $t \in [0,T)$, where the process $Y$ solves
the following BSDE
\be  \lab{pcxxdmc}
dY_t = r Y_t \, dt + \sum_{i=1}^d Z^{i}_t S^i_t\sigma^i_t \, d\wt W^i_t + (r_t - r^{\pCc}_t )
\big( \bar \delta^1_t Y_t^-  - \bar \delta^2_t  Y_t^+ \big) dt
\ee
with the terminal condition $Y_T=-X$. Consequently, the price $S(x,A,C) = S(A,C)$ is independent of the hedger's
initial endowment $x$.
\ep

\proof
The proof is similar to the proof of Proposition \ref{prox4} and thus we omit the details.
Let us only observe that the uniqueness of a solution to BSDE \eqref{pcxxdmc} follows from
the general theory of BSDEs with Lipschitz continuous coefficients (see, e.g., \cite{EKH}).
\endproof

In view of \eqref{cc3c}, equation \eqref{mexx3} may also be represented as follows
\be  \lab{pcc3db-1}
dV_t (\phi) = r^{\pCc}_t V_t (\phi) \, dt  + \sum_{i=1}^d \xi^i_t S^i_t\sigma^i_t \, d\wt W^i_t  + (r_t - r^{\pCc}_t )(\pC_t+V_t(\phi) - xB_t ) \, dt + d A_t .
\ee
This yield the following representation (note that we may write here $V(\phi ) = V(\xi^x)$)
\be \lab{mmK3}
S_t (A,C) =  - B^{\pCc}_t \, \wt \E_t \bigg( (B^{\pCc}_T)^{-1} X  + \int_t^T (B^{\pCc}_u)^{-1} ( r_u - r^{\pCc}_u )(\pC_u+V_u(\xi^x ) - xB_u ) \, du \bigg).
\ee
Furthermore, in the case of the fully collateralized contract, we postulate that $\delta_t^1 = \delta_t^2 = 0$ so that
the equalities
$$
\pC_t = - \wh{V}_t(\xi^x ) =  xB_t - V_t(\xi^x )
$$
are satisfied for all $t \in [0,T]$. Hence \eqref{pcc3db-1} reduces to
\bde
dV_t (\phi) = r^{\pCc}_t V_t (\phi) \, dt  + \sum_{i=1}^d \xi^i_t S^i_t\sigma^i_t \, d\wt W^i_t + d A_t
\ede
and this in turn yields the following explicit representation for the ex-dividend price of the fully collateralized contract, for
every $t \in [0,T)$,
\be \lab{xK3T}
S_t (A,C) =  - B^{\pCc}_t \, \wt \E_t \big( (B^{\pCc}_T)^{-1} X \big).
\ee
Note that the price given by equation \eqref{xK3T} not only does not depend on the initial endowment $x$, but it is also
linear as a mapping from the space of contingent claims to real numbers.

\ssc{Expected Cash Flows Approach}\label{P}
%%%%%%%%%%%%%%%%%%%%%%%%%%%%%%%%%%%%%%%%%%%%%%%%%%%%%%%%%%%%%%%%%%%%%%%%%%%%%%%%%%%%%%%%%

We conclude this paper by offering a brief analysis of mathematical underpinning for an alternative pricing method under collateralization, which was proposed in a recent work by Pallavicini et al.~\cite{PPB12}. We propose to dub their approach the {\it expected cash flows approach} since it primarily hinges on the computation of the expected value of discounted cash flows of a contract. This should be contrasted with the more sound {\it hedging based} approach advocated in the present paper,
where an arbitrage price is defined in terms of the initial wealth of a replicating (or superhedging) strategy and the concept of a martingale measure (if needed) only appears at a later stage as a computational tool. Of course, the methodology presented here
also starts with a careful delineation and analysis of cash flows associated the contract, but in the next step a systematic approach to trading strategies is applied.

We do not analyze in detail the successive cash flow transformations discussed in \cite{PPB12}, but we instead focus on analysis of the underlying paradigm implemented in \cite{PPB12}, that is, the possibility of defining the `price' as the expected
value of discounted cash flows under a `martingale measure', even when the `discount factor' is not postulated to be a traded asset. Let us first summarize the main steps in the approach proposed in \cite{PPB12}.
The authors start there by introducing a fictitious risk-free short-term
interest rate as an `instrumental variable' without assuming that this rate corresponds to any traded asset.
Next, they formally postulate the existence of a `martingale measure' associated with
discounting of prices of all traded risky assets using the fictitious cash account. More importantly,
they also make an ansatz that the price of any derivative contract is given by the conditional expectation
of `discounted cash flows with costs' using this martingale measure (see formula (1) in \cite{PPB12}).
Of course, this valuation recipe would be manifestly flawed if someone would attempt to apply it directly to the contractual cash flows of a given contract. Several non-trivial adjustments of cash flows are obviously needed in order to account for the actual funding costs, margin account, closeout payoff, etc..  For instance, to deal with the funding costs, the authors propose to use formula (17) in \cite{PPB12} as a plausible valuation tool. We contend that all formulae on pages 1--26 in \cite{PPB12} should rather be seen as (equivalent) definitions, which describe either the actual or transformed cash flows of a contract, rather than strict pricing results derived from sound fundamentals.

For this reason, we will henceforth focus on the most intriguing result from \cite{PPB12}, namely, Theorem 4.3. It shows that, by changing the probability measure, one can in fact circumvent the need to use the fictitious risk-free rate altogether, which justifies the term `instrumental variable' attributed to this rate. Let us make clear that we do not question the validity of the statement of Theorem 4.3 in \cite{PPB12}. However, as we will argue below, the approach proposed in \cite{PPB12} is somewhat artificial, since it heavily relies on making a right guess regarding a suitable adjustment to contractual cash flows. More importantly, rather cumbersome arguments used in \cite{PPB12} can be avoided, since it is always enough to focus directly on hedging arguments in a correctly specified market model with funding costs, instead of postulating a priori the validity of some form of a `risk-neutral pricing formula'.

To clarify the arguments underpinning the approach proposed by Pallavicini et al. \cite{PPB12}, we consider here a market model with non-dividend paying risky assets $S^1, S^2, \dots , S^d$ and the cash account $B$ such that $dB_t = r_t B_t \, dt$. Although dividends, margin account, and closeout  payoff can also be covered by the foregoing analysis, to illustrate the rationale
for the cash flows based approach, it suffices to consider the issue of funding costs only.

\bhyp \lab{ahh1}
We assume that the model is arbitrage-free, so that the martingale measure $\PT$ for the process $\wt S = B^{-1} S$ exists.
\ehyp

Let $V (\phi )$ be the wealth of a self-financing trading portfolio $\phi = (\xi , \psi^0)$.
The following lemma is well known and thus its proof is omitted.

\bl \lab{oop}
The discounted wealth process $\wt V(\phi ) := B^{-1} V(\phi )$ satisfies  $d\wt V_t (\phi ) =
\sum_{i=1}^d \xi^i_t \, d\wt S^i_ t$ and thus it is a $\PT$-local martingale (or a $\PT$-martingale under suitable integrability assumptions).
\el

% \newpage

Let us now define an arbitrary process of finite variation, say $B^{\gamma },$ such that
$dB^{\gamma}_t = \gamma_t B^{\gamma }_t \, dt$ and $B^\gamma_0>0 $.
It is crucial to stress that it is not postulated that the process $B^{\gamma }$ represents the price of a traded
asset or indeed has anything to do with the market model at hand.
Nevertheless, it still makes sense to work under the following formal postulate.

\bhyp \lab{ahh2}
There exists a probability measure $\P^{\gamma }$ such that the process
$\bar S := (B^{\gamma })^{-1} S$ is a $\P^{\gamma }$-local martingale.
\ehyp

\brem In a typical market model (say, the Black and Scholes model), this assumption will
be satisfied, due to Girsanov's theorem. This does not imply, however, that the process $B^{\gamma }$ has any well defined financial interpretation relative to our market model. When referring to results from \cite{PPB12}, we will sometimes following their convention to refer to $\gamma $ as a fictitious risk-free rate, although this terminology is in fact arbitrary and it does not have any bearing on the validity of results presented below (even worse, it could be misleading when a risk-free rate is already one of the instruments in a market model).
\erem

We now define an auxiliary process $V^{\gamma } (\phi ) $, which can be formally associated to any self-financing trading strategy $\phi $. Obviously, there is no reason to expect that the process $V^{\gamma }_t (\phi )$ represents the wealth of a self-financing strategy, in general.

\bd
Let $\phi $ be a self-financing trading portfolio with the wealth process $V(\phi )$.
Then the process $V^{\gamma } (\phi ) $ is defined by the following equality
\be \lab{gamm1}
V^{\gamma }_t (\phi ) := V_t (\phi ) + B^{\gamma}_t \int_0^t (\gamma_u - r_u) \psi^0_u B_u (B^{\gamma }_u)^{-1} \, du
\ee
or, equivalently,
\be \lab{gamm1x}
V^{\gamma }_t (\phi ) := V_t (\phi ) + B^{\gamma}_t \int_0^t (\gamma_u - r_u)(B^{\gamma }_u)^{-1} \Big( V_u (\phi )
 - \sum_{i=1}^d \xi^i_u S^i_u \Big)  \, du .
\ee
\ed

Unless $\gamma $ coincides with the risk-free rate $r$, so that $V^{\gamma }(\phi )= V(\phi )$, no financial interpretation for the difference  $V^{\gamma }(\phi )- V(\phi )$ can be offered.
Nevertheless, it is possible show that the martingale measure $\P^{\gamma }$ can be used to compute the price, which is
defined by means of replication, if the cash flows of a contract are suitably (but artificially from the financial perspective) transformed to account for the fact that the process $B^{\gamma }$ does not the price of a traded asset in our economy (see equations \eqref{gamm5a} and \eqref{gamm5}).

As a first step towards our goal, we define the $B^{\gamma }$-relative process $\bar V^{\gamma } (\phi )$ by setting $\bar V^{\gamma } (\phi ) = V^{\gamma } (\phi )/B^{\gamma }$. The following proposition shows that, for any self-financing trading strategy $\phi $, the process  $\bar V^{\gamma } (\phi )$ enjoys the martingale property under the probability measure $\P^{\gamma }$. This is a purely mathematical result and thus it would be unjustified to claim that the probability measure $\P^{\gamma }$ can be interpreted as a `risk-neutral probability' (of course, unless $\gamma =r$).

\bl \lab{gamprof}
Let $\phi $ be a self-financing trading strategy and let the process $V^{\gamma } (\phi )$ be given by \eqref{gamm1}.
Then the process $\bar V^{\gamma } (\phi )$ is a $\P^{\gamma }$-local martingale.
\el

\proof
For the sake of brevity, we drop $\phi $ from the notation $V(\phi )$ and $V^\gamma (\phi )$.
It suffices to show that
$
d\bar V^{\gamma }_t = \sum_{i=1}^d \xi^i_t \, d \bar S^i_t
$
or, equivalently,
\be \lab{gamm2}
dV^{\gamma }_t  - \gamma_t V^{\gamma }_t \, dt = \sum_{i=1}^d \xi^i_t \,  \big( dS^i_t - \gamma_t S^i_t \, dt \big).
\ee
By applying the It\^o formula to \eqref{gamm1}, we get
\begin{align*}
dV^{\gamma }_t &= dV_t + (\gamma_t - r_t)  \psi^0_t B_t \, dt +\frac{ V^{\gamma}_t - V_t }{B^{\gamma}_t } \, dB^{\gamma }_t  \\
&= dV_t + (\gamma_t - r_t) \Big( V_t - \sum_{i=1}^d \xi^i_t S^i_t \Big) \, dt + \gamma_t ( V^{\gamma}_t - V_t ) \, dt .
\end{align*}
Therefore, using the self-financing property of $\phi $, we obtain
\begin{align*}
dV^{\gamma }_t  - \gamma_t V^{\gamma }_t \, dt  &= dV_t - r_t V_t \, dt - \sum_{i=1}^d (\gamma_t - r_t) \xi^i_t S^i_t \, dt \\
&= dV_t - r_t \Big( \sum_{i=1}^d \xi^i_t S^i_t + \psi^0_t B_t \Big) \, dt - \sum_{i=1}^d (\gamma_t - r_t) \xi^i_t S^i_t \, dt
\\ &=  dV_t - r_t \psi^0_t B_t \, dt  - \sum_{i=1}^d \gamma_t \xi^i_t S^i_t \, dt
\\ &= \sum_{i=1}^d \xi^i_t \, dS^i_t + \psi^0_t \, dB_t  - r_t \psi^0_t B_t \, dt   - \sum_{i=1}^d \gamma_t \xi^i_t S^i_t \, dt
\\ &= \sum_{i=1}^d \xi^i_t \, \big( dS^i_t - \gamma_t S^i_t \, dt \big)
\end{align*}
as was required to show. \endproof

Of course, if the equality $\gamma = r$ holds, then Lemma  \ref{gamprof} reduces to Lemma \ref{oop}.  Lemma  \ref{gamprof} can
thus be seen as an extension of  Lemma \ref{oop} to the general case when the `virtual discount factor' does not necessarily
correspond to a traded asset. In the final step, we are going to illustrate Theorem 4.3 in \cite{PPB12}. Note, however, that a counterpart of
formula \eqref{gamm5a} was postulated in \cite{PPB12}, whereas it is here derived from fundamentals.

\bhyp \lab{ahh3}
Assume that a contract has a single cash flow $X$ at time $T$ and a replicating self-financing portfolio $\phi $ for $X$ exists.
\ehyp

Under suitable integrability assumption, the discounted wealth process $\wt V(\phi )$ is a $\PT$-martingale and the process $\bar V^{\gamma } (\phi )$ is a $\P^{\gamma }$-martingale. Consequently, the classic arbitrage price of a contingent claim $X$ can be computed using the standard version of the risk-neutral valuation formula under $\PT$, specifically,
\be \lab{gamm4}
\pi_t (X) = - B_t \, \EPT ( X B_T^{-1} \, | \, \F_t )
\ee
where the minus sign is due to our convention regarding the financial interpretation of $X$.

Let us now take any process $B^{\gamma }$ such that the probability measure $\P^{\gamma }$ is well defined. From Proposition  \ref{gamprof}, we deduce the following corollary showing that  $\P^{\gamma }$ can also be used
as a `pricing measure' after a suitable transformation of cash flows. Note that the additional integral term that appears
in  \eqref{gamm5a}  (or \eqref{gamm5}) has no financial interpretation, since $\gamma $ is here an arbitrary process
unrelated to any financial quantity.

\bcor
If Assumptions \ref{ahh1}-\ref{ahh3} are satisfied, then the price $\pi_t(X) = V_t(\phi )$ is also given by the following expression
\be  \lab{gamm5a}
\pi_t (X) = -  B^{\gamma}_t \, \EPG \Big( X(B^{\gamma }_T)^{-1}
+  \int_t^T (r_u - \gamma_u ) \psi^0_u B_u (B^{\gamma }_u)^{-1} \, du  \, \Big| \, \F_t \Big).
\ee
Since, for every $0 \leq t < T$,
\bde
\psi^0_t B_t =  V_t (\phi ) - \sum_{i=1}^d \xi^i_t S^i_t = \pi_t (X) - \sum_{i=1}^d \xi^i_t S^i_t ,
\ede
equality \eqref{gamm5a} may also be rewritten as follows
\be \lab{gamm5}
\pi_t (X) = - B^{\gamma}_t \, \EPG \Big( X  (B^{\gamma }_T)^{-1}
+ \int_t^T (r_u - \gamma_u ) (B^{\gamma }_u)^{-1} \Big(\pi_u (X) - \sum_{i=1}^d \xi^i_u S^i_u \Big)  \, du  \, \Big| \, \F_t \Big).
\ee
\ecor

\proof
It suffices to show that the right-hand side in  \eqref{gamm5a} coincides with $V(\phi )$ where a strategy $\phi $ replicates $X$, in the sense of Definition \ref{def:replicate}.
The martingale property of $\bar V^{\gamma } (\phi )$ under $\P^{\gamma }$ means that, for all $t \in [0,T]$,
\be \lab{uu8}
\bar V^{\gamma }_t (\phi ) = \EPG \big( \bar V^{\gamma }_{T-} (\phi ) \, | \, \F_t \big).
\ee
In view of \eqref{gamm1} and the equality $V_{T-}(\phi )=- X$, equality \eqref{uu8} implies that
\begin{align*}
V_t (\phi )(B^{\gamma}_t)^{-1} &+ \int_0^t (\gamma_u - r_u) \psi^0_u B_u (B^{\gamma }_u)^{-1} \, du \\
&=\EPG \Big( - X (B^{\gamma}_T)^{-1} + \int_0^T (\gamma_u - r_u) \psi^0_u B_u (B^{\gamma }_u)^{-1} \, du \, \Big| \, \F_t \Big).
\end{align*}
This immediately yields the asserted formula.
\endproof

As was already mentioned, a version of formula \eqref{gamm5} was postulated in  \cite{PPB12} as an ad hoc valuation recipe under funding costs (see the first formula in Section 4.5.1 in \cite{PPB12}). We contend that the arguments put forward in \cite{PPB12} in support of their approach, although they may yield correct valuation results, are too contrived and may require some guesswork (or, at least, tedious computations) leading to a suitable adjustment of cash flows. It is definitely more natural to start with a market model in which an `instrumental variable' (e.g., a non-traded risk-free short-term rate) is not introduced and all processes that are modeled have a clear financial interpretation.  To conclude, formula \eqref{gamm4} is much easier to establish and implement than \eqref{gamm5}, so there is no practical advantage of using the latter representation even for the purpose of numerical computations.

\vskip 10 pt

%%%%%%%%%%%%%%%%%%%%%%%%%%%%%%%%%%%%%%%%%%%%%%%%%%%%%%%%%%%%%%%%%%%%%%%%%%%%%

\noindent {\bf Acknowledgements.}
The research of Tomasz R. Bielecki was supported under NSF grant DMS-1211256.
The research of Marek Rutkowski was supported under Australian Research Council's
Discovery Projects funding scheme (DP120100895). The authors express their gratitude to Damiano Brigo, 
St\'ephane Cr\'epey, Ivan Guo, Tianyang Nie and Desmond Ng  for valuable discussions.

%%%%%%%%%%%%%%%%%%%%%%%%%%%%%%%%%%%%%%%%%%%%%%%%%%%%%%%%%%%%%%%%%%%%%%%%%%%%%

%%%%%%%%%%%%%%%%%%%%%%%%%%%%%%%%%%%%%%%%%%%%%%%%%%%%%%%%%%%%%%%%%%%%%%%%%%%%%%%%%%%%%%%%%%%%%%%%%%%%
%%%%%%%%%%%%%%%%%%%%%%%%%%%%%%%%%%%%%%     REFERENCES      %%%%%%%%%%%%%%%%%%%%%%%%%%%%%%%%%%%%%%%%%
%%%%%%%%%%%%%%%%%%%%%%%%%%%%%%%%%%%%%%%%%%%%%%%%%%%%%%%%%%%%%%%%%%%%%%%%%%%%%%%%%%%%%%%%%%%%%%%%%%%%

\end{document}